\documentclass[twocolumn,prd,aps,superscriptaddress,preprintnumbers,tightenlines,showpacs,nofootinbib,eqsecnum,amsfonts,amsmath,longbibliography]{revtex4-2}
\usepackage{placeins}
\usepackage{amsmath}
\numberwithin{equation}{section}
\usepackage{mathrsfs}
\usepackage{slashed}
\usepackage{amsfonts}
\usepackage{txfonts}
\usepackage{mathrsfs}
\usepackage{bbding}
\usepackage{graphicx}
\usepackage{color}
\usepackage[colorlinks=true, linkcolor=blue, citecolor=cyan]{hyperref}
\newcommand{\bee}{{\bf{e}}}
\newcommand{\be}{\begin{equation}}
\newcommand{\ee}{\end{equation}}
\newcommand{\dd}{\text{d}}
\usepackage{amsmath}
\graphicspath{{./plots/}}

\usepackage{xcolor}
\colorlet{BLUE}{blue}
\newcommand{\beq}{\begin{equation}}
\newcommand{\eeq}{\end{equation}}
\newcommand{\bea}{\begin{eqnarray}}
\newcommand{\eea}{\end{eqnarray}}

\newcommand{\Gal}{{\rm G}}

\newcommand{\nn}{{\nonumber}}

\usepackage[stable]{footmisc}

\newcommand\jcap{{\it JCAP}}

\begin{document}



\title{Measurement of the Cross-Correlation Angular Power Spectrum Between the Stochastic Gravitational Wave Background and Galaxy Over-Density}

\author{Kate Z.Yang}
\email{yang5991@umn.edu}
\affiliation{University of Minnesota, School of Physics and Astronomy, Minneapolis, MN 55455, USA}
\author{Jishnu Suresh}
\affiliation{Centre for Cosmology, Particle Physics and Phenomenology (CP3), Universit\'e Catholique de Louvain, Louvain-la-Neuve, B-1348, Belgium}
\author{Giulia Cusin}
\affiliation{Sorbonne Université, CNRS, UMR 7095, Institut d'Astrophysique de Paris, 75014 Paris, France}
\affiliation{Universit\'e de Gen\'eve, D\'epartement de Physique Th\'eorique and Centre for Astroparticle Physics, 24 quai Ernest-Ansermet, CH-1211 Gen\'eve 4, Switzerland}
\author{Sharan Banagiri}
\affiliation{Center for Interdisciplinary Exploration and Research in Astrophysics, Northwestern University, 1800 Sherman Ave, Evanston, IL 60201, USA }
\affiliation{University of Minnesota, School of Physics and Astronomy, Minneapolis, MN 55455, USA}
\author{Noelle Feist}
\affiliation{University of Minnesota, School of Physics and Astronomy, Minneapolis, MN 55455, USA}
\author{Vuk Mandic}
\affiliation{University of Minnesota, School of Physics and Astronomy, Minneapolis, MN 55455, USA}
\author{Claudia Scarlata}
\affiliation{University of Minnesota, School of Physics and Astronomy, Minneapolis, MN 55455, USA}
\author{Ioannis Michaloliakos}
\affiliation{Department of Physics, University of Florida, Gainesville, Florida 32611, USA}

	\begin{abstract}
	    We study the cross-correlation between the stochastic gravitational-wave background (SGWB) generated by binary black hole (BBH) mergers across the universe and the distribution of galaxies across the sky. We use the anisotropic SGWB measurement obtained using data from the third observing run (O3) of Advanced LIGO detectors and galaxy over-density obtained from the Sloan Digital Sky Survey (SDSS) spectroscopic catalog. We compute, for the first time, the angular power spectrum of their cross-correlation. Instead of integrating the SGWB across frequencies, we analyze the cross-correlation in 10 Hz wide SGWB frequency bands to study the frequency dependence of the cross-correlation angular power spectrum. Finally, we compare the observed cross-correlation to the spectra predicted by astrophysical models. We apply a Bayesian formalism to explore the parameter space of the theoretical models, and we set constraints on a set of (effective) astrophysical parameters describing the galactic process of gravitational wave (GW) emission. Parameterizing with a Gaussian function the astrophysical kernel describing the local process of GW emission at galactic scales, we find the 95\% upper limit on kernel amplitude to be $2.7 \times 10^{-32}$ erg cm$^{-3}$s$^{-1/3}$ when ignoring the shot noise in the GW emission process, and $2.16 \times 10^{-32}$ erg cm$^{-3}$s$^{-1/3}$ when the shot noise is included in the analysis. As the sensitivity of the LIGO-Virgo-KAGRA network improves, we expect to be able to set more stringent bounds on this kernel function and constrain its parameters. 
	\end{abstract}
	
	\maketitle
	\section{Introduction}
	%
The first three observing runs of Advanced LIGO \cite{aligo}, Advanced Virgo \cite{avirgo}, and KAGRA \cite{KAGRA:2020agh} gravitational wave (GW) detectors have resulted in detections of nearly a hundred mergers of compact binary systems \cite{GWTC3}: binary black holes (BBH), binary neutron stars (BNS), and binary systems composed of one neutron star and one black hole (NSBH). These discoveries have enabled a series of investigations including measurements of the rate and distributions of these binary systems \cite{O3ratepop}, tests of General Relativity \cite{O3TGR}, independent measurements of the Hubble constant \cite{O3H0}, tests of the neutron star equation of state \cite{GW170817_EOS}, and others. This trend is expected to continue in the upcoming observation runs O4 and O5 \cite{LVKlivingreview} of LIGO Scientific, Virgo, and KAGRA Collaboration. 
	
One of the prime targets of the upcoming observing runs will be the stochastic gravitational wave background (SGWB), which arises as a superposition of uncorrelated signals of many different GW sources \cite{maggiore,regimbau_review}. The SGWB is expected to include contributions from many different production processes in the early Universe, including models of amplification of primordial tensor vacuum fluctuations~\cite{grishchuk,barkana,starob,turner}, inflationary models that include back-reaction of gauge fields~\cite{peloso_parviol,seto}, parametric resonances in the preheating stage following inflation~\cite{eastherlim}, models of additional "stiff'' energy components in the early universe~\cite{boylebuonanno}, phase transitions models \cite{witten,hogan,kosowsky,caprini1,binetruy,caprini2,margotpaper}, and cosmic (super)strings models \cite{caldwellallen,DV1,DV2,cosmstrpaper,olmez1,polchinski,siemens,ringeval,olum,O3cosmstr,jenkins_cosmstr}. On the other hand, the SGWB of astrophysical origin is given by the superposition of GW signals emitted by different populations of astrophysical sources, from the onset of stellar activity until today. In the frequency band of current Earth-based observatories, sources include BBH and BNS systems
\cite{regfrei,zhu_cbc,marassi_cbc,rosado,regman,wu_cbc,GW150914stoch,GW170817stoch}, rotating neutron stars~\cite{cutler,bonazzola,marassi_magnetar,owen,regman,wu_mag},
and supernovae~\cite{regimbau_review,SNe,marassi_cc, firststars,buonanno_cc,crocker1,crocker2,finkel}. Both cosmological and astrophysical SGWB components are expected to be anisotropic. A primary source of anisotropy in the received flux is due to the anisotropic distribution of emitting sources and the anisotropic emission process. A secondary source of anisotropy is due to propagation: even if a given SGWB component is isotropic at the time of emission, anisotropies are created due to the fact that GW signals propagate in a Universe where cosmic structures are present, hence they feel the effects of the gravitational potential of matter structures (in the form of lensing, time-delay, integrated time delay, see e.g. \cite{Contaldi:2016koz, Cusin:2017fwz, Cusin:2017mjm}). We also note that kinematic anisotropies are expected, due to the relative motion of our rest frame on Earth with respect to the emission rest frame \cite{ Cusin:2022cbb, Chung:2022xhv}. 

This raises a distinct possibility that the SGWB energy density is correlated with the anisotropy in other (electromagnetic) observables, such as galaxy counts (GC), gravitational (weak) lensing, cosmic microwave background, cosmic infrared background, and others. Measurements of these correlations would provide new ways to study the distribution of matter in our Universe, and its evolution. 

Typical cosmological background components are expected to have the same level of anisotropy as the CMB: a scale-invariant spectrum $\ell(\ell+1) C_{\ell}\propto {\rm constant}$, and a level of anisotropy of the order of $\sim10^{-5}$ with respect to the monopole \cite{Geller:2018mwu}. 
In contrast, extra-galactic astrophysical SGWBs have a scaling given by clustering, resulting in $(\ell+1) C_{\ell}\propto {\rm constant}$, and a higher level of anisotropy, at the level of $\sim10^{-2}$ with respect to the monopole \cite{Cusin:2017mjm, Cusin:2017fwz, Cusin:2018rsq, Cusin:2019jpv, Jenkins:2018uac, Jenkins:2018kxc, Jenkins:2018lvb, Cusin:2019jhg, Jenkins:2019uzp, Jenkins:2019nks, Cusin:2018avf, Pitrou:2019rjz, Alonso:2020mva}.

While most of the cosmological SGWB components are expected to be stationary and continuous over the observation time (hence representing \emph{irreducible} background components), the astrophysical background in the frequency band of ground-based detectors is expected to have \emph{popcorn-like nature} due to the discreteness of emissions in time. 
As a consequence, the angular power spectrum of the SGWB from mergers of compact binaries has an important Poisson shot-noise component, which adds to the clustering one \cite{Cusin:2019jpv, Jenkins:2019uzp, Jenkins:2019nks, Alonso:2020mva}. Formally, the total SGWB angular power spectrum is given by $C_{\ell}^{\rm tot}=C_{\ell}+N^{\rm shot}$,
where the first term on the right hand side is the contribution from clustering while the second component represents shot noise. This latter is flat in $\ell$-space (it is just an offset) and it is expected to dominate over the clustering contribution, see \cite{Cusin:2019jpv, Jenkins:2019uzp, Jenkins:2019nks, Alonso:2020mva}. Even if shot noise contains astrophysical information, it does not provide any information about the spatial distribution of sources. A possible way to overcome this problem, i.e. to separate the clustering part from the shot noise, is to consider cross-correlations between a SGWB map and electromagnetic tracers of structure, such as galaxy distributions~\cite{Alonso:2020mva}. 
In addition to serving as independent observables of structure in the universe, cross-correlations provide one with powerful SGWB anisotropy detection tools, as they typically have a higher signal-to-noise ratio (SNR) than the SGWB auto-correlation---see e.g. \cite{Cusin:2018rsq, Alonso:2020mva, Cusin:2019jpv,Yang:2020usq, Capurri:2021zli} in the context of the extra-galactic astrophysical background.
We are aware that cross-correlating EM tracers with individual events of compact binary coalescence is also used, such as in the calculation of the Hubble constant $H_0$ \cite{Mukherjee:2022afz}.

In this paper, we focus on correlations between the SGWB (as measured in the recent observing runs of Advanced LIGO, Advanced Virgo, and KAGRA) and the distribution of galaxies across the sky (from SDSS). 
We assume that the dominant background components in the $\sim 100$ Hz band is coming from mergers of extragalactic compact objects, and we use the astrophysical model of \cite{Cusin:2018rsq, Cusin:2019jpv, Cusin:2019jhg} to describe the galactic process of GW emission. We compute the corresponding angular power spectrum of the cross-correlation, and we compare it with the angular power spectrum extracted from data. Our final goal is to perform a parameter estimation: we introduce an effective parameterization for the astrophysical model describing GW production and propagation, and we study the constraints that can be set on these effective model parameters from a comparison with data. We stress that the methods developed here can also be applied to cross-correlations between SGWB and other electromagnetic tracers of structure in the universe. 

The paper is structured as follows. 
In Section \ref{sec:Model} we will review the model predictions for the angular power spectrum of the cross-correlation between SGWB and the galaxy counts distribution. In Section \ref{sec:GWdata}, we will present the frequency-dependent anisotropic SGWB search results using the latest data from terrestrial GW detectors. In Section \ref{sec:Galaxy} we review galaxy catalog that will be used in our study, namely from the Sloan Digital Sky Survey. In Section \ref{sec:Cross-corr} we present the measured SGWB-GC angular power spectra. In Section \ref{sec:PE} we use the measured angular power spectra to make estimates of model parameters introduced in Section \ref{sec:Model}. A discussion and our final remarks are presented in Section \ref{sec:Conclusion}.

\section{Modeling SGWB-Galaxy Count Angular Power Spectra}\label{sec:Model}
\subsection{Astrophysical models of angular power spectra}
\begin{figure}[ht]
	        \centering
            \includegraphics[width=\columnwidth]{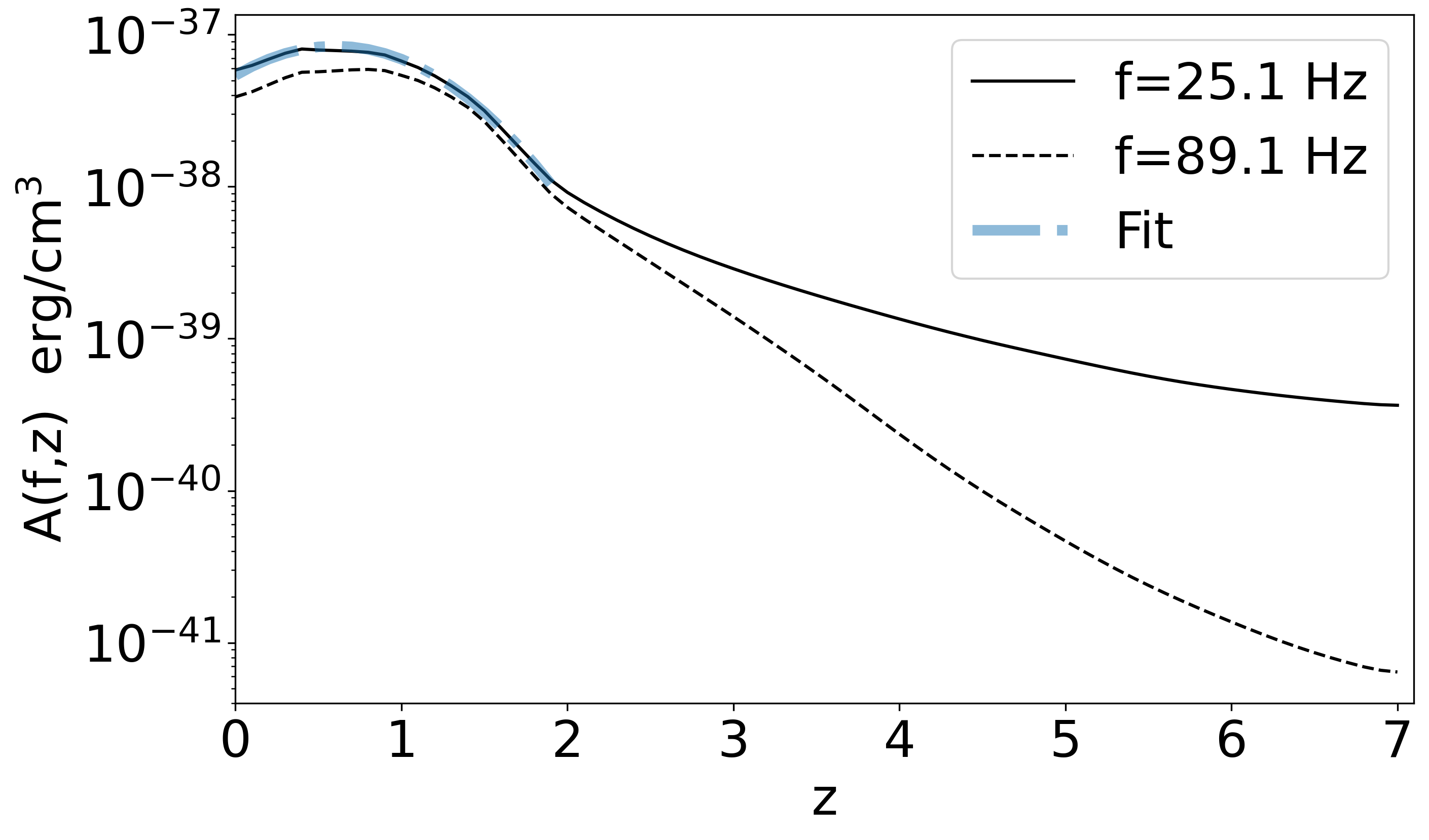}
	        \caption{Astrophysical kernel for the astrophysical model used as a reference in \cite{Cusin:2019jpv}, as function of redshift and for frequencies 25.1 and 89.1 Hz and angular power spectrum, with a power-law-Gaussian fit of $A_{\rm max}=4\times 10^{-37}$erg cm$^{-3}$s$^{-1/3},z_c=0.6,\sigma_z=0.9$.} 
	    \label{fig:KernelCl}
	\end{figure}

	The observed GW energy density parameter, $\Omega_{\text{GW}}$ is defined as the background energy density per units of logarithmic frequency $f$ and solid angle $\bf{e}$, normalized by the critical density of the Universe today $\rho_c$. It can be divided into an isotropic background contribution $\bar{\Omega}_{\text{GW}}$ and a contribution from anisotropic perturbations $\delta \Omega_{\text{GW}}$ \cite{Cusin:2017fwz, Cusin:2017mjm, Cusin:2019jpv}:
	\begin{align}
	    \Omega_{\text{GW}}({\bf{e}},f)=\frac{f}{\rho_c}\frac{\dd^3 \rho_{\text{GW}}}{\dd^2 {\bf{e}}df}({\bf{e}},f)=\frac{\bar{\Omega}_{\text{GW}}(f)}{4\pi}+\delta \Omega_{\text{GW}}({\bf{e}},f),
	\end{align}
	where the background power can be written as the integral over conformal distance $r$ (we assume speed of light $c$=1):
	\begin{align}\label{BackandPert}
	    &\bar{\Omega}_{\text{GW}}(f) = \int \dd r \, \partial_r \bar{\Omega}_{\text{GW}}(f,r),\\
	    &\partial_r \bar{\Omega}_{\text{GW}}(f,r)= \frac{f}{\rho_c}\,\mathcal{A}(f,r)\,,
	\end{align}
	and the function $\mathcal{A}(f,r)$ is an astrophysical kernel that contains information on the local production of GWs at galaxy scales. Schematically this kernel can be parameterized as \cite{Alonso:2020mva}
	\be
	\mathcal{A}(f,r)=\frac{a^4}{4\pi} \int \dd \mathcal{L}_{\text{GW}}\, \bar{n}_{G}(\mathcal{L}_{\text{GW}}, r) \mathcal{L}_{\text{GW}}\,,
	\ee
	where $a$ is the universe scale factor and $\bar{n}_{G}$ is the average physical number density of galaxies at distance $r$ with gravitational wave luminosity $\mathcal{L}_{\text{GW}}$. 
	Different astrophysical models give quite different predictions for this kernel, see e.g. \cite{Cusin:2019jpv} for an explorative approach. 
	 For the SGWB due to mergers of compact objects such as BBH and BNS, the low frequency band ($f \lesssim 100$ Hz) is dominated by the inspiral phase contributions and follows a simple power law $\Omega_{\rm GW} \sim f^{2/3}$.
    Looking at predictions of different astrophysical models, see e.g. \cite{Cusin:2019jhg, Cusin:2019jpv}, one can recognize some common features in the redshift dependence of the kernel which in first approximation can be captured by the following Gaussian parameterization 
	\begin{align}\label{eq:A_z_f}
	    \mathcal{A}(f, z)=\mathcal{A}(f) \,e^{-(z-z_c)^2/2\sigma_z^2}=A_{\text{max}} \, f^{-1/3} \, e^{-(z-z_c)^2/2\sigma_z^2}\,,
	\end{align}
 where we used $z=z(r)$ to express the astrophysical kernel as a function of redshift and frequency. In FIG. \ref{fig:KernelCl} we present the astrophysical kernel as a function of redshift for several representative frequencies in the band of terrestrial GW detectors~\cite{Cusin:2019jpv}.
When making use of the parameterization in Eq. \eqref{eq:A_z_f}, we have three parameters in total $\theta = (A_{\text{max}}, z_c, \sigma_z)$: a constant value $A_{\text{max}}$ setting the amplitude of the kernel, redshift peak $z_c$, and peak width $\sigma_z$. 
Typically the peak of the astrophysical kernel follows the peak of star formation rate (i.e. $z_c\lesssim 1$) and the width $\sigma_z$ depends on the astrophysical model chosen and it is typically of the order $\sigma_z\sim 0.5$, see \cite{Cusin:2019jpv} for details.
As we can see in FIG. \ref{fig:KernelCl}, the Gaussian approximation is valid for redshifts between 0 to 2. This applies to our analysis below which extends up to $z=0.8$. 

Since $\delta\Omega_{\text{GW}}$ is a stochastic quantity, it can correlate with other cosmological stochastic observables. An interesting observable to look at is the cross-correlation of the SGWB with the distribution of galaxies, i.e. with the galaxy number counts $\Delta$ defined as the overdensity of the number of galaxies per unit of redshift and solid angle
\be\label{Ncounts}
\Delta(\bee, z)\equiv \frac{N(z, \bee)-\bar{N}(z)}{\bar{N}(z)}\,.
\ee
First, if astrophysical GW sources are located in galaxies, we would expect the SGWB and the galaxy distribution to have a high correlation level. Second, cross-correlating with galaxies helps to mitigate the problem of shot noise and to possibly extract the clustering information out of the shot noise threshold \cite{Alonso:2020mva, Cusin:2019jpv, Jenkins:2019uzp, Jenkins:2019nks}. Finally, by cross-correlating with the galaxy distribution at different redshifts, one could try to get a tomographic reconstruction of the redshift distribution of sources. In this article, to maximize the SNR of the cross-correlation we do not bin the galaxy distribution in redshift, but we rather integrate the number counts (\ref{Ncounts}) over the redshift range covered by our catalog.

The angular power spectrum of the GW and galaxy counts auto-correlations and for their cross-correlations are defined as
\begin{align}\label{ClGW}
(2\ell+1)\,C^{\text{GW}}_{\ell}(f; \theta)&\equiv \sum_{m=-\ell}^{\ell} \langle a_{\ell m}(f; \theta)\,a^*_{\ell m}(f; \theta)\rangle\,,\\
(2\ell+1)\,C^{\text{GC}}_{\ell}&\equiv \sum_{m=-\ell}^{\ell} \ \langle b_{\ell m} \,b^*_{\ell m}\rangle\,,\\
(2\ell+1)\,C^{\text{cross}}_{\ell}(f; \theta)&\equiv \sum_{m=-\ell}^{\ell} \ \langle a_{\ell m}(f; \theta)\,b^*_{\ell m}\rangle\,,
\end{align}
where the bracket denotes an ensemble average and $a_{\ell m}(f)$ and $b_{\ell m}$ are the coefficients of the spherical harmonics decomposition of the SGWB energy density and galaxy number counts, respectively. Explicitly 
\begin{eqnarray}
\label{Eq:Sec2Sph}
   \delta \Omega_{\text{GW}}({\bf{e}},f; \theta) & = & \sum_{\ell=0}^{\infty} \sum_{m=-\ell}^{\ell} a_{\ell m}(f; \theta) \, Y_{\ell m}({\bf{e}})\,,\nn\\
    \Delta({\bf{e}}) & = & \sum_{\ell=0}^{\infty} \sum_{m=-\ell}^{\ell} b_{\ell m} \, Y_{\ell m}({\bf{e}})\,.
\end{eqnarray}

It can be shown that the angular power spectra of the auto- and cross-correlation are given by \cite{Cusin:2017fwz}: 
	\begin{align}\label{eq:modelCl}
 	    C_{\ell}^{\text{GW}}(f; \theta)&= \frac{2}{\pi}\int \dd k\ k^2 \, |\delta \Omega_{\text{GW}\,,{\ell}}(k,f; \theta)|^2\,,\\
      	    C_{\ell}^{\text{GC}}&= \frac{2}{\pi}\int \dd k\ k^2 \, |\Delta_{\ell}(k)|^2\,,\\
	    C_{\ell}^{\text{cross}}(f; \theta)&= \frac{2}{\pi}\int \dd k\ k^2 \, \delta \Omega^{*}_{\text{GW}\,,{\ell}}(k,f; \theta) \, \Delta_{\ell}(k)\,,
	\end{align}
	where $k$ is the wave-number. Keeping only the leading-order contribution to the anisotropy given by clustering (neglecting line of sight effects), we have 
		\begin{align}\label{eq:deltaOmegaGW}
	    \delta \Omega_{\text{GW}\,,\ell}(k,f; \theta)=\frac{f}{4\pi \rho_c} \int \dd r \, \mathcal{A}(r,f; \theta) \, \big[b(r)\,\delta_{m,k}(r)j_{\ell}(k r) \big]\,,
	\end{align}
where $j_{\ell}$ are spherical Bessel functions, while $\delta_{m}$ is the dark-matter over-density, related to galaxy overdensity via the bias factor that we assume to be scale-independent and with redshift evolution given by $b(z)=b_0 \sqrt{1+z}$ and $b_0=1.5$ \cite{WiggleZ:2013kor, Rassat:2008ja}. The corresponding contribution from galaxy overdensities reads 
		\begin{align}\label{eq:DeltaGC}
	    \Delta_{\ell}(k)=\int \dd r W(r)\, \big[b(r)\,\delta_{m,k}(r)j_{\ell}(k r) \big]\,, 
	\end{align}
	where $W(r)$ is a window function normalized to one which selects the redshift bin in the galaxy catalog we want to consider in the cross-correlations. As already mentioned, in our analysis we do not bin in redshift in order to maximize the SNR, hence the function $W(r)$ extends to the entire redshift range of the galaxy catalog. 	
	%
	\subsection{Shot Noise}\label{subsec:shotnoise}
	
	Up to now, in our description of GW sources we implicitly introduced two assumptions: we assumed that astrophysical sources are located in galaxies, distributed in space as a continuous field, and we assumed that the GW emission is continuous and stationary over the observation time. However, when considering the SGWB due to BBH mergers, the realization of the BBH mergers during the observation period is subject to Poisson (shot, or popcorn) noise in both space and time \cite{Jenkins:2019uzp,Alonso:2020mva}. This shot noise introduces additional angular structure in both the SGWB and the galaxy distribution, and therefore has to be accounted for in both the prediction of $C_\ell$'s (GW, GC, and cross) and in their covariance matrices. As shown in \cite{Alonso:2020mva}, the shot noise contribution to the cross-correlation angular power spectrum is independent of $\ell$ but still dependent on astrophysical parameters $\theta = (A_{\text{max}}, z_c, \sigma_z)$. That is, the shot noise offsets the clustering values given in Eq.\,(\ref{eq:modelCl}),
	\be\label{Eq:Cshot}
	C_{\ell}^{{\text{cross}}, \text{tot}}(\theta) = C_{\ell}^{\text{cross}}(\theta) + N_{\rm shot}^{\text{cross}}(\theta)\,.
	\ee
	
	Hence, while shot noise may (partly) mask the clustering contribution, it still carries astrophysical information that can be measured. Further, as discussed in detail in \cite{Alonso:2020mva}, the shot noise associated with the cross-correlation is much smaller than the one associated with the SGWB auto-correlation, which is why cross-correlating is a very promising method to get a first detection of the SGWB anisotropy. Indeed, assuming that shot noise is the only noise component (i.e. considering a perfect instrument with infinite sensitivity) one has that the signal-to-noise ratio (SNR) of the cross-correlation scales as \cite{Alonso:2020mva}
	\be\label{SNR}
	\left(\frac{S}{N}\right)^2_{\text{cross}}=\sum_{\ell}\frac{(2\ell+1) C_{\ell}^{\text{cross}}}{(C_{\ell}^{\text{cross}}+N_{\rm shot}^{\text{cross}})^2+(C_{\ell}^{\text{GW}}+N_{\rm shot}^{\text{GW}})(C_{\ell}^{\text{GC}}+N_{\rm shot}^{\text{GC}})}\,,
	\ee
	where $C_{\ell}^{\text{GW}}$ and $N_{\rm shot}^{\text{GW}}$ denote the angular power spectrum and the shot noise of the SGWB map, respectively, while $C_{\ell}^{\text{GC}}$ and $N_{\rm shot}^{\text{GC}}$ are the angular power spectrum and shot noise of the galaxy map (we have suppressed their dependencies on parameters $\theta$). 
	The three noise contributions are given by~\cite{Alonso:2020mva}
		\begin{align}
        N_{\rm shot}^{\text{GW}}(\theta)&=\left(1+\frac{1}{\beta_T}\right)\left(\frac{f}{4\pi \rho_c}\right)^2\int\frac{\dd r}{r^2} \frac{1}{a^3\bar{n}_\Gal} \mathcal{A}^2(r, f; \theta)\,,\label{popGW}\\
        N_{\rm shot}^{\text{cross}}(\theta)&=\frac{f}{4\pi \rho_c}\int\frac{\dd r}{r^2}\frac{1}{a^3\bar{n}_\Gal} W(r) \, \mathcal{A}(r, f; \theta)\,,\\
        N_{\rm shot}^{\text{GC}}&=\int\frac{\dd r}{r^2}\frac{1}{a^3\bar{n}_\Gal} W^2(r)\,,
    \end{align}
    where $a^{3} \bar{n}_{\Gal}$ is the comoving number density of galaxies and we defined 
    \begin{align}\label{beta}
        \beta_T\equiv\frac{T}{a^3 \bar{n}_{\Gal}} \frac{\dd^2 \mathcal{N}}{\dd t \, \dd V}\,,
    \end{align}
        where $T$ is the observation time and $d^2\mathcal{N}/dVdt$ denotes the local merger rate. To get these expressions we have assumed a monochromatic GW luminosity function and that all galaxies emit GW.

      To get an estimate for the prefactor (\ref{beta}), we can use the observed local rate of BBH mergers, 
      $\dd^2\mathcal{N}/\dd V \dd t\sim 100\;\text{Gpc}^{-3} \text{yr}^{-1}$. This estimate for the merger rate, which neglects the contribution of BNS mergers, provides a lower bound for the total merger rate in the $\sim 100$ Hz band, and hence leads to a conservative estimate for the GW shot noise. We also assume a constant comoving galaxy density $a^{3} \bar{n}_{\Gal}\sim 0.1$ Mpc$^{-3}$. For LIGO-Virgo-KAGRA O3, the observation time period $T \sim 1$ yr, so finally $\beta_T\sim 10^{-6}$. This leads to a large prefactor $\propto \beta_T^{-1}$ when evaluating the shot noise for the GW map (Eq. \ref{popGW}), much larger than the ones of cross-correlation and of galaxies alone. Since the denominator of (Eq. \ref{SNR}) scales linearly with the GW shot noise $N_{\rm shot}^{\text{GW}}$ (as opposed to scaling quadratically in the SNR of SGWB auto-correlation), the SNR of the cross-correlation is typically much larger than the one of the SGWB auto-correlation (see \cite{Alonso:2020mva} for a detailed analysis). 

\section{Measurement of SGWB Angular Power Spectra}\label{sec:GWdata}
In this section, we review how the SGWB anisotropy is measured using GW data. We use the publicly available folded data set~\cite{folding,folded_data} from the third observing run (O3) of Advanced LIGO detectors located in Hanford, WA and Livingston, LA. In order to capture the frequency dependence of the model presented in Section \ref{sec:Model}, we analyze the data in 10 Hz frequency bands and build an unbiased estimator of the SGWB angular power spectrum. 

\subsection{Basic concepts: dirty and clean maps}\label{subsec:GWbasic}

From an observational point of view, a SGWB is typically estimated by cross-correlating the output of two different detectors
located at two different points on Earth and assuming that the noise and the noise-signal in the two detectors are not correlated. 

Assuming that the SGWB is unpolarized, Gaussian, and stationary, the quadratic expectation value of the GW strain $h_{A} (f,\bf{e})$ across different sky positions and frequencies can be expressed as
	\begin{align}
	\label{DATA:expect_value}
	    \langle h_{A}^{*} (f,{\bf{e}}) \, h_{A'} (f',{\bf{e'}})\rangle = \frac{1}{4}\, P(f,{\bf{e}})\,\delta_{AA'}\,\delta{(f-f')}\,\delta{({\bf{e}}, {\bf{e'}})}\,,
	\end{align}
where $A$ denotes the GW polarization and $P(f,{\bf{e}})$ encodes the contribution from all parts of the sky and frequency to the total SGWB. 
Given these assumptions, one can express the anisotropy of the SGWB as
	\begin{align}
	\label{DATA:Omega_SGWB}
	    \Omega_{\text{GW}}(f, {\bf{e}})=\frac{f}{\rho_c}\frac{\dd^3 \rho_{\text{GW}}}{\dd f d^2 \bf{e}}=\frac{2\pi^2 f^3}{3H_0^2}\, P(f,{\bf{e}}) \, ,
	\end{align}
where $H_0$ is the Hubble constant taken to be $H_{0} = 67.9 \, \text{km}\, \text{s}^{-1}\, \text{Mpc}^{-1}$. 
In what follows, we further assume that $\Omega_{\text{GW}}$ can be factorized into frequency and direction-dependent terms by separating $P(f,{\bf{e}})$ as~\cite{Thrane:2009fp,romanocornish}\footnote{The factorization does not amount to a loss of generality when conducting stochastic search analysis in small frequency bands, as we expect the signal to have a smooth power spectral profile.}
	\begin{align}
	\label{DATA:aniso_factorize}
	    P(f,{\bf{e}}) = P({\bf{e}})\, H(f)\,,
	\end{align}
In our analysis, we model the spectral dependence $H(f)$ as a power law,
	\begin{align}
	    \label{DATA:power_law} 
	    H(f) =  \Big(\frac{f}{f_{\text{ref}}}\Big)^{\alpha-3}\,,
	\end{align}
where $\alpha$ is the spectral index and $f_{\text{ref}}$ denotes a reference frequency. Throughout this analysis, we set the reference frequency to 25 Hz and choose the power-law index $\alpha=2/3$ as predicted for a compact binary coalescence SGWB. 
The angular distribution $P({\bf{e}})$ can be expanded in terms of any set of basis functions defined on the two-sphere. The choice of this basis will not affect the physical search results. However, to reduce the computational burden and ease the interpretation of the results, one usually chooses either pixel or spherical harmonic basis for the analysis, depending on the sky distribution of sources. A spherical harmonic (SpH) basis is suitable for searching for a diffuse background considered in this work. In SpH basis, one can expand the anisotropy map over the basis functions $Y_{\ell m}$ as
	\begin{align}
	    \label{DATA:SHD}
	    P({\bf{e}})=\sum_{\ell=0}^{\ell_{\rm max}} \sum_{m=-\ell}^{\ell}  P_{\ell m}\,Y_{\ell m}(\bf{e})\,.
	\end{align}
We will discuss the choice of $\ell_{\rm max}$ below. 
Following the maximum-likelihood (ML) method for mapping the GW anisotropy~\cite{Thrane:2009fp,Mitra:2007mc}, a standard ML solution for $P({\bf{e}})$ in SpH basis can be written as (in the limit of low signal-to-noise ratio)
    \begin{align}
       \label{DATA:MLsolution} 
       \hat{P}_{\ell m} = \sum_{\ell^{'},m'} \left(\Gamma ^{-1}\right)_{\ell m,\ell^{'} m'} \, \hat{X}_{\ell^{'} m'} \, ,
    \end{align}
where 
    \begin{align}
    \label{DATA:DirtyMap}
        \hat{X}_{\ell m} = \sum_t \sum_f \gamma_{\ell m}(f,t)\, \frac{H(f)}{P_1 (f,t) \, P_2 (f,t)}\, \hat{C}(f,t) \,,
    \end{align}
    \begin{align}
    \label{DATA:FisherMatrix}
        \Gamma_{\ell m,\ell^{'}m'} = \sum_t \sum_f \gamma^{*} _{\ell m}(f,t)\, \frac{H^{2}(f)}{P_1 (f,t) \, P_2 (f,t)}\, \gamma_{\ell^{'}m'}(f,t) \,,
    \end{align}
where $\hat{C}(f,t)$ is the cross-correlation spectrum computed by multiplying Fourier transforms of the strain time-series from the two GW detectors used in the analysis~\cite{Thrane:2009fp}. The summation is done over time-segments denoted by $t$ (typically data is divided into short segments, each lasting 1-3 minutes and over frequency bins denoted by $f$ (typically 1/4 Hz or 1/32 Hz binning is used). As discussed below, we will repeat the analysis in 10 Hz wide bands, summing over all frequency bins between 20-30 Hz, 30-40 Hz, etc. 

The quantity $\hat{X}_{\ell m}$ is usually referred to as the \textit{dirty map} (it represents the SGWB sky seen through the response matrices of a baseline created by a pair of detectors) whereas the $\Gamma_{\ell m,\ell^{'} m'}$ is called the \textit{Fisher information matrix} (it encodes the uncertainty associated with the dirty map measurement). In both equations, $P_{i} (f,t)$ is the noise power spectral density of detector $i$, and $\gamma_{\ell m}$ captures the geometrical factors associated with the two geographically separated detectors with different orientations (usually referred to as the overlap reduction function~\cite{allenromano,Thrane:2009fp}). 

The observed $\hat{P}_{\ell m}$, the \textit{clean map}, obtained through the \textit{deconvolution} shown in Eq. ~\eqref{DATA:MLsolution} is an unbiased estimator of the angular distribution of the SGWB, $\langle \hat{P}_{\ell m} \rangle = P_{\ell m}$. It is also worth noting that, in the weak-signal limit, one can show that 
	\begin{align}\label{DATA:expXP}
	\langle \hat{X}_{\ell m} \, \hat{X}^*_{\ell^{'} m'} \rangle - \langle \hat{X}_{\ell m} \rangle \, \langle \hat{X}^*_{\ell^{'} m'} \rangle \approx \Gamma_{\ell m,\ell^{'} m'} \, , \nonumber \\
	\langle \hat{P}_{\ell m} \,\hat{P}^*_{\ell^{'} m'} \rangle - \langle \hat{P}_{\ell m} \rangle \, \langle \hat{P}^*_{\ell^{'} m'} \rangle \approx (\Gamma^{-1})_{\ell m,\ell^{'} m'}\,.
	\end{align}
The above equation implies that $\Gamma_{\ell m,\ell^{'} m'}$ is the covariance matrix of the dirty map, and $(\Gamma^{-1})_{\ell m,\ell^{'} m'}$ is the covariance matrix of the clean map. In particular, since the dirty map is obtained by averaging over many time segments and frequency bins, by Central Limit Theorem the resulting $\hat{X}_{\ell m}$'s are multi-variate Gaussian variables with zero means and the covariance matrix given by $\Gamma_{\ell m,\ell^{'} m'}$. Further, since the clean map is obtained by a linear transformation of the dirty map, the $\hat{P}_{\ell m}$'s are also multi-variate Gaussian variables with zero means and the covariance matrix given by $(\Gamma^{-1})_{\ell m,\ell^{'} m'}$.

One can then introduce an estimator for the SGWB
angular power spectrum, which describes the angular scale of structure in the clean map as,
    \begin{align}
    \label{AngPowerSpectra}
        \hat{C}_{\ell}=  \frac{1}{2\ell+1}\sum_{m=-\ell}^{\ell} \lvert \hat{P}_{\ell m} \rvert^2 \, .
    \end{align}	
We will see in the next section that this estimator is biased, and we will describe how one can obtain an unbiased estimator from it. Also, note that by conducting the analysis in narrow frequency bands (10 Hz wide in our case), these estimators will encode frequency dependence. 

	%
\subsection{Unbiased regularized estimator}
\begin{figure}[h]
	        \centering
            \includegraphics[width=\columnwidth]{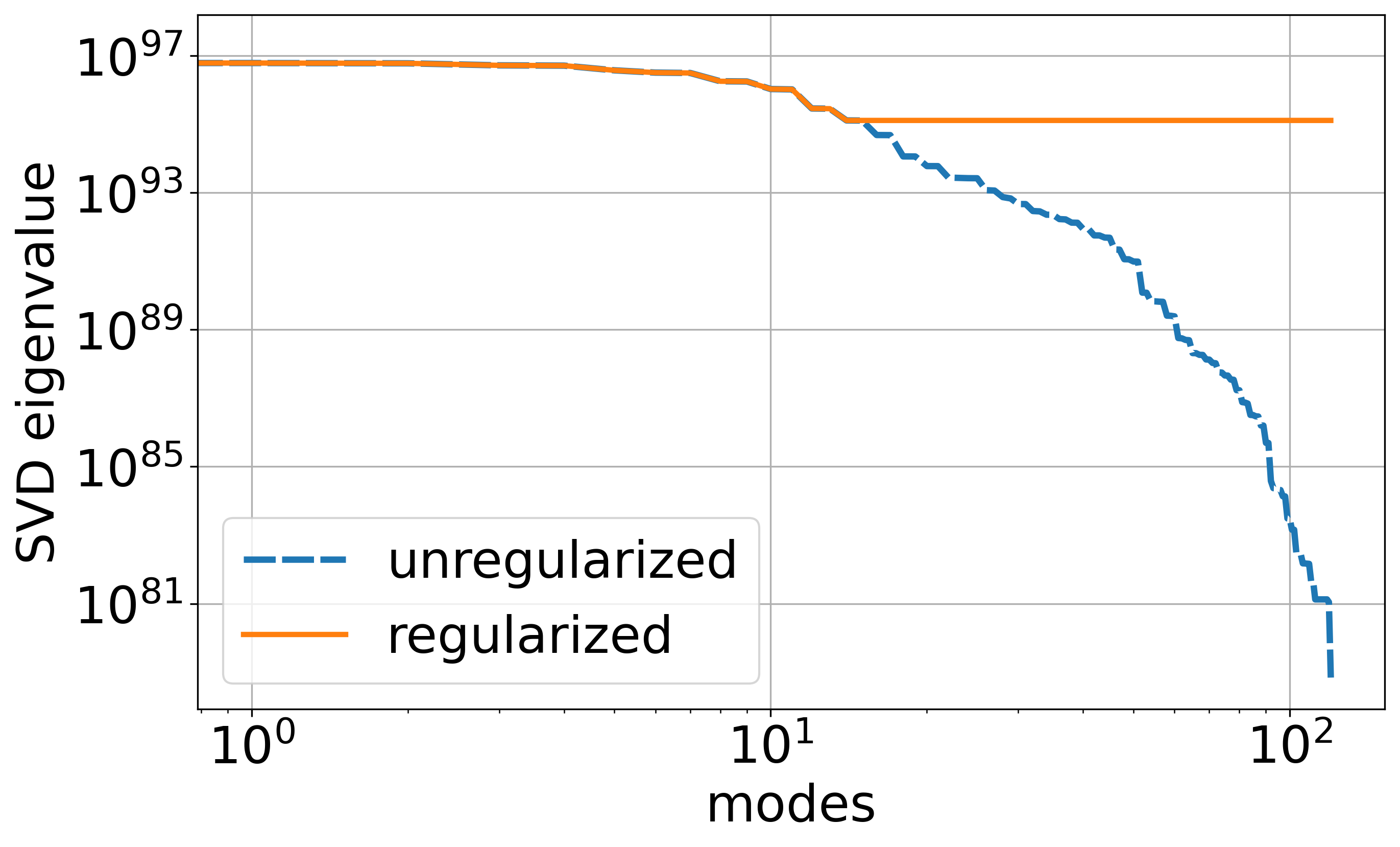}
	        \caption{SVD eigenvalues of the Fisher matrix in the 20-30 Hz band are shown. Regularization of the Fisher matrix is accomplished by replacing eigenvalues smaller than $S_{\text{min}}$ with $S_{\text{min}}$, where $S_{\text{min}}$ is defined to be $10^{-2}$ of the maximum eigenvalue and is depicted by the horizontal part of the orange line.}
	    \label{fig:FisherMatrixReg}
	\end{figure}

In practice, the Fisher matrices are degenerate owing to the existence of blind directions of GW detectors. This implies that the Fisher matrix is insensitive to certain $\ell m$
modes. Hence a full inversion cannot be performed. Therefore we use a regularized pseudoinverse, which conditions the original matrix to circumvent other numerical errors, to obtain our estimators. One can employ different regularization techniques to perform this pseudoinversion~\cite{Thrane:2009fp,Panda:2019hyg,Agarwal:2021gvz,Floden:2022scq}. One of the most common regularization procedures used in the literature is the Singular Value Decomposition (SVD) technique. In the SVD procedure one can decompose $\Gamma_{\ell m,\ell^{'} m'}$ (which is a Hermitian matrix) as
    \begin{align}
        \label{DATA:svd}
        \Gamma = U S V^{*} \, ,
    \end{align}
where $U$ and $V$ are unitary matrices and $S$ is a diagonal matrix whose non-zero elements are the positive and real eigenvalues of the Fisher matrix. Then the problematic $\ell m$ modes will correspond to the smallest elements of $S$. To illustrate the general nature of these eigenvalues, we have plotted the relative size of the eigenvalues for a typical Fisher matrix (computed from the 20-30 Hz GW data set with $\ell_{\text{max}} = 10$) in FIG. \ref{fig:FisherMatrixReg}. Then to condition the ill-conditioned matrix, a threshold on the eigenvalue ($S_{\text{min}}$) is chosen. The choice is made by considering the proper trade-off between the quality of the deconvolution and the increase in numerical noise from less sensitive modes\footnote{It is worth noting that the regularization problem becomes severe as one considers smaller frequency bands. A proper trade-off between the variance of the estimator and the subsequent biases needs to be thoroughly explored in such cases.}. Any values below this cutoff are considered too small, and one can replace them either with infinity or with the smallest eigenvalue above the cutoff. Throughout this work, we will set the threshold $S_{\text{min}}$ to be $10^2$ times smaller than the largest eigenvalue; all eigenvalues smaller than $S_{\text{min}}$ are replaced by $S_{\text{min}}$. These choices and the subsequent regularized eigenvalues are also illustrated in FIG. \ref{fig:FisherMatrixReg}.

Given the regularized inverse Fisher matrix $\Gamma ^{-1} _{R}$, the ML solution in Eq. ~\eqref{DATA:MLsolution} takes the form,
    \begin{align}
        \label{DATA:RegCleanMap} 
       \hat{P}^{R} _{\ell m} = \sum_{\ell^{'} ,m'} \left(\Gamma ^{-1} _{R} \right)_{\ell m,\ell^{'} m'} \, \hat{X}_{\ell^{'} m'} \,,
    \end{align}
still obeying multi-variate Gaussian distribution. 
The covariance matrix of this clean map (under weak-signal approximation) also takes a slightly different form compared to the one given in Eq. ~\eqref{DATA:expXP}. It can be written as (we have dropped the indices for the Fisher matrix for simplicity)
	\begin{align}
 \label{Eq:KGW}
	    K_{\ell m, \ell' m'} = \langle \hat{P}^{R}_{\ell m} \hat{P}^{R*}_{\ell^{'} m'} \rangle - \langle \hat{P}^{R}_{\ell m} \rangle \langle \hat{P}^{R*}_{\ell^{'} m'} \rangle = \Gamma^{-1}_R \, \Gamma \, \Gamma^{-1}_R\,.
	\end{align}
From the expectation value and uncertainty in the estimators defined in Eq. \eqref{DATA:expXP}, one can show that the regularized SGWB angular power spectrum estimators obey
    \begin{align}\label{DATA:CellBias}
	\langle \hat{C}_{\ell}^R \rangle \approx C_{\ell} + \frac{1}{2\ell + 1} \sum_{m} ( \Gamma^{-1}_R )_{\ell m,\ell m}, \\\label{eq:GWClvar}
	\langle (\hat{C}_{\ell}^R)^2 \rangle - \langle \hat{C}_{\ell}^R \rangle^2 \approx \frac{2}{(2\ell + 1)^2} \sum_{mm'} \lvert (\Gamma^{-1}_R)_{\ell m,\ell m'} \rvert^2\,.
	\end{align}
One can see from the expressions of estimators of the clean map and the angular power spectra that both depend on inverting the Fisher information matrix $\Gamma_{\ell m,\ell^{'} m'}$. Thus our estimators are biased. The unbiased estimators of the SGWB angular power spectrum are given by
	\begin{align}\label{DATA:unbiasedCell}
	\hat{C}^{'}_{\ell}= \hat{C}_{\ell} - \frac{1}{2\ell+1} \sum_{m} ( \Gamma^{-1}_R )_{\ell m,\ell m}\,.
	\end{align}
\subsection{Choice of $\ell_{\rm max}$}
    The choice of $\ell_{\rm max}$ in the expansion in Eq. \eqref{DATA:SHD} is ultimately determined by the detector sensitivity and the frequency dependence of the searched SGWB model~\cite{Floden:2022scq}. However, when the Fisher matrix is ill-defined, the regularization procedure introduces a bias that increases with $\ell_{\rm max}$. In particular, larger $\ell_{\rm max}$ implies a larger Fisher matrix, regularization of a larger number of eigenvalues, and hence larger bias. 

\begin{figure}[h]
	   \centering
        \includegraphics[width=\columnwidth]{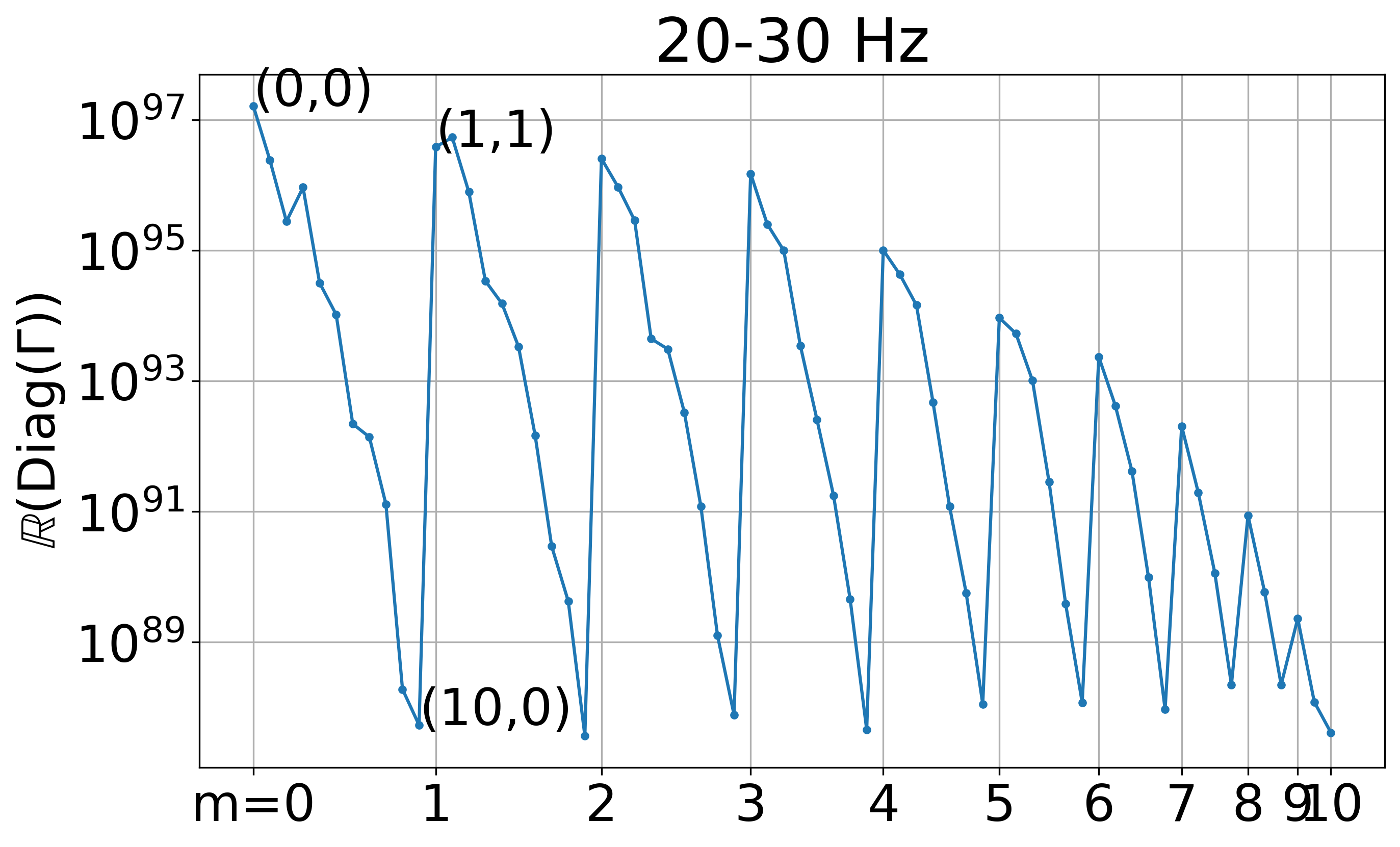} \\
	    \includegraphics[width=\columnwidth]{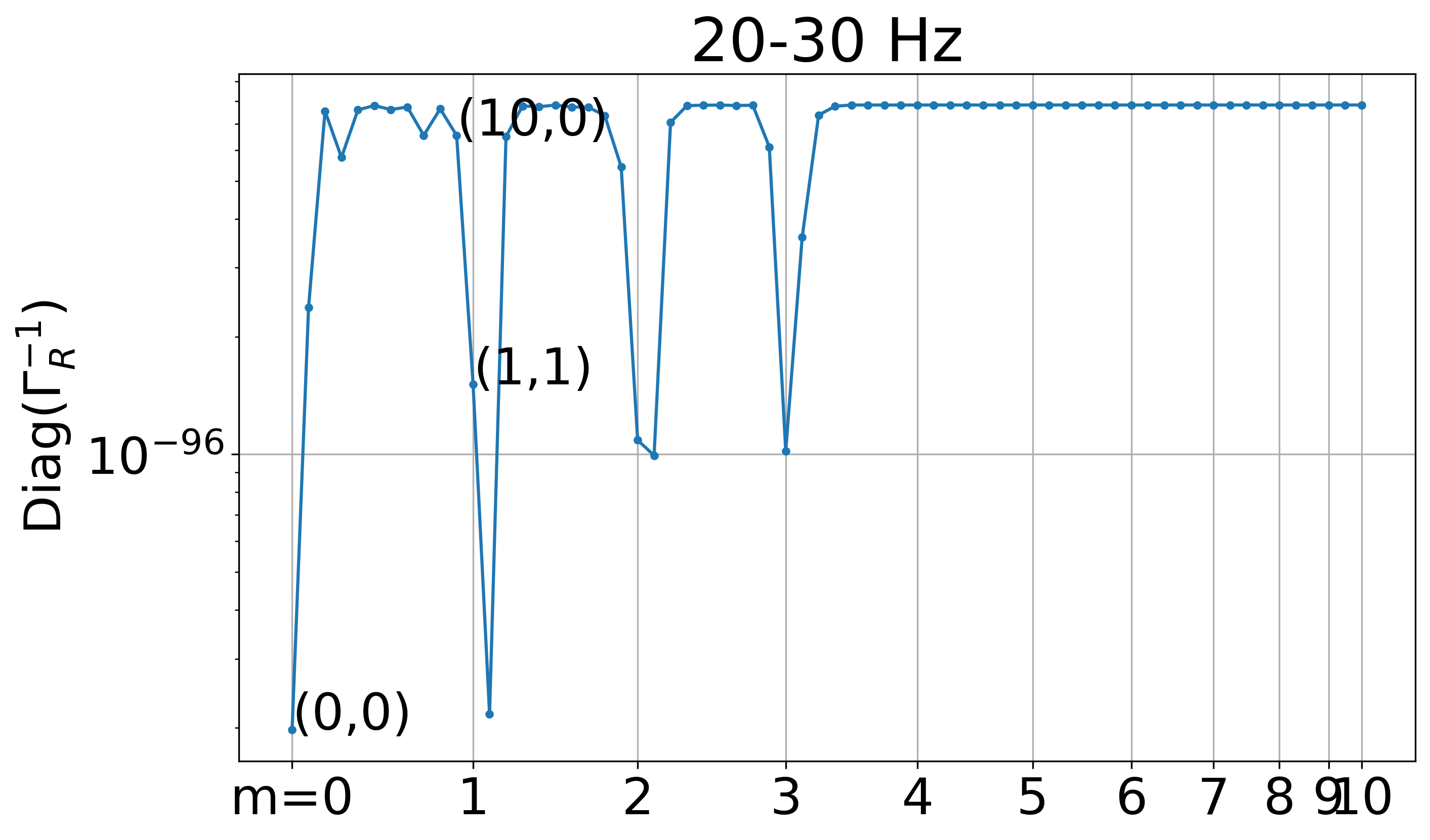}
	    \caption{Diagonal entries of the SGWB Fisher matrix and its inverse Fisher matrix are shown for the 20-30 Hz band, with regularization defined in the text. The indices along the x-axis are $(\ell,m)$=(0,0),(1,0)...,($\ell_{\text{max}}$,0),(1,1),..., ($\ell_{\text{max}}$,1),...,($\ell_{\text{max}}$-1,$\ell_{\text{max}}$-1),($\ell_{\text{max}}$,$\ell_{\text{max}}$), with $\ell_{\text{max}}$=10.}
	    \label{fig:FisherInverse}
	\end{figure}

    One way to assess this is to examine the diagonal entries of the Fisher and regularized inverse Fisher matrices, as in FIG. \ref{fig:FisherInverse}. The Fisher matrix diagonal elements decrease significantly as $\ell$ increases for the same $m$. If the Fisher matrix could be inverted, the diagonal elements of the inverse Fisher matrix would correspondingly increase with $\ell$ for a fixed $m$. FIG. \ref{fig:FisherInverse} (bottom) indeed shows this increasing trend, but the trend saturates (reaches a plateau) after $\ell = 5$ because of the regularization. Propagating this to $K^{GW}$ in Eq. \eqref{Eq:KGW} implies that the covariance matrix for the clean map could have artificially low values (implying artificially good sensitivity) if one uses too large value of $\ell_{\rm max}$. We therefore choose $\ell_{\rm max} = 5$ in our analysis to avoid this regularization bias.

\subsection{Final angular power spectrum estimator} \label{subsec:gwautopower}

We note that the definition of the SGWB anisotropy in the theoretical model of Section \ref{sec:Model} (c.f. Eq. \ref{Eq:Sec2Sph}) and in the SGWB search formalism (Eqs. \ref{DATA:Omega_SGWB}-\ref{DATA:power_law}) have different normalizations. To build estimators that are directly compatible with the model prediction, Eq. (\ref{ClGW}), note that the frequency dependence of the angular power defined in the SGWB search is (cf. Eqs.\,\ref{DATA:Omega_SGWB}-\ref{DATA:power_law}):
    \begin{align}\label{eq:KHf}
        \mathcal{K}\equiv\frac{2\pi^2 f^3}{3H_0^2}\left(\frac{f}{f_{\text{ref}}}\right)^{\alpha}.
    \end{align}
We can then define frequency dependent estimators of the spherical harmonic coefficients of the clean map, whose expectation values are consistent with their theoretical counterparts in Eq. \eqref{Eq:Sec2Sph}:
	\begin{align}
	    \hat{a}_{\ell m}(f)=\mathcal{K} \hat{P}_{\ell m}.
	\end{align}	
The covariance matrix for these coefficients is given by a similar scaling,
	\begin{align}
	    K^{\rm GW}_{\ell m, \ell' m'} = \mathcal{K}^2 K_{\ell m, \ell' m'}.
	\end{align}	
    We then introduce the properly normalized, frequency dependent estimators of the SGWB angular power spectrum, 
    \begin{align}
    \label{DATA:AngPowerSpectra}
        \hat{C}_{\ell}^{GW} (f)=  \frac{1}{2\ell + 1}\sum_{m=-l}^l \lvert \hat{a}_{\ell m} (f) \rvert^2 \, .
    \end{align}	
    Referring to Eq. \eqref{DATA:unbiasedCell}, the unbiased angular power spectrum of the SGWB auto-correlation is then:
    \begin{align}
        \hat{C}_{\ell}^{'GW} (f)= \hat{C}_{\ell}^{GW} (f)  - \frac{\mathcal{K}^2}{2\ell + 1} \sum_{m} ( \Gamma^{-1}_R )_{\ell m,\ell m} \, .
    \end{align}	

\begin{figure}[h]
	    \centering
        \includegraphics[width=\columnwidth]{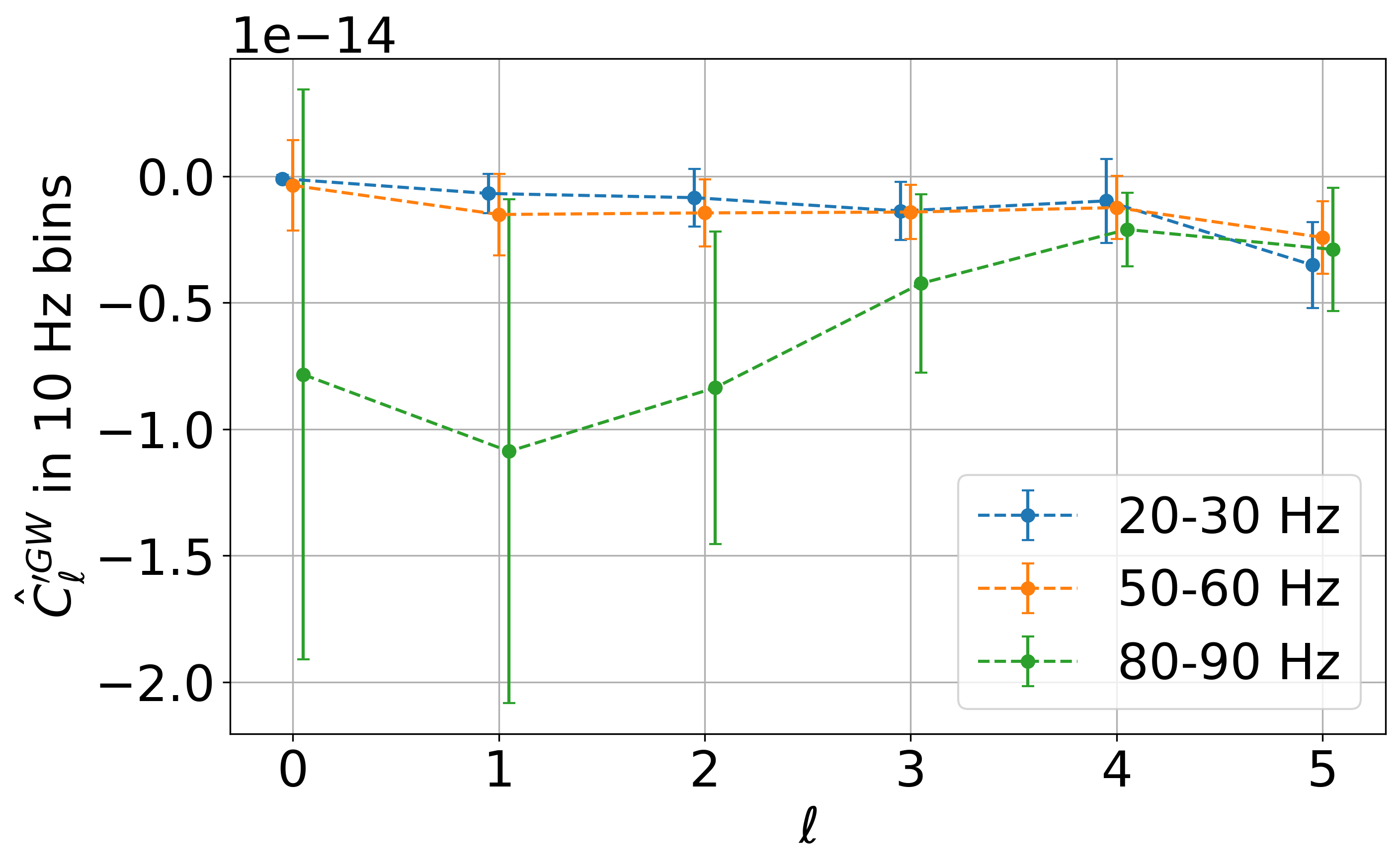}
	    \includegraphics[width=\columnwidth]{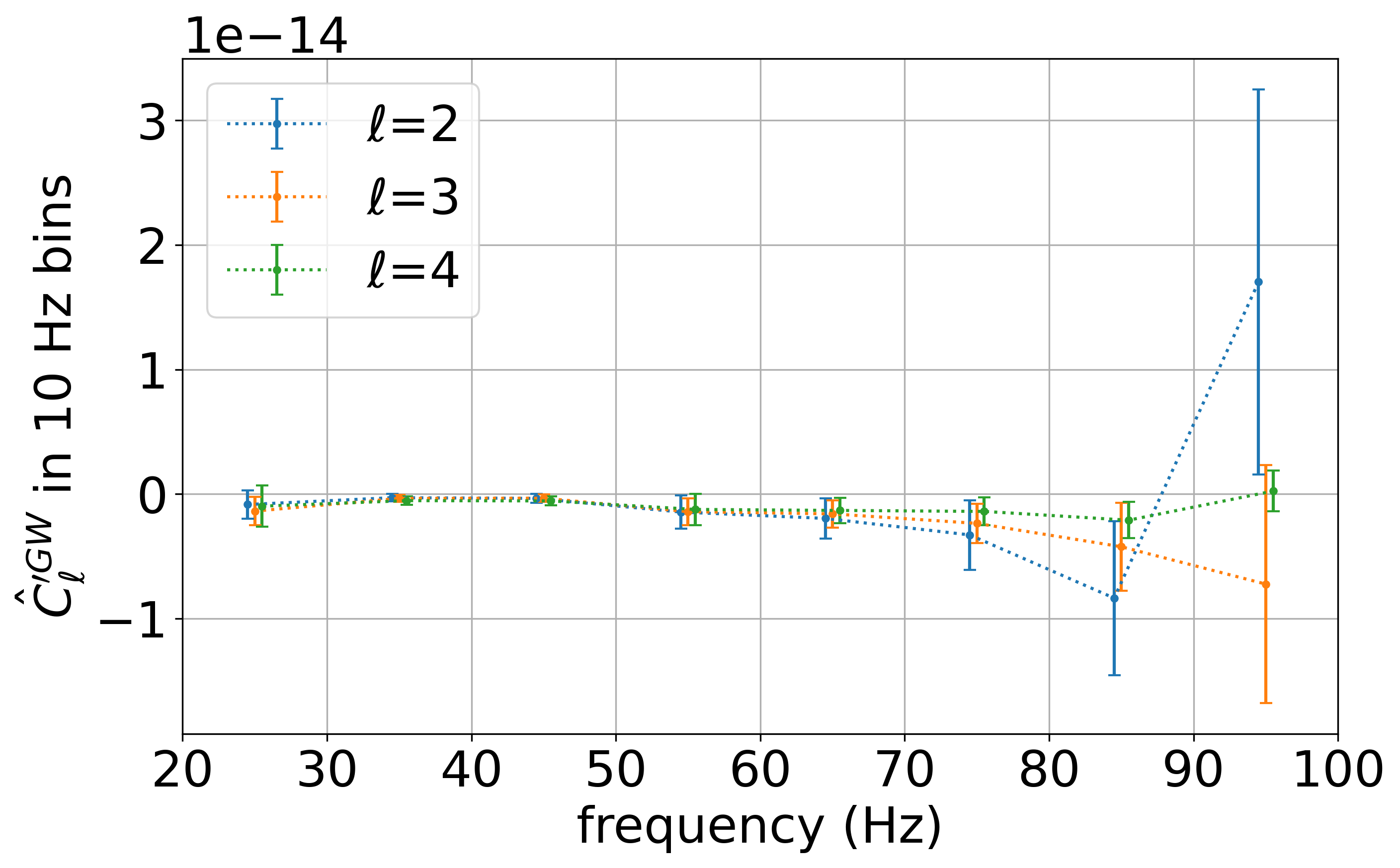}
	    \caption{Unbiased $\hat{C}^{'GW}_\ell$ estimators with standard deviation error-bars of the SGWB angular power spectrum are shown in 10 Hz wide frequency bands (20-30 Hz, 50-60 Hz, and 80-90 Hz) as a function of $\ell$ (top) and as a function of frequency for  $\ell=2,3,4$ (bottom). }
	    \label{fig:GWCl}
	\end{figure}

We apply these definitions to the publicly available folded data set~\cite{folding,folded_data} from the third observing run (O3) of Advanced LIGO detectors located in Hanford, WA (H) and Livingston, LA (L). We perform the analysis in 10 Hz frequency bands from 20 Hz to 100 Hz with $\ell_{\text{max}}=5$, and use the {\tt{PyStoch}} pipeline~\cite{pystoch,pystoch_sph} to compute the unbiased $\hat{C}^{'GW}_{\ell}$ estimators of the angular power spectra and the corresponding $\hat{a}_{\ell m}$.
The $\hat{C}^{' GW}_{\ell}$ estimators and their variance (calculated from Eq. \eqref{eq:GWClvar} times $\mathcal{K}^2$) in these frequency bands are shown in FIG. \ref{fig:GWCl}, as a function of $\ell$ for different frequencies and as a function of frequency for various values of the multipole $\ell$ (top and bottom panels respectively). 
This Figure shows that the SGWB auto power in all frequency bins and at all $\ell$s is consistent with zero, implying there is no evidence for an anisotropic SGWB in these data. Note that the error bars increase at higher frequencies, which is a consequence of the lower strain sensitivity of LIGO detectors at higher frequencies and of the power law frequency dependence in Eq. \eqref{eq:KHf}. It is worth noting here that the SGWB auto power and the error bars are consistent with the noise as in the case of results published in ~\cite{O3directional}. It is not straightforward to have a one-to-one comparison, given our analysis is performed in 10 Hz frequency bands in contrast to the broadband one shown in~\cite{O3directional}. However, the SGWB auto power is in good agreement with the all-sky all-frequency SGWB angular power spectra shown in~\cite{Agarwal:2023lzz}.

	\section{Measurement of Galaxy Overdensity Angular Power Spectra}\label{sec:Galaxy}
	In our study, we use the galaxy number count from the Sloan Digital Sky Survey (SDSS)~\cite{SDSS_DR16} for computing the galaxy over-density angular power spectra. The Sloan Digital Sky Survey (SDSS) imaging data covers around 1.5$\times 10^4$~deg$^2$, or one-third of the sky. Within the range of $r$-band magnitude between 17 to 21 ($17<m_r\le21$), after removing quasars and stars, there are 52.4 million galaxies in its photometric catalog and 2.8 million galaxies in its spectroscopic catalog. We remove stripe No. 82, which is scanned many more times compared to other stripes in the survey and is hence much brighter. This leaves us with 43.4 and 1.7 million galaxies in the two catalogs, respectively.
	
\begin{figure}[h]
	    \centering
	    \includegraphics[width=0.45\textwidth]{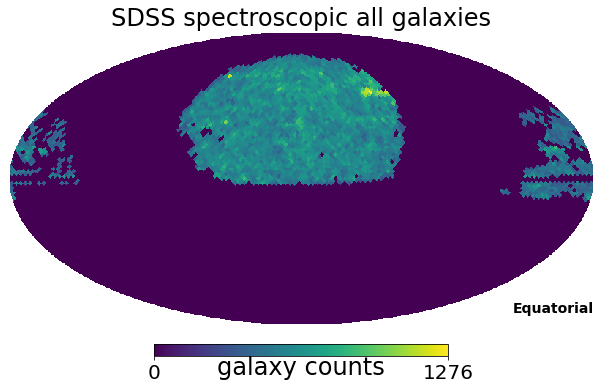}
	    \caption{Galaxy number count sky map in equatorial coordinates from the SDSS spectroscopic catalog. We have selected galaxies with $r$-band magnitude between 17 to 21 after removing quasars and stars and applied a mask to correct for systematics. The color bar stands for galaxy count in each HEALPix basis pixel with resolution of $N_{\text{side}}$=32.}
	    \label{fig:sdss}
	\end{figure}

	We use the galaxies in the SDSS spectroscopic catalog, whose redshift range extends to 0.8, with a median redshift of 0.39. We address systematic issues in the survey following \cite{Yang:2020usq}. In particular, we select only galaxies with $r$-band seeing $<$1.5 and extinction $<$0.13. Galaxy counts in pixels that are affected by these data quality cuts are replaced by the average galaxy counts of the remaining unaffected neighboring pixels. This leads to the final sky map of the galaxy number count in HEALPix-based representation \cite{Gorski:1999rt}, with the systematic effects accounted for. This sky map is shown in equatorial coordinates in FIG. \ref{fig:sdss}. The pixels with information cover around 20\% of the full sky.
 
\begin{figure}[h]
	    \centering
	    \includegraphics[width=0.45\textwidth]{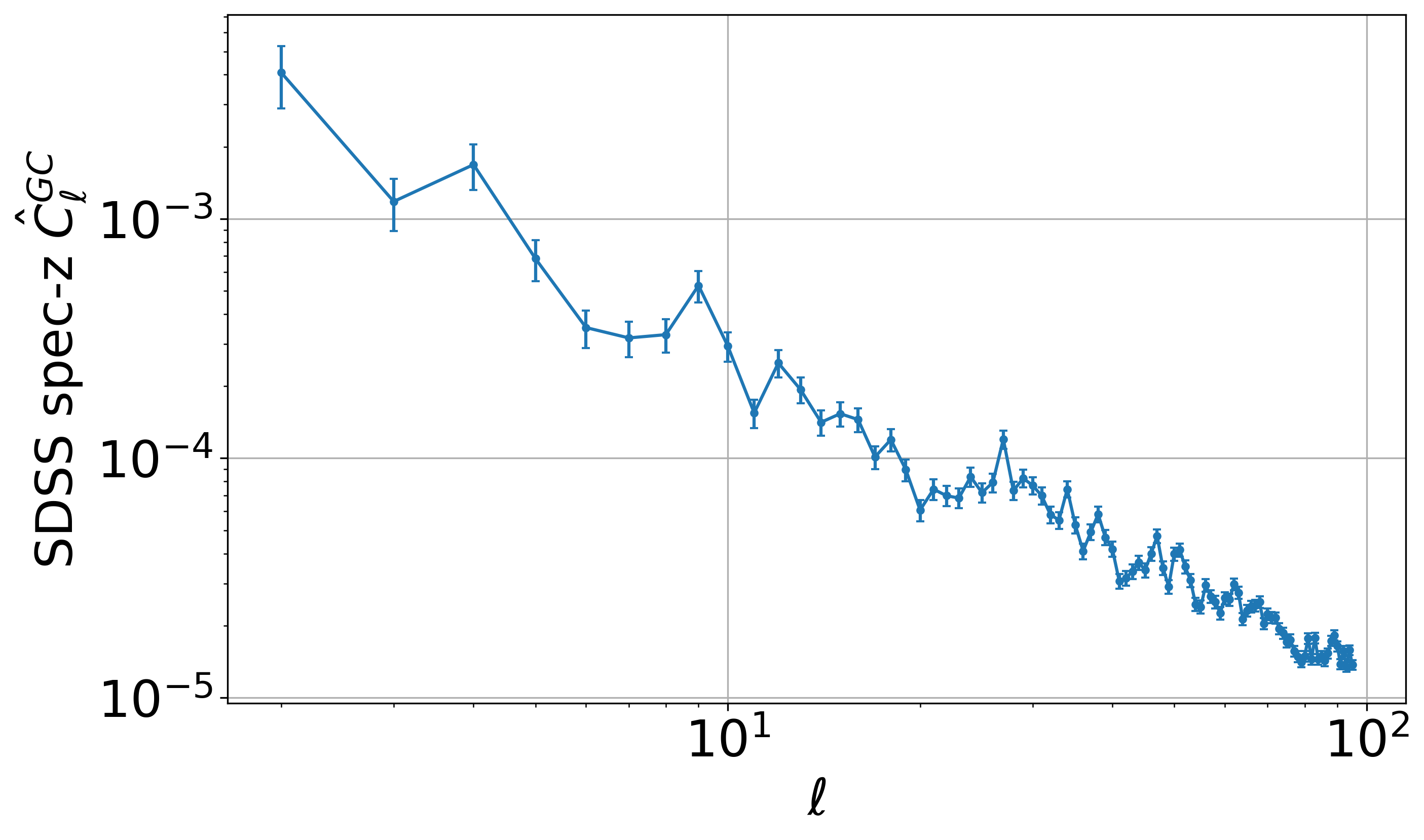}
	    \caption{The angular power spectrum $\hat{C}_{\ell}^{\text{GC}}$ for galaxy count overdensity, corrected for the partial-sky coverage, from the SDSS spectroscopic catalog of $2\le \ell \le95$. Uncertainties associated with the cosmic variance are shown.} 
	    \label{fig:galCl}
	\end{figure}

Based on this galaxy count sky map, we calculate the galaxy over-density as a function of the sky direction and expand it in spherical harmonics as defined in Eq. \eqref{Eq:Sec2Sph}. To account for the pixels with missing information, we apply a
binary mask to the galaxy over-density sky map in pixel basis, where we mask out every pixel without information (due to no observations or high systematics), before applying the spherical harmonic transformation. The obtained spherical harmonic coefficient estimators, $\hat{b}_{\ell m}$'s, are then used to compute the angular power spectrum for the galaxy overdensity auto-correlation:
    \begin{align}\label{eq:GCCl}
        \hat{C}_{\ell}^{\text{GC}}=\frac{1}{f_\text{sky}}\frac{1}{2\ell + 1}\sum_{m=-l}^l \lvert \hat{b}_{\ell m} \rvert^2\,.
    \end{align}
Here, the factor $f_{\rm sky}$ denotes the fraction of the sky covered by the survey, and is needed to account for the missing power in the sky map when performing the spherical harmonic transformation. We note that the same scaling must also be applied when computing the cross-correlation angular power spectrum between SGWB and GC partial sky maps. The resulting GC angular power spectrum of the SDSS spectroscopic catalog is shown in FIG. \ref{fig:galCl}, including uncertainties defined by the cosmic variance. The maximum $\ell$ used in this Figure is determined by the angular resolution in FIG. \ref{fig:sdss} and is larger than the maximum $\ell$ obtained from the SGWB analysis above. Furthermore, due to the partial sky coverage, there is a lower limit on $\ell$ that can be estimated as $\ell_{\text{min}}=\pi/\theta$, where $\theta$ is the spot size in the sky in radians. Hence we will use $\ell\ge 2$ for the SDSS spectroscopic catalog sky map (FIG. \ref{fig:sdss}). 
%
%

	\section{Measurement of cross-correlation Angular Power Spectra}\label{sec:Cross-corr}
	We now introduce an unbiased estimator for the angular power spectrum of the cross-correlation. 
 We use the frequency-dependent SGWB multipoles $\hat{a}_{\ell m}(f)$ (estimated in 10 Hz frequency bins and introduced in Section \ref{subsec:gwautopower}), and the SDSS sky map multipoles, $\hat{b}_{\ell m}$, introduced in Section \ref{sec:Galaxy}. We define the estimator of their cross correlation angular power spectrum as 
	\begin{align}\label{eq:crossCl}
	    \hat{C}_{\ell}^{\text{cross}}=\frac{1}{f_\text{sky}} \; \frac{1}{2\ell + 1}  
 \; \sum_{m=-\ell}^{\ell} \hat{b}_{\ell m}^{*} \hat{a}_{\ell m}\,.
	\end{align}
As noted above, the $1/f_\text{sky}$ factor accounts for the incomplete sky coverage of the SDSS survey. To compute the covariance of this estimator, $K_C$, we assume that the galaxy map multipoles have much smaller uncertainties than their SGWB counterparts. This is a safe assumption since each pixel in the SDSS map in FIG. \ref{fig:sdss} counts thousands of galaxies (implying uncertainties at the level of a few percent), while the SGWB sky map is dominated by detector noise and shows no evidence of a signal. Consequently, Eq. \eqref{eq:crossCl} can be regarded as a linear transformation of the SGWB multipoles $\hat{a}_{\ell m}$, implying that the resulting $\hat{C}_{\ell}^{\text{cross}}$ are also multi-variate Gaussian with the covariance matrix given by the appropriate propagation of the covariance of the SGWB multipoles $K^{\text{GW}}$:

 \begin{align}\label{eq:crosscov}
	    (K_C)_{\ell,\ell '} =\frac{1}{f^2_\text{sky}} \frac{1}{(2\ell + 1)(2 \ell' + 1)} \sum_{m,m'} \hat{b}_{\ell m}^{*} \, K^{\text{GW}}_{\ell m \ell' m'} \,\hat{b}_{\ell' m'} \,.
\end{align}

This covariance matrix does not take into account the cosmic variance or the shot noise contributions discussed in Section \ref{subsec:shotnoise}, c.f. Eq. \eqref{SNR}. Following \cite{Alonso:2020rar,Alonso:2020mva}, these contributions are diagonal and should be added to the above covariance matrix.
%
Our final covariance is therefore given by
\begin{widetext}
 \begin{align}\label{final_cov}
(K_C^{\text{tot}})_{\ell\ell^{'}}=(K_C)_{\ell\ell^{'}}
+\frac{\delta_{\ell \ell^{'}}}{(2\ell+1)} \,\left[ \left( C_{\ell}^{\text{GW}}(\theta) + N_{\rm shot}^{\text{GW}}(\theta) \right) \left(C_{\ell}^{\text{GC}} + N_{\rm shot}^{\text{GC}} \right) + \left (C_{\ell}^{\text{cross}}(\theta) + N_{\rm shot}^{\text{cross}}(\theta) \right)^2\right]\,.
 \end{align}
\end{widetext}
%
We note that the shot noise is Poissonian in origin, so it can spoil the multi-variate Gaussian nature of the $\hat{C}_{\ell}^{\text{cross}}$ estimators. In the limit when the cross-correlated signal is small, the shot noise contribution to the covariance matrix will be relatively small compared to the SGWB instrumental noise contribution, and the distribution will be approximately Gaussian. This will be the case in our simulation analyses presented below. It is important to note, however, that as the SGWB instrumental noise improves and the cross-correlated signal becomes more significant, the shot noise contribution will alter the $\hat{C}_{\ell}^{\text{cross}}$ distribution away from Gaussian. The parameter estimation scheme presented below will have to be correspondingly adapted. 

\begin{figure}[ht]
	    \centering
	    \includegraphics[width=\columnwidth]{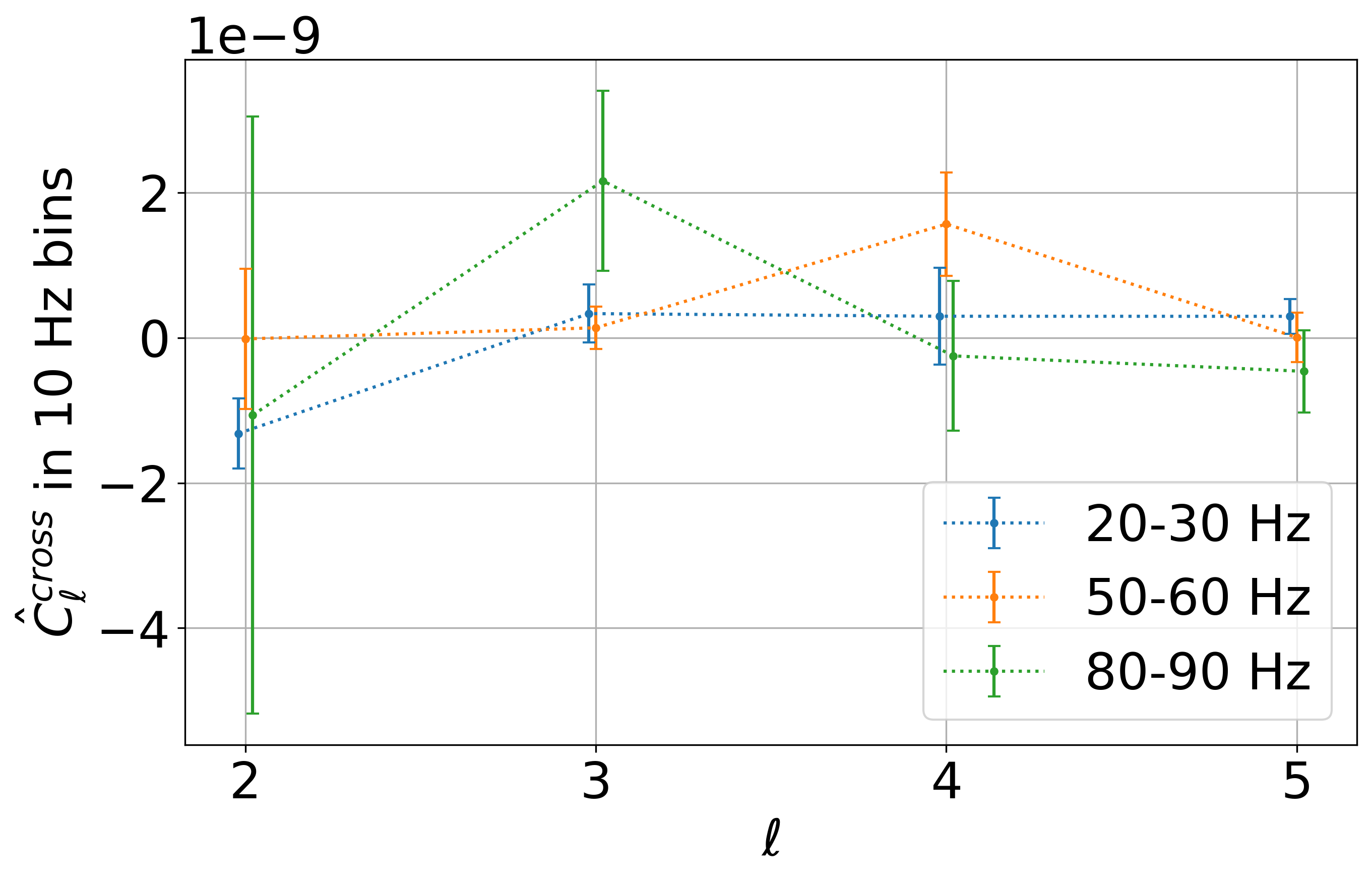}\\
	    \includegraphics[width=\columnwidth]{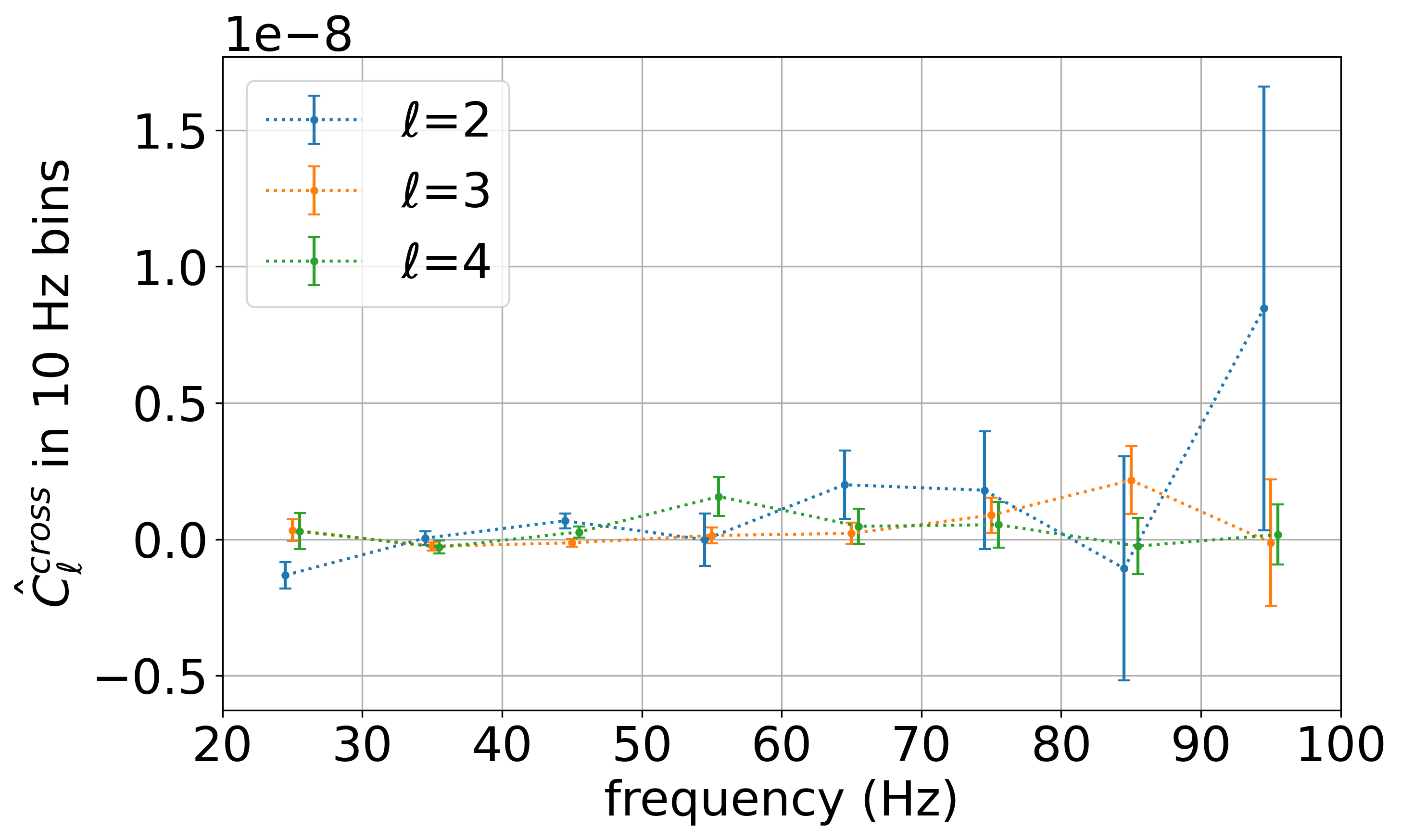}
	    \caption{Angular power spectra $\hat{C}_{\ell}^{\text{cross}}$ and standard deviation error-bars of the cross correlation between the measured SGWB sky-maps (in 10 Hz wide frequency bands) and the galaxy over-density of the SDSS spectroscopic catalog, with $\ell\ge 2$ up to $\ell_{\text{max}}=5$ are shown for several example frequency bands (top) and for several values of $\ell$ (bottom).} 
	    \label{fig:GWgalCl}
    \end{figure}

The angular power spectra of the cross-correlation between the measured SGWB sky-maps (in 10 Hz width frequency bands from 20 to 100 Hz) and galaxy over-density in the SDSS spectroscopic catalog are shown in FIG. \ref{fig:GWgalCl}. Error bars are also shown, defined as the square root of the diagonal terms of the $K_C$ matrix, indicating no evidence for a cross correlation signal. 
We observe that the noise level of cross-correlation $\hat{C}_\ell^{\text{cross}}$ increases with frequency. This is not surprising---in absence of a cross correlation signal, the covariance of $\hat{C}_\ell^{\text{cross}}$ is given by the SGWB covariance in Eq. \eqref{eq:crosscov}, which also increases with frequency, c.f. FIG. \ref{fig:GWCl}. 

	\section{Parameter Estimation}\label{sec:PE}
Having measured the angular power spectra $\hat{C}_{\ell}^{\rm cross}$ of the cross correlation between the SGWB sky-maps and the galaxy over-density of the SDSS spectroscopic catalog, we now turn our attention to extracting the astrophysical information from these measurements. We implement a Bayesian inference framework, where the posterior distribution of the astrophysical model parameters $\theta$ is given by
\begin{equation}
\label{Eq:Bayes}
p(\theta|\hat{C}_\ell^{\text{cross}}) \propto \mathcal{L}(\hat{C}_{\ell}^{\text{cross}} | C_{\ell}^{\rm cross} (\theta)) \, \pi(\theta),
\end{equation}
where $\pi(\theta)$ denotes the prior distribution of the model parameters, and $\mathcal{L}$ denotes the likelihood of observing the data for given model parameters. As discussed above, in the limit when the cross correlation signal is small, the $\hat{C}_{\ell}^{\text{cross}}$'s will approximately follow the multivariate Gaussian distribution with the covariance matrix given by either $K_C$ if shot noise is ignored or by $K_C^{\rm tot}$ if shot noise is included. In particular, if shot noise is ignored 
\begin{align}\label{eq:lnlk}
	\ln \mathcal{L}(\hat{C}_{\ell} | C_{\ell}(\theta)) =\frac{1}{2} \ln |K_C|-\frac{1}{2}(\hat{C}_{\ell}-C_{\ell}(\theta))^T \, K_C^{-1} \,(\hat{C}_{\ell}-C_{\ell}(\theta))\,.
\end{align}
Notice that here we are omitting the superscript `cross' to simplify the notation. If shot noise is included in the analysis, $K_C$ is replaced by $K_C^{\rm tot}$. This likelihood is GW frequency dependent since it can be computed for each 10 Hz wide band in the SGWB analysis. The overall likelihood is obtained by multiplying the likelihoods for individual frequency bands (or equivalently by summing up the individual log-likelihoods). Finally, since $K_C$ does not depend on model parameters, the first term dependent on $|K_C|$ can be dropped. This is not the case when using $K_C^{\rm tot}$, which depends on model parameters through the shot noise terms. 

Our astrophysical model describing the galactic process of GW emission is given in Section \ref{sec:Model}, with the shot noise terms defined in Section \ref{subsec:shotnoise}. This model is parameterized by an astrophysical Gaussian kernel that has three parameters $\theta = (A_{\text{max}}, z_c, \sigma_z)$, with the Gaussian peak appearing near $z=1$. Since the SDSS galaxy catalog used in our analysis extends only up to redshift $z\sim 0.8$, our analysis will not be able to assess the Gaussian peak: at small redshift, the kernel can be approximated by a linear function monotonically increasing with redshift \cite{Cusin:2019jpv}. In other words, the parameters $z_c$ and $\sigma_z$ will appear degenerate in our analysis, since increasing the mean $z_c$ or decreasing the variance $\sigma_z$ both result in a faster increase of the linear function. We, therefore, choose to fix $\sigma_z = 0.7$, which is compatible with the astrophysical model predictions, and our parameter space becomes 2-dimensional: $\theta = (A_{\text{max}},z_c)$. 

In the following analyses, we will choose parameters $\theta$, compute the corresponding $C_\ell^{\rm cross}(\theta)$'s and add them to the measured $\hat{C}_\ell^{\rm cross}$ presented in Section \ref{sec:Cross-corr}. We will then scan the parameter space $\theta$ and compute the posterior distribution given by Eq. \eqref{Eq:Bayes}, for both cases when we ignore the presence of shot noise and when we include the shot noise. Our goal is to demonstrate that our formalism correctly recovers the simulated parameters, and to study how inclusion of shot noise impacts the recovery. We also perform the recovery without adding a simulated signal, which yields the first upper limits on the astrophysical kernel parameters.

	
 \subsection{Results without shot noise}
As a first step, we calculate the posterior distribution using the likelihood of Eq. \eqref{eq:lnlk}, ignoring the shot noise contribution (in both the signal and in the covariance matrix) and without adding any simulated signals. As noted above, we evaluate the likelihood in every 10 Hz wide frequency bin between 20-100 Hz, and then multiply these likelihoods to obtain the overall likelihood. We assume uniform prior distributions in the two parameters: $A_{\text{max}}\in [1\times 10^{-38}, 5\times 10^{-32}]$ erg cm$^{-3}$s$^{-1/3}$ and $z_c\in [0, 1]$ \cite{Cusin:2019jpv}. These ranges are both astrophysically well motivated and consistent with the sensitivity of our $\hat{C}_{\ell}^{\rm cross}$ measurements. We define a uniform linear grid in this parameter space, and evaluate the model $C_{\ell}$'s, the likelihood, and the posterior at each grid point. The result (for the entire 20-100 Hz band) is shown in the upper-left panel of FIG. \ref{fig:PE-SDSS}. While there is a slight preference for larger values of $z_c$, no constraint can be placed on this parameter. However, a 95\% confidence upper limit on $A_{\text{max}}$ can be placed, $A_{\text{max}}^{95\%}=2.7\times 10^{-32}$ erg cm$^{-3}$s$^{-1/3}$.
We next add a simulated signal to the measured $\hat{C}_{\ell}^{\rm cross}$'s. The simulated signal is computed for $A_{\text{max}}=2\times 10^{-32}$ erg cm$^{-3}$s$^{-1/3}$ and $z_c=0.6$. We again evaluate the posterior distribution, with the linear grid adjusted to be around these simulated values. The recovery (for the entire 20-100 Hz band) is shown in the lower-left panel of FIG. \ref{fig:PE-SDSS}. Note that the simulated parameter point is well within the recovered 2-dimensional 68\% and 95\% contours, and the one-dimensional distributions include the simulated values within 95\% confidence, even though the $z_c$ posterior is not very informative. Hence, our framework successfully recovers the simulated signal in the absence of shot noise.
\subsection{Results with shot noise}
Inclusion of shot noise requires two modifications. First, shot noise adds an offset to the angular power spectrum, as in Eq. \eqref{Eq:Cshot}. This offset is independent of $\ell$, but it is dependent on the astrophysical model parameters $\theta$. Second, the shot noise also modifies the covariance matrix, as in Eq. \eqref{final_cov}. This modification is also dependent on astrophysical parameters $\theta$. Consequently, while shot noise will make harder the recovery of clustering anisotropy, it may actually improve the accuracy for estimation of astrophysical parameters. 

\begin{figure*}[ht]
	        \centering
	        \includegraphics[width=\columnwidth]{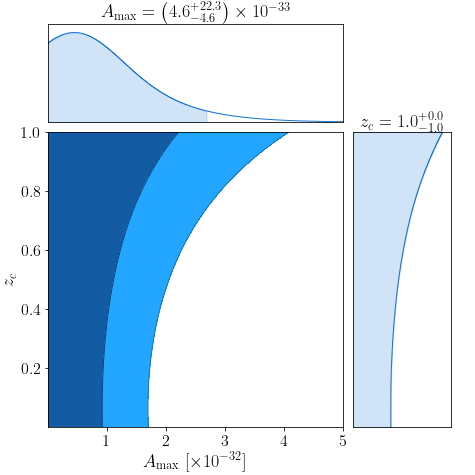}
	        \includegraphics[width=\columnwidth]{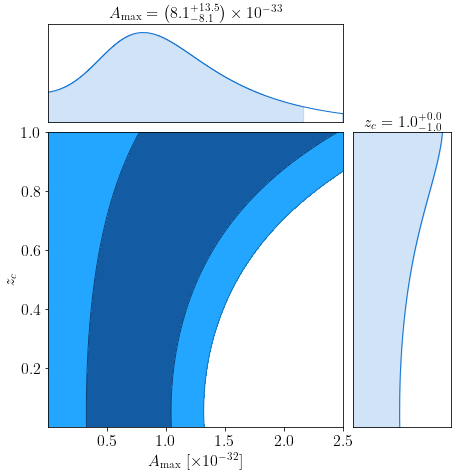}\\
	        \includegraphics[width=\columnwidth]{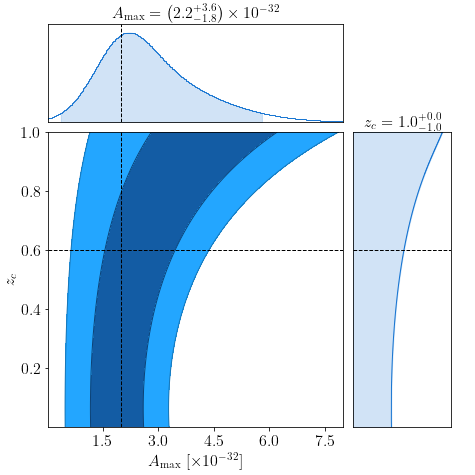}
	        \includegraphics[width=\columnwidth]{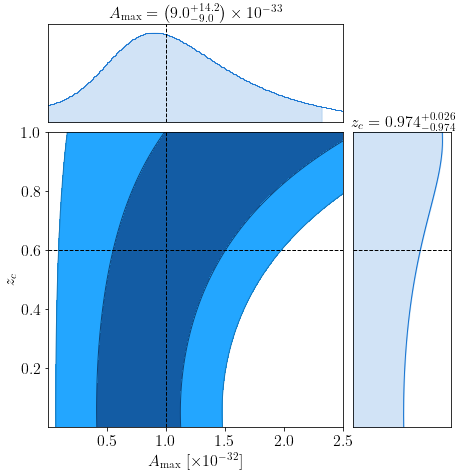}
	    \caption{Results of the parameter estimation for the cross-correlation between the SGWB (20 to 100 Hz) and the galaxy over-density from the SDSS spectroscopic catalog. Each panel shows 2-dimensional posterior with 65\% and 95\% confidence contours, as well as 1-dimensional marginalized posteriors with 95\% confidence intervals for the two model parameters $A_{\text{max}}$ in units of erg cm$^{-3}$s$^{-1/3}$ and $z_c$. Left column panels correspond to the no-shot-noise case, while the right column panels include the shot noise. The upper row panels present upper limits on model parameters (no simulated signal is added). The lower panels show recoveries when a simulated signal is added to the data. The dashed lines indicate the values of simulated parameters. See text for further details.}
	    \label{fig:PE-SDSS}
    \end{figure*}

As in the no-shot-noise case, we start by computing the posterior distribution in Eq. \eqref{Eq:Bayes} using the measured $\hat{C}_{\ell}^{\rm cross}$, and replacing $C_{\ell}^{\rm cross}(\theta)\rightarrow C_{\ell}^{\rm cross, tot}(\theta)$ and $K_C \rightarrow K_C^{\rm tot}$ in Eq. \eqref{eq:lnlk}. The results are shown in the upper-right panel of FIG. \ref{fig:PE-SDSS}. Again, there is no evidence of signal, even though there is a small (statistically insignificant) preference for higher values of $A_{\text{max}}$. While the $z_c$ posterior is again not informative, we can place a 95\% confidence upper limit on $A_{\text{max}}$: $A_{\text{max}}^{95\%}=2.16\times10^{-32}$ erg cm$^{-3}$s$^{-1/3}$. Note that this upper limit is stronger than in the no-shot-noise case, indicating that addition of shot noise actually improves the sensitivity of this analysis.

We next include a simulated signal. In order to keep the shot noise contribution small (so as to maintain the approximate multi-variate Gaussian distribution of $\hat{C}_{\ell}^{\rm cross}$'s) we choose 2 times smaller value of $A_{\text{max}}=1\times10^{-32}$ erg cm$^{-3}$s$^{-1/3}$ for this simulation. We keep the peak redshift the same as in the no-shot-noise case, $z_c=0.6$. The lower-right panel of FIG. \ref{fig:PE-SDSS} shows the recovery results. While this signal is not fully resolvable at 95\% significance, the $A_{\text{max}}$ posterior distribution peaks at $9.0\times 10^{-33}$ erg cm$^{-3}$s$^{-1/3}$, which is very close to the injection amplitude. While the $z_c$ posterior is still not informative, it does indicate slight preference for larger values of $z_c$, consistently with the simulated value of 0.6. We note that this level of $A_{\text{max}}$ is below the 95\% upper limit on $A_{\text{max}}$ from the no-shot-noise analysis, indicating that it would not have been observable in the no-shot-noise analysis. This is another indication that inclusion of shot noise in the analysis improves the sensitivity to $A_{\text{max}}$.


\section{Conclusion and Discussion}\label{sec:Conclusion}
In this paper, we have studied the cross-correlation between the SGWB (as measured in the recent observing runs of Advanced LIGO, Advanced Virgo, and KAGRA) and the distribution of galaxies across the sky measured by the SDSS survey, and we have extracted for the first time the angular power spectrum of the cross-correlation, in different GW frequency bands. In our study, we assumed that the dominant contribution to the SGWB in the 100 Hz band comes from mergers of extragalactic compact objects. The resulting cross-correlation angular power spectrum is noise dominated (we do not have a detection yet). However, this spectrum can be compared with predictions from an astrophysical model of the SGWB due to BBH mergers, allowing us to set bounds on model parameters. 
We assumed that the GW emission is well-captured by the quadrupole formula, hence when modeling the angular power spectrum we could factorize out the frequency dependence. We then introduced a simplified parameterization for the redshift-dependent astrophysical kernel characterizing GW emission at galactic scales: we described this kernel in terms of a global amplitude and a peak position, corresponding to the redshift bin that contributes the most to the total background budget. We explored this 2D parameter space in a Bayesian inference framework, and we found an upper bound for the amplitude of the kernel to be $A_{\text{max}}=2.7\times 10^{-32}$ erg cm$^{-3}$s$^{-1/3}$ while the peak redshift is left unconstrained. We demonstrated that while including shot noise in the analysis reduces the ability to recover clustering contributions to the anisotropy, it actually improves the sensitivity to the astrophysical kernel parameters. We checked the robustness of our analysis via injection-recovery tests.

We stress that in our modeling of the cross-correlation, we assumed that the dominant contribution to the anisotropy comes from clustering, i.e. we assumed that the cross-correlation is dominated by the overdensity term (which we will refer to as $\delta\delta$-term where $\delta$ stands for galaxy overdensity).\footnote{In other words, we neglected line of sight effects in the expression of GW overdensity and galaxy number counts, see e.g. \cite{Cusin:2017fwz} for details and derivations.} This is a safe assumption in the redshift range that we considered in this work $[0, 0.8]$ and as long as we do not slice it into smaller bins: the clustering term gives indeed the dominant contribution to the anisotropic part of the GW energy density \cite{Cusin:2019jhg}. 
However, if one wants to take a tomographic approach and try to better reconstruct the redshift dependence of the astrophysical kernel by cross-correlating with a redshift-binned galaxy catalog, some additional care is needed. 
The reason why is the following: at the angular scales we have access to, the anisotropic part of $\Omega_{\text{GW}}$ is dominated by contributions of low redshift sources, e.g. sources at $z<0.1$. Then if we cross-correlate it with galaxies in a low redshift bin e.g. $z= 0.05$ the dominant contribution in the cross-correlation comes from the clustering ($\delta\delta$) term because the two over-density terms appearing in the expressions of galaxy number counts and GW energy density have the same support.
However, if we correlate with a high redshift bin e.g. $z=2$, then the density terms in the two observables do not have the same support. In this case, the dominant contribution to the cross-correlation comes from the (de)magnification term in the galaxy number counts, because it includes contributions of gravitational potentials integrated along the line of sight (this term corresponds to the $\kappa$ term in Eq. (13) of \cite{Bruni:2011ta}). 
This term has a minus sign, hence the $\delta\kappa$ term in the cross-correlation, if dominant, will give rise to an anti-correlation.
However, if one integrates the galaxy catalog over redshift, the term $\delta\delta$ (on the redshift range where the two have same support) gives the dominant contribution to the cross correlation.
It follows that considering only clustering is enough for our current purposes.
We stress that the method developed here can be easily adapted and applied to cross-correlations between SGWB and other electromagnetic tracers of structure in the universe. In particular, it can be interesting to perform a joint study of cross-correlation of the SGWB with galaxy counts, weak lensing, Cosmic Microwave Background (CMB), Cosmic Infrared Background (CIB), and others. A statistical formalism that enables a joint analysis of these datasets may improve the overall sensitivity of the approach and enable distinguishing different contributions to the BBH SGWB model (e.g. stellar and primordial contributions that may be correlated with different electromagnetic tracers).
It would also be interesting to test our pipeline using realistic simulations of the GW sky (simulating the galaxy field and GW emission on galactic scales). Such simulations could enable studies of multiple BBH contributions to the SGWB.

As the sensitivity of the GW detector network improves, the sensitivity of the approach presented here will also improve, enabling more stringent constraints on the astrophysical kernel and its effective parameters. Advanced LIGO, Advanced Virgo, and KAGRA will conduct the fourth observing run in 2023-2024, to be followed with the fifth observing run in 2026-2028. The improved detector sensitivity, extended observation time, and availability of multiple detector pairs for the analysis, will improve the sensitivity to $\hat{C}_{\ell}^{\rm cross}$ by $20-30$ times relative to this work. The next generation of ground based detectors, such as Einstein Telescope~\cite{ET} and Cosmic Explorer~\cite{CE} will enable another $\sim 1000$ times improvement in sensitivity. Yet another approach could be to use the Bayesian Search to estimate the BBH SGWB anisotropy~\cite{Sharan}, which could also lead to $\sim 1000$ times sensitivity improvements relative to the approach presented here. These improvements are expected to reach and explore the astrophysically interesting region of the parameter space.

Finally, we note that a significant constraint in this work came from the need to regularize the Fisher matrix in order to invert it. This regularization introduces a potential bias in our analysis, which forced us to constrain the analysis to a relatively small number of spherical harmonics coefficients, and to $\ell_{\rm max} = 5$. There are two ways to remedy this situation in the future. First, availability of more than two GW detectors for this analysis can naturally regularize the Fisher matrix---different baseline pairs have different blind directions, effectively complementing each other and removing the zero-eigenvalues of the Fisher matrix. Second, it may be possible to conduct this analysis using the dirty SGWB sky-maps. This approach would avoid inverting the Fisher matrix, but it comes with the challenge of mapping the model $C_{\ell}^{\rm cross}$'s into the "dirty space" to enable defining a likelihood function. Studies of this approach are ongoing.

\section*{Acknowledgment}
This material is based upon work supported by NSF's LIGO Laboratory, which is a major facility fully funded by the National Science Foundation. The authors are grateful for computational resources provided by the LIGO Laboratory (CIT) supported by National Science Foundation Grants PHY-0757058 and PHY-0823459, and Inter-University Center for Astronomy and Astrophysics (Sarathi). This research has made use of data or software obtained from the Gravitational Wave Open Science Center (gw-openscience.org), a service of LIGO Laboratory, the LIGO Scientific Collaboration, the Virgo Collaboration, and KAGRA. LIGO Laboratory and Advanced LIGO are funded by the United States National Science Foundation (NSF) as well as the Science and Technology Facilities Council (STFC) of the United Kingdom, the Max-Planck-Society (MPS), and the State of Niedersachsen/Germany for support of the construction of Advanced LIGO and construction and operation of the GEO600 detector. Additional support for Advanced LIGO was provided by the Australian Research Council. Virgo is funded, through the European Gravitational Observatory (EGO), by the French Centre National de Recherche Scientifique (CNRS), the Italian Istituto Nazionale di Fisica Nucleare (INFN) and the Dutch Nikhef, with contributions by institutions from Belgium, Germany, Greece, Hungary, Ireland, Japan, Monaco, Poland, Portugal, Spain. The construction and operation of KAGRA are funded by Ministry of Education, Culture, Sports, Science and Technology (MEXT), and Japan Society for the Promotion of Science (JSPS), National Research Foundation (NRF) and Ministry of Science and ICT (MSIT) in Korea, Academia Sinica (AS) and the Ministry of Science and Technology (MoST) in Taiwan.

This paper is assigned the LIGO (Laser Interferometer Gravitational-Wave Observatory Laboratory) document control number LIGO-P2200370.

The work for GC is founded by CNRS and by the Swiss National Foundation "Ambizione" grant \emph{Gravitational wave propagation in the clustered universe}. We are very grateful to David Alonso and Fabien Lacasa for interesting exchanges and discussions during this project. JS is supported by a Actions de Recherche Concertées (ARC) grant. KY, SB, VM, and CS were in part supported by the NSF grant PHY-2011675.

\section*{Data Availability}
The gravitational wave data that support the findings of this study are openly available at (\url{https://doi.org/10.5281/zenodo.6326656}).

The SDSS data are available at the SDSS official website, where the photometric catalog ("photoObj") is under the "imaging data" (\url{https://www.sdss.org/dr16/imaging/catalogs/}), and the spectroscopic catalog is under the "Optical Spectra" (\url{https://www.sdss.org/dr16/spectro/}).
\appendix
\section{Parameter Estimation in 10 Hz frequency bins}
\begin{figure*}[ht]
	\centering
	    \includegraphics[width=0.32\textwidth]{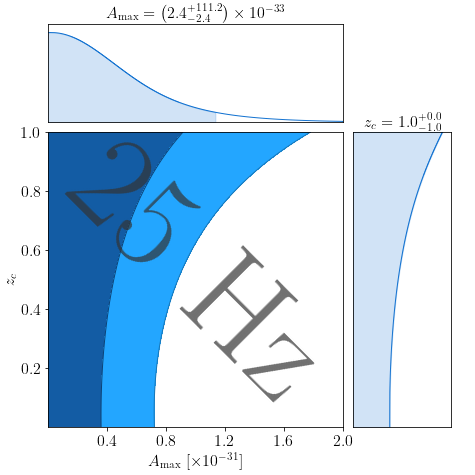}
	    \includegraphics[width=0.32\textwidth]{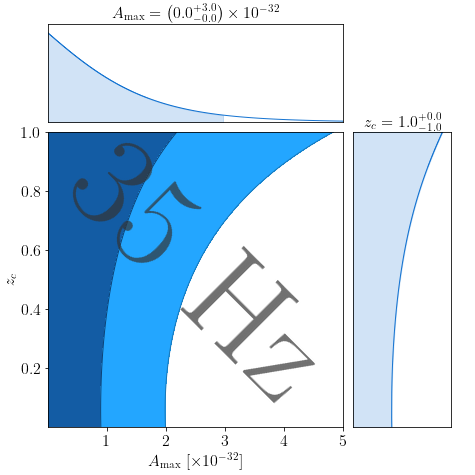}
	    \includegraphics[width=0.32\textwidth]{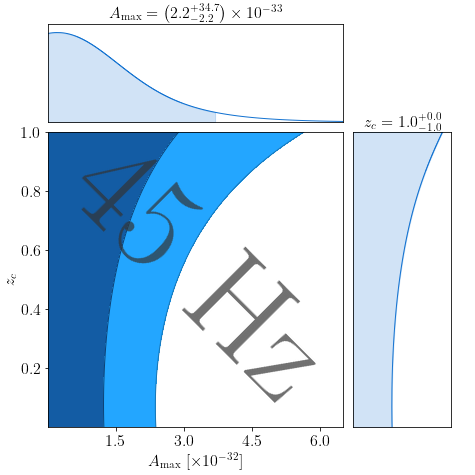}\\
	    \includegraphics[width=0.32\textwidth]{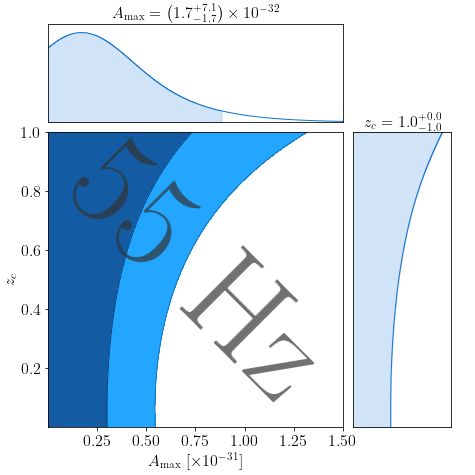}
	    \includegraphics[width=0.32\textwidth]{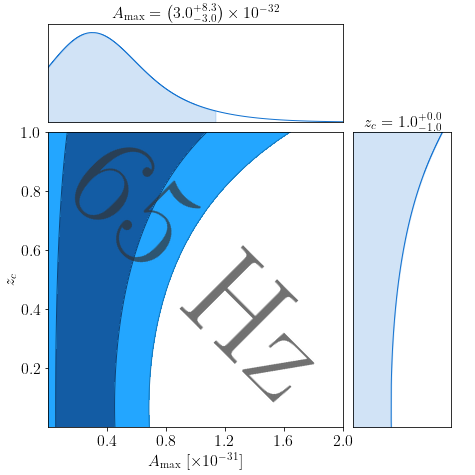}
	    \includegraphics[width=0.32\textwidth]{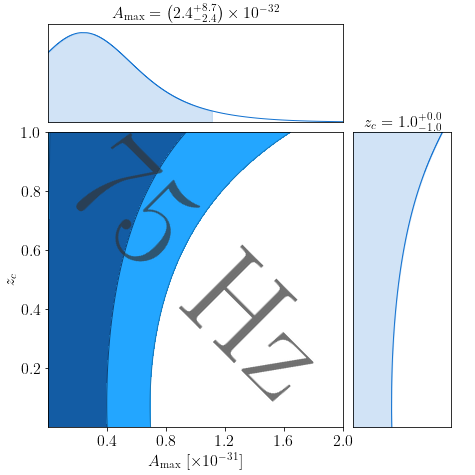}\\
	    \includegraphics[width=0.32\textwidth]{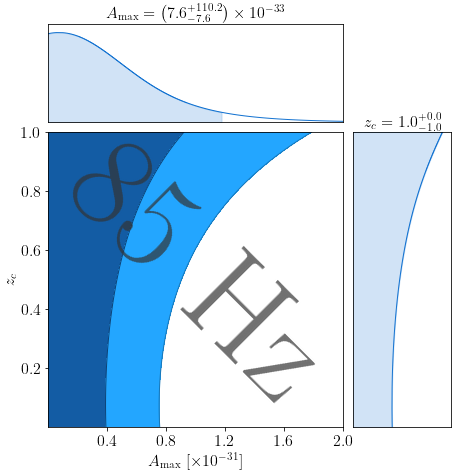}
	    \includegraphics[width=0.32\textwidth]{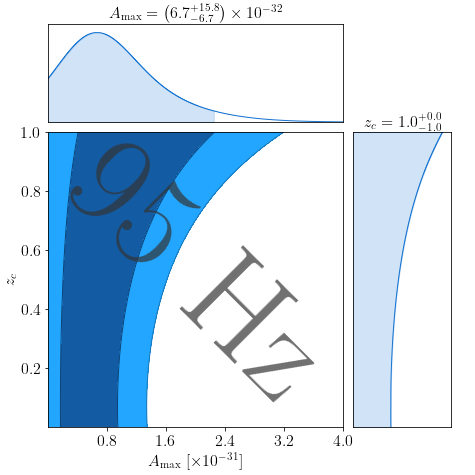}
	    \caption{Upper limits of parameters $A_{\text{max}}$ in units of erg cm$^{-3}$s$^{-1/3}$ and $z_c$ for SGWB measured in different frequency bands, without including the shot noise.}
	    \label{fig:PE-UL-10Hz}
	\end{figure*}

\begin{figure*}[ht]
	 \centering
	    \includegraphics[width=0.32\textwidth]{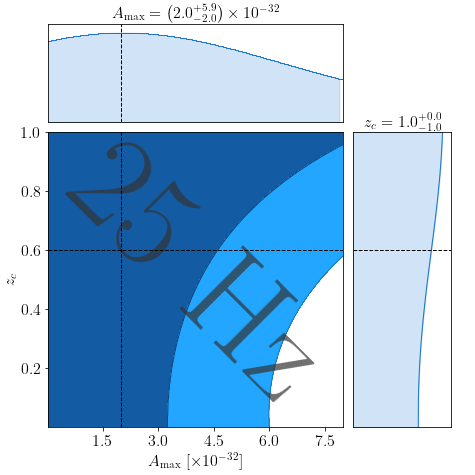}
	    \includegraphics[width=0.32\textwidth]{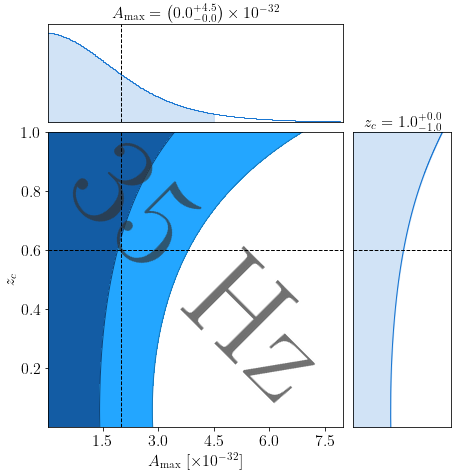}
	    \includegraphics[width=0.32\textwidth]{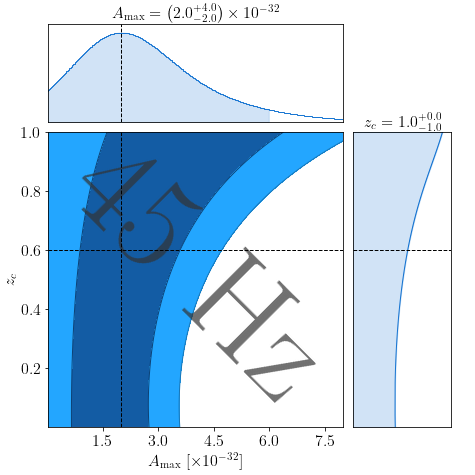}\\
	    \includegraphics[width=0.32\textwidth]{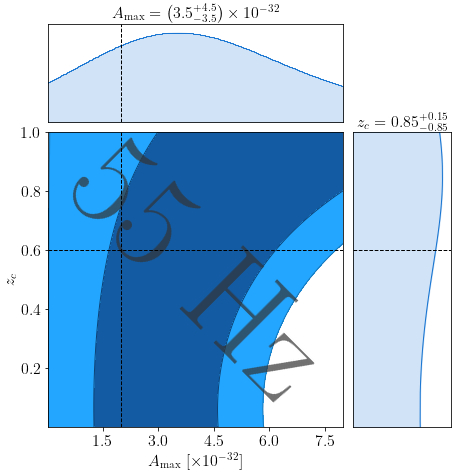}
	    \includegraphics[width=0.32\textwidth]{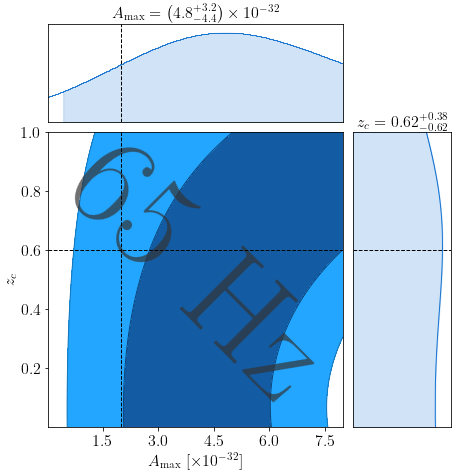}
	    \includegraphics[width=0.32\textwidth]{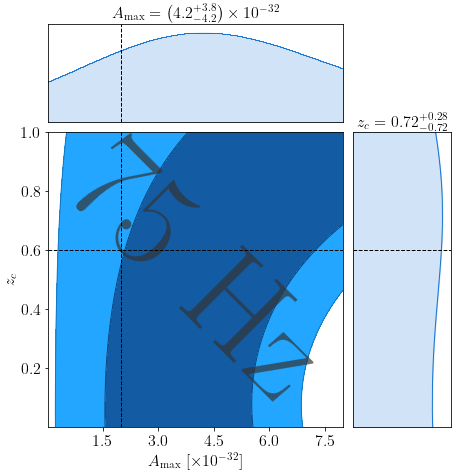}\\
	    \includegraphics[width=0.32\textwidth]{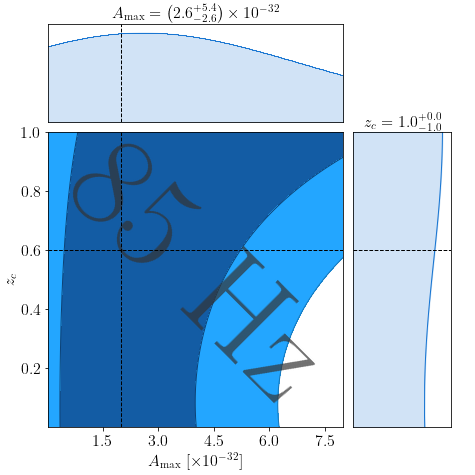}
	    \includegraphics[width=0.32\textwidth]{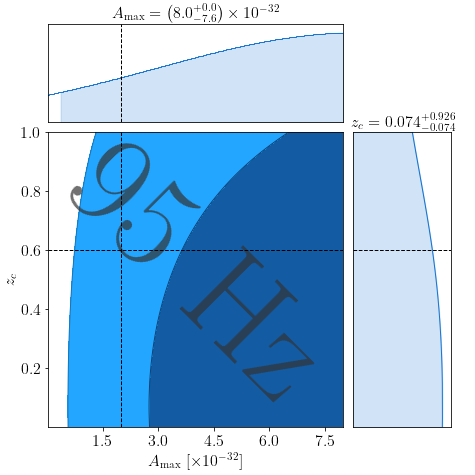}
	    \caption{Recovery of a simulated signal with parameters $A_{\text{max}}=2 \times 10^{-32}$ erg cm$^{-3}$s$^{-1/3}$ and $z_c=0.6$, using SGWB measured in different frequency bands and without including shot noise.}
	    \label{fig:PE-IR-10Hz}
	\end{figure*}

\begin{figure*}[ht]
	\centering
	    \includegraphics[width=0.32\textwidth]{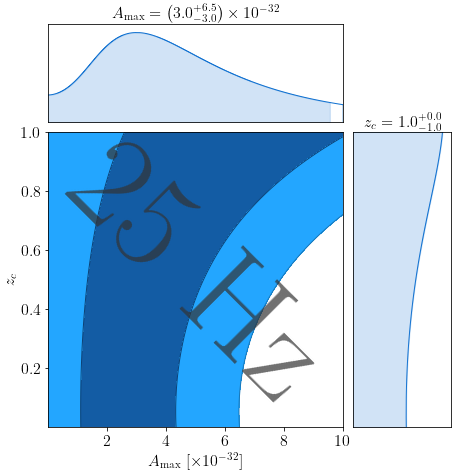}
	    \includegraphics[width=0.32\textwidth]{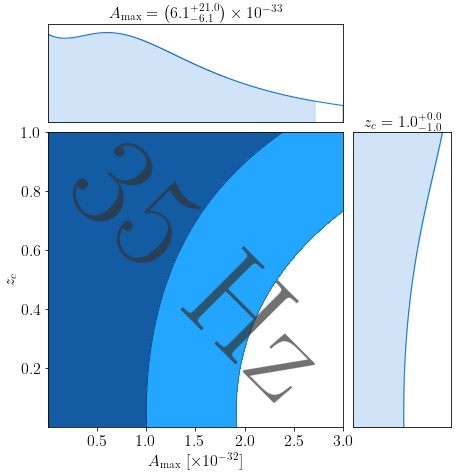}
	    \includegraphics[width=0.32\textwidth]{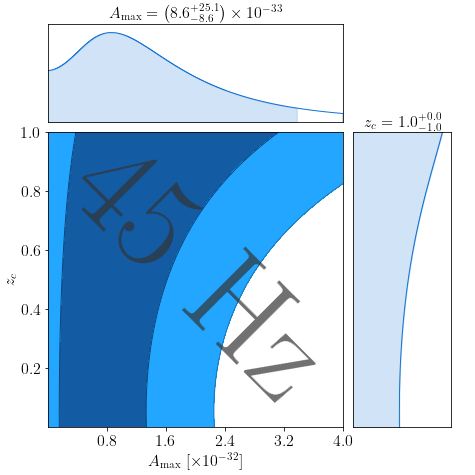}\\
	    \includegraphics[width=0.32\textwidth]{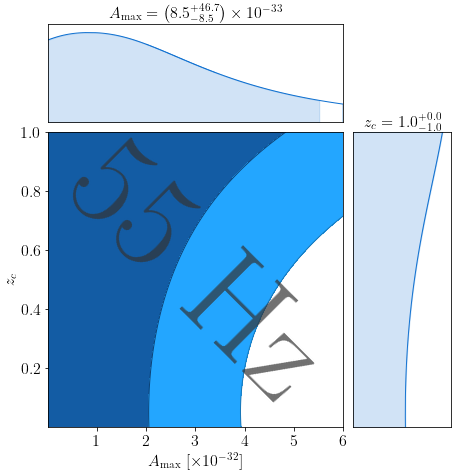}
	    \includegraphics[width=0.32\textwidth]{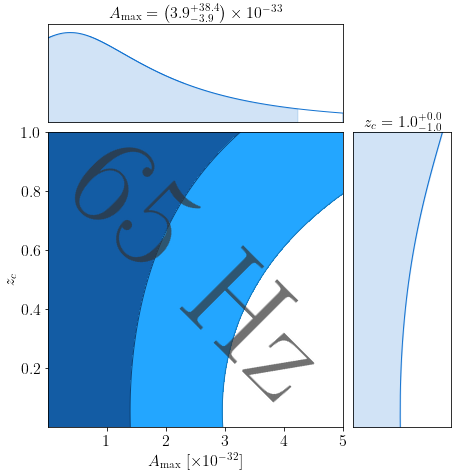}
	    \includegraphics[width=0.32\textwidth]{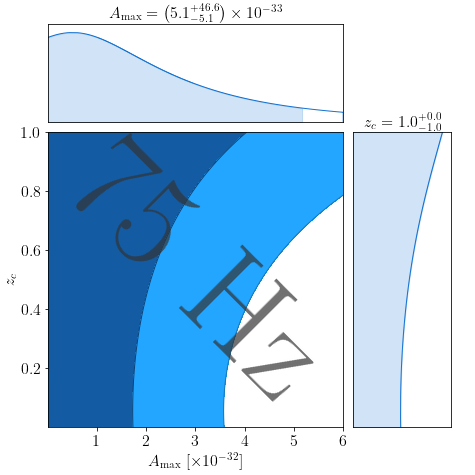}\\
	    \includegraphics[width=0.32\textwidth]{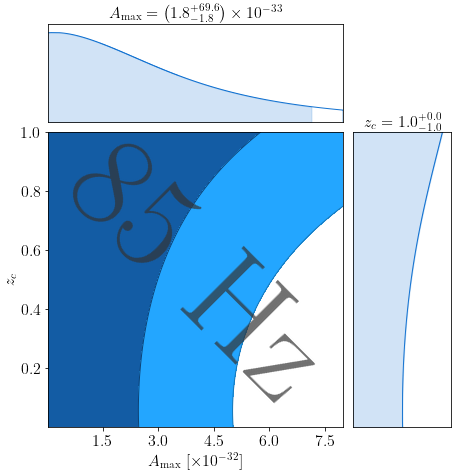}
	    \includegraphics[width=0.32\textwidth]{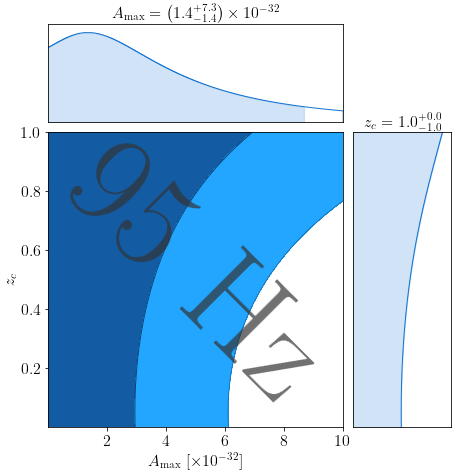}
	    \caption{Upper limits on parameters $A_{\text{max}}$ in units of erg cm$^{-3}$s$^{-1/3}$ and $z_c$ for SGWB measured in different frequency bands, including the shot noise.}
	    \label{fig:PE-UL-SN-10Hz}
	\end{figure*}

\begin{figure*}[ht]
	  \centering
	    \includegraphics[width=0.32\textwidth]{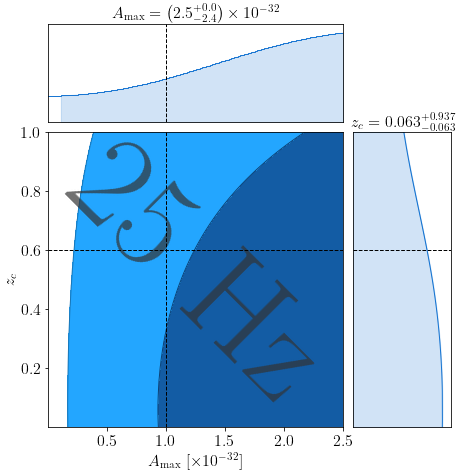}
	    \includegraphics[width=0.32\textwidth]{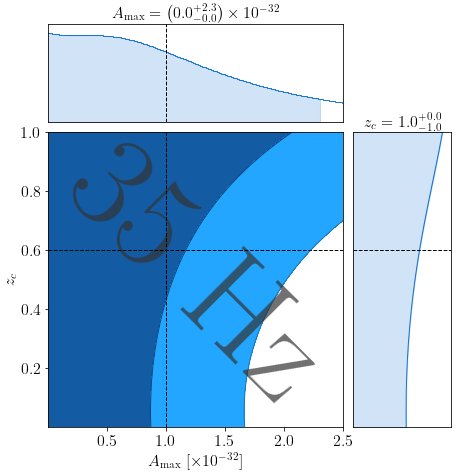}
	    \includegraphics[width=0.32\textwidth]{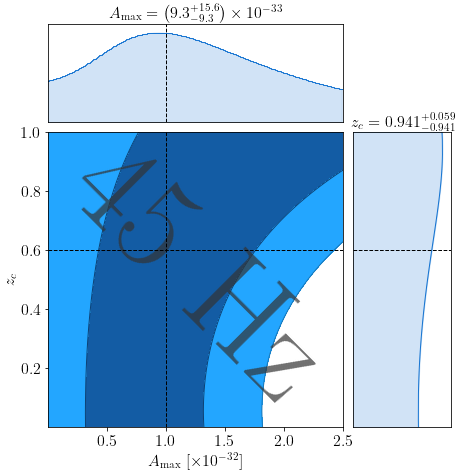}\\
	    \includegraphics[width=0.32\textwidth]{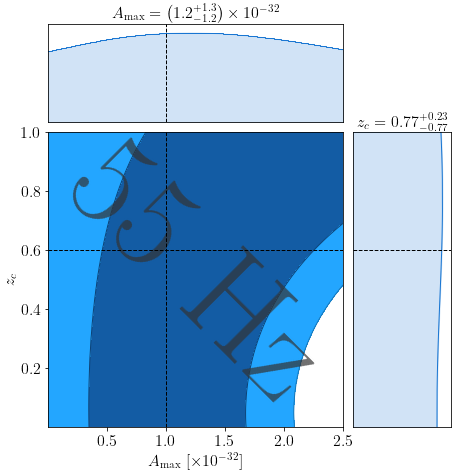}
	    \includegraphics[width=0.32\textwidth]{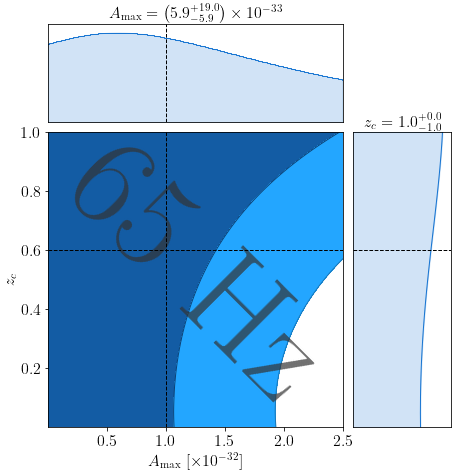}
	    \includegraphics[width=0.32\textwidth]{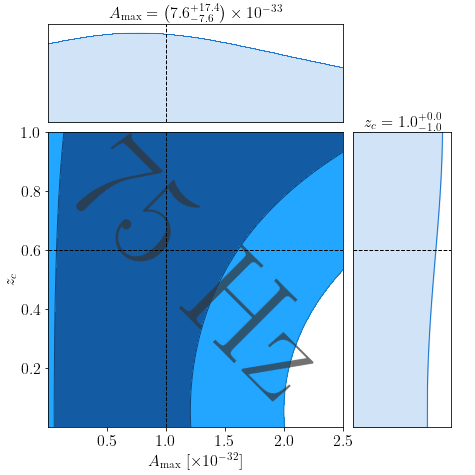}\\
	    \includegraphics[width=0.32\textwidth]{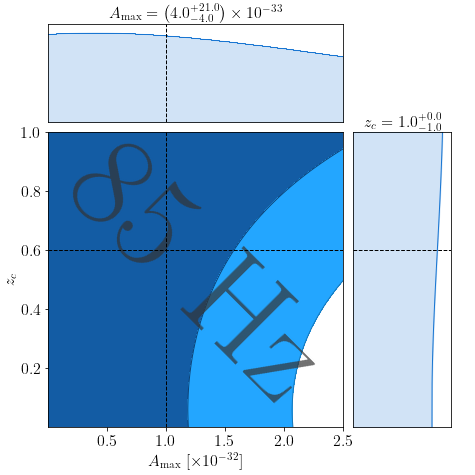}
	    \includegraphics[width=0.32\textwidth]{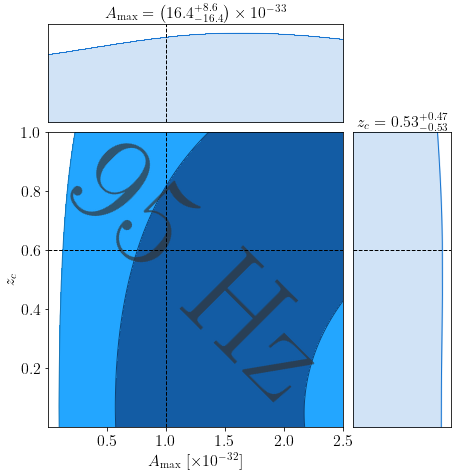}
	    \caption{Recovery of a simulated signal with parameters $A_{\text{max}}=1 \times 10^{-32}$ erg cm$^{-3}$s$^{-1/3}$ and $z_c=0.6$, using SGWB measured in different frequency bands and including shot noise.}
	    \label{fig:PE-IR-SN-10Hz}
	\end{figure*}

We present parameter estimation results in each 10 Hz wide frequency bin from 20 to 100 Hz (labeled by their center frequency) in the following figures: Upper limits of parameters without and with shot noise effects (FIG. \ref{fig:PE-UL-10Hz} and \ref{fig:PE-UL-SN-10Hz}, respectively); injection recovery without and with shot noise effects (FIG. \ref{fig:PE-IR-10Hz} and \ref{fig:PE-IR-SN-10Hz}, respectively). 
For FIG. \ref{fig:PE-IR-10Hz}, we simulated the signal with $A_{\text{max}}=2\times 10^{-32}$ erg cm$^{-3}$s$^{-1/3}$ and $z_c=0.6$, and successfully recover it in most of the frequency bands except 65 and 95 Hz. For FIG. \ref{fig:PE-IR-SN-10Hz}, we simulated the signal with $A_{\text{max}}=1\times 10^{-32}$ erg cm$^{-3}$s$^{-1/3}$ and $z_c=0.6$ is still successful at most frequencies except 25 Hz. While the recovered contours are still consistent with the simulated parameter values, the contours are rather large due to the smallness of the simulated $A_{\text{max}}$. Combining all frequency bands gives a much stronger estimate of $A_{\text{max}}$ as shown in the lower-right panel of FIG \ref{fig:PE-SDSS}. 

\bibliography{GW_EM.bib}

\begin{thebibliography}{99}%
\makeatletter
\providecommand \@ifxundefined [1]{%
 \@ifx{#1\undefined}
}%
\providecommand \@ifnum [1]{%
 \ifnum #1\expandafter \@firstoftwo
 \else \expandafter \@secondoftwo
 \fi
}%
\providecommand \@ifx [1]{%
 \ifx #1\expandafter \@firstoftwo
 \else \expandafter \@secondoftwo
 \fi
}%
\providecommand \natexlab [1]{#1}%
\providecommand \enquote  [1]{``#1''}%
\providecommand \bibnamefont  [1]{#1}%
\providecommand \bibfnamefont [1]{#1}%
\providecommand \citenamefont [1]{#1}%
\providecommand \href@noop [0]{\@secondoftwo}%
\providecommand \href [0]{\begingroup \@sanitize@url \@href}%
\providecommand \@href[1]{\@@startlink{#1}\@@href}%
\providecommand \@@href[1]{\endgroup#1\@@endlink}%
\providecommand \@sanitize@url [0]{\catcode `\\12\catcode `\$12\catcode
  `\&12\catcode `\#12\catcode `\^12\catcode `\_12\catcode `\%12\relax}%
\providecommand \@@startlink[1]{}%
\providecommand \@@endlink[0]{}%
\providecommand \url  [0]{\begingroup\@sanitize@url \@url }%
\providecommand \@url [1]{\endgroup\@href {#1}{\urlprefix }}%
\providecommand \urlprefix  [0]{URL }%
\providecommand \Eprint [0]{\href }%
\providecommand \doibase [0]{https://doi.org/}%
\providecommand \selectlanguage [0]{\@gobble}%
\providecommand \bibinfo  [0]{\@secondoftwo}%
\providecommand \bibfield  [0]{\@secondoftwo}%
\providecommand \translation [1]{[#1]}%
\providecommand \BibitemOpen [0]{}%
\providecommand \bibitemStop [0]{}%
\providecommand \bibitemNoStop [0]{.\EOS\space}%
\providecommand \EOS [0]{\spacefactor3000\relax}%
\providecommand \BibitemShut  [1]{\csname bibitem#1\endcsname}%
\let\auto@bib@innerbib\@empty
\bibitem [{\citenamefont {{LIGO Scientific Collaboration}}(2015)}]{aligo}%
  \BibitemOpen
  \bibfield  {author} {\bibinfo {author} {\bibnamefont {{LIGO Scientific
  Collaboration}}},\ }\bibfield  {title} {\bibinfo {title} {{{A}dvanced
  {LIGO}}},\ }\href {https://doi.org/10.1088/0264-9381/32/7/074001} {\bibfield
  {journal} {\bibinfo  {journal} {Classical and Quantum Gravity}\ }\textbf
  {\bibinfo {volume} {32}},\ \bibinfo {eid} {074001} (\bibinfo {year}
  {2015})},\ \Eprint {https://arxiv.org/abs/1411.4547} {arXiv:1411.4547
  [gr-qc]} \BibitemShut {NoStop}%
\bibitem [{\citenamefont {{F Acernese et al.}}(2015)}]{avirgo}%
  \BibitemOpen
  \bibfield  {author} {\bibinfo {author} {\bibnamefont {{F Acernese et al.}}},\
  }\bibfield  {title} {\bibinfo {title} {{A}dvanced {Virgo}: a
  second-generation interferometric gravitational wave detector},\ }\href
  {http://stacks.iop.org/0264-9381/32/i=2/a=024001} {\bibfield  {journal}
  {\bibinfo  {journal} {Classical and Quantum Gravity}\ }\textbf {\bibinfo
  {volume} {32}},\ \bibinfo {pages} {024001} (\bibinfo {year}
  {2015})}\BibitemShut {NoStop}%
\bibitem [{\citenamefont {Akutsu}\ \emph {et~al.}(2021)\citenamefont {Akutsu}
  \emph {et~al.}}]{KAGRA:2020agh}%
  \BibitemOpen
  \bibfield  {author} {\bibinfo {author} {\bibfnamefont {T.}~\bibnamefont
  {Akutsu}} \emph {et~al.} (\bibinfo {collaboration} {KAGRA}),\ }\bibfield
  {title} {\bibinfo {title} {{Overview of {KAGRA}: Calibration, detector
  characterization, physical environmental monitors, and the geophysics
  interferometer}},\ }\href {https://doi.org/10.1093/ptep/ptab018} {\bibfield
  {journal} {\bibinfo  {journal} {PTEP}\ }\textbf {\bibinfo {volume} {2021}},\
  \bibinfo {pages} {05A102} (\bibinfo {year} {2021})},\ \Eprint
  {https://arxiv.org/abs/2009.09305} {arXiv:2009.09305 [gr-qc]} \BibitemShut
  {NoStop}%
\bibitem [{\citenamefont {Abbott}\ \emph
  {et~al.}(2022{\natexlab{a}})\citenamefont {Abbott} \emph {et~al.}}]{GWTC3}%
  \BibitemOpen
  \bibfield  {author} {\bibinfo {author} {\bibfnamefont {R.}~\bibnamefont
  {Abbott}} \emph {et~al.},\ }\bibfield  {title} {\bibinfo {title} {{GWTC}-3:
  Compact binary coalescences observed by {LIGO} and {Virgo} during the second
  part of the third observing run},\ }\href@noop {} {\bibfield  {journal}
  {\bibinfo  {journal} {arXiv:2111.03606, submitted for publication in Phys.
  Rev. X}\ } (\bibinfo {year} {2022}{\natexlab{a}})}\BibitemShut {NoStop}%
\bibitem [{\citenamefont {Abbott}\ \emph
  {et~al.}(2022{\natexlab{b}})\citenamefont {Abbott} \emph
  {et~al.}}]{O3ratepop}%
  \BibitemOpen
  \bibfield  {author} {\bibinfo {author} {\bibfnamefont {R.}~\bibnamefont
  {Abbott}} \emph {et~al.},\ }\bibfield  {title} {\bibinfo {title} {The
  population of merging compact binaries inferred using gravitational waves
  through {GWTC}-3},\ }\href@noop {} {\bibfield  {journal} {\bibinfo  {journal}
  {arXiv:2111.03634, submitted for publication in Phys. Rev. X}\ } (\bibinfo
  {year} {2022}{\natexlab{b}})}\BibitemShut {NoStop}%
\bibitem [{\citenamefont {Abbott}\ \emph
  {et~al.}(2022{\natexlab{c}})\citenamefont {Abbott} \emph {et~al.}}]{O3TGR}%
  \BibitemOpen
  \bibfield  {author} {\bibinfo {author} {\bibfnamefont {R.}~\bibnamefont
  {Abbott}} \emph {et~al.},\ }\bibfield  {title} {\bibinfo {title} {Tests of
  general relativity with {GWTC}-3},\ }\href@noop {} {\bibfield  {journal}
  {\bibinfo  {journal} {arXiv:2112.06861, accepted for publication in Phys.
  Rev. D}\ } (\bibinfo {year} {2022}{\natexlab{c}})}\BibitemShut {NoStop}%
\bibitem [{\citenamefont {Abbott}\ \emph
  {et~al.}(2022{\natexlab{d}})\citenamefont {Abbott} \emph {et~al.}}]{O3H0}%
  \BibitemOpen
  \bibfield  {author} {\bibinfo {author} {\bibfnamefont {R.}~\bibnamefont
  {Abbott}} \emph {et~al.},\ }\bibfield  {title} {\bibinfo {title} {Constraints
  on the cosmic expansion history from {GWTC}-3},\ }\href@noop {} {\bibfield
  {journal} {\bibinfo  {journal} {arXiv:2111.03604, accepted for publication
  Ap.J.}\ } (\bibinfo {year} {2022}{\natexlab{d}})}\BibitemShut {NoStop}%
\bibitem [{\citenamefont {Abbott}\ \emph
  {et~al.}(2018{\natexlab{a}})\citenamefont {Abbott} \emph
  {et~al.}}]{GW170817_EOS}%
  \BibitemOpen
  \bibfield  {author} {\bibinfo {author} {\bibfnamefont {B.~P.}\ \bibnamefont
  {Abbott}} \emph {et~al.} (\bibinfo {collaboration} {The LIGO Scientific
  Collaboration and the Virgo Collaboration}),\ }\bibfield  {title} {\bibinfo
  {title} {{GW170817}: Measurements of neutron star radii and equation of
  state},\ }\href {https://doi.org/10.1103/PhysRevLett.121.161101} {\bibfield
  {journal} {\bibinfo  {journal} {Phys. Rev. Lett.}\ }\textbf {\bibinfo
  {volume} {121}},\ \bibinfo {pages} {161101} (\bibinfo {year}
  {2018}{\natexlab{a}})}\BibitemShut {NoStop}%
\bibitem [{\citenamefont {Abbott}\ \emph {et~al.}(2020)\citenamefont {Abbott}
  \emph {et~al.}}]{LVKlivingreview}%
  \BibitemOpen
  \bibfield  {author} {\bibinfo {author} {\bibfnamefont {B.}~\bibnamefont
  {Abbott}} \emph {et~al.},\ }\bibfield  {title} {\bibinfo {title} {Prospects
  for observing and localizing gravitational-wave transients with {A}dvanced
  {LIGO}, {A}dvanced {Virgo} and {KAGRA}},\ }\href@noop {} {\bibfield
  {journal} {\bibinfo  {journal} {Living Rev Relativ}\ }\textbf {\bibinfo
  {volume} {23}},\ \bibinfo {pages} {3} (\bibinfo {year} {2020})}\BibitemShut
  {NoStop}%
\bibitem [{\citenamefont {Maggiore}(2000)}]{maggiore}%
  \BibitemOpen
  \bibfield  {author} {\bibinfo {author} {\bibfnamefont {M.}~\bibnamefont
  {Maggiore}},\ }\bibfield  {title} {\bibinfo {title} {Gravitational wave
  experiments and early universe cosmology},\ }\href
  {https://doi.org/https://doi.org/10.1016/S0370-1573(99)00102-7} {\bibfield
  {journal} {\bibinfo  {journal} {Physics Reports}\ }\textbf {\bibinfo {volume}
  {331}},\ \bibinfo {pages} {283} (\bibinfo {year} {2000})}\BibitemShut
  {NoStop}%
\bibitem [{\citenamefont {Regimbau}(2011)}]{regimbau_review}%
  \BibitemOpen
  \bibfield  {author} {\bibinfo {author} {\bibfnamefont {T.}~\bibnamefont
  {Regimbau}},\ }\bibfield  {title} {\bibinfo {title} {The astrophysical
  gravitational wave stochastic background},\ }\href
  {https://doi.org/10.1088/1674-4527/11/4/001} {\bibfield  {journal} {\bibinfo
  {journal} {Research in Astronomy and Astrophysics}\ }\textbf {\bibinfo
  {volume} {11}},\ \bibinfo {pages} {369} (\bibinfo {year} {2011})}\BibitemShut
  {NoStop}%
\bibitem [{\citenamefont {{Grishchuk}}(1975)}]{grishchuk}%
  \BibitemOpen
  \bibfield  {author} {\bibinfo {author} {\bibfnamefont {L.~P.}\ \bibnamefont
  {{Grishchuk}}},\ }\bibfield  {title} {\bibinfo {title} {{Amplification of
  gravitational waves in an isotropic universe}},\ }\href
  {https://ui.adsabs.harvard.edu/abs/1975JETP...40..409G} {\bibfield  {journal}
  {\bibinfo  {journal} {Soviet Journal of Experimental and Theoretical
  Physics}\ }\textbf {\bibinfo {volume} {40}},\ \bibinfo {pages} {409}
  (\bibinfo {year} {1975})}\BibitemShut {NoStop}%
\bibitem [{\citenamefont {{Bar-Kana}}(1994)}]{barkana}%
  \BibitemOpen
  \bibfield  {author} {\bibinfo {author} {\bibfnamefont {R.}~\bibnamefont
  {{Bar-Kana}}},\ }\bibfield  {title} {\bibinfo {title} {{Limits on direct
  detection of gravitational waves}},\ }\href
  {https://doi.org/10.1103/PhysRevD.50.1157} {\bibfield  {journal} {\bibinfo
  {journal} {\prd}\ }\textbf {\bibinfo {volume} {50}},\ \bibinfo {pages} {1157}
  (\bibinfo {year} {1994})},\ \Eprint {https://arxiv.org/abs/astro-ph/9401050}
  {astro-ph/9401050} \BibitemShut {NoStop}%
\bibitem [{\citenamefont {{Starobinski{\v i}}}(1979)}]{starob}%
  \BibitemOpen
  \bibfield  {author} {\bibinfo {author} {\bibfnamefont {A.~A.}\ \bibnamefont
  {{Starobinski{\v i}}}},\ }\bibfield  {title} {\bibinfo {title} {{Spectrum of
  relict gravitational radiation and the early state of the universe}},\ }\href
  {https://ui.adsabs.harvard.edu/abs/1979JETPL..30..682S} {\bibfield  {journal}
  {\bibinfo  {journal} {Soviet Journal of Experimental and Theoretical Physics
  Letters}\ }\textbf {\bibinfo {volume} {30}},\ \bibinfo {pages} {682}
  (\bibinfo {year} {1979})}\BibitemShut {NoStop}%
\bibitem [{\citenamefont {Turner}(1997)}]{turner}%
  \BibitemOpen
  \bibfield  {author} {\bibinfo {author} {\bibfnamefont {M.}~\bibnamefont
  {Turner}},\ }\bibfield  {title} {\bibinfo {title} {Detectability of
  inflation-produced gravitational waves},\ }\href@noop {} {\bibfield
  {journal} {\bibinfo  {journal} {\prd}\ }\textbf {\bibinfo {volume} {55}},\
  \bibinfo {pages} {R435} (\bibinfo {year} {1997})}\BibitemShut {NoStop}%
\bibitem [{\citenamefont {{Barnaby}}\ \emph {et~al.}(2012)\citenamefont
  {{Barnaby}}, \citenamefont {{Pajer}},\ and\ \citenamefont
  {{Peloso}}}]{peloso_parviol}%
  \BibitemOpen
  \bibfield  {author} {\bibinfo {author} {\bibfnamefont {N.}~\bibnamefont
  {{Barnaby}}}, \bibinfo {author} {\bibfnamefont {E.}~\bibnamefont {{Pajer}}},\
  and\ \bibinfo {author} {\bibfnamefont {M.}~\bibnamefont {{Peloso}}},\
  }\bibfield  {title} {\bibinfo {title} {{Gauge field production in axion
  inflation: Consequences for monodromy, non-Gaussianity in the {CMB}, and
  gravitational waves at interferometers}},\ }\href
  {https://doi.org/10.1103/PhysRevD.85.023525} {\bibfield  {journal} {\bibinfo
  {journal} {\prd}\ }\textbf {\bibinfo {volume} {85}},\ \bibinfo {eid} {023525}
  (\bibinfo {year} {2012})},\ \Eprint {https://arxiv.org/abs/1110.3327}
  {arXiv:1110.3327 [astro-ph.CO]} \BibitemShut {NoStop}%
\bibitem [{\citenamefont {{Seto}}\ and\ \citenamefont {{Taruya}}(2007)}]{seto}%
  \BibitemOpen
  \bibfield  {author} {\bibinfo {author} {\bibfnamefont {N.}~\bibnamefont
  {{Seto}}}\ and\ \bibinfo {author} {\bibfnamefont {A.}~\bibnamefont
  {{Taruya}}},\ }\bibfield  {title} {\bibinfo {title} {{Measuring a
  Parity-Violation Signature in the Early Universe via Ground-Based Laser
  Interferometers}},\ }\href {https://doi.org/10.1103/PhysRevLett.99.121101}
  {\bibfield  {journal} {\bibinfo  {journal} {Physical Review Letters}\
  }\textbf {\bibinfo {volume} {99}},\ \bibinfo {eid} {121101} (\bibinfo {year}
  {2007})},\ \Eprint {https://arxiv.org/abs/0707.0535} {arXiv:0707.0535}
  \BibitemShut {NoStop}%
\bibitem [{\citenamefont {{Easther}}\ and\ \citenamefont
  {{Lim}}(2006)}]{eastherlim}%
  \BibitemOpen
  \bibfield  {author} {\bibinfo {author} {\bibfnamefont {R.}~\bibnamefont
  {{Easther}}}\ and\ \bibinfo {author} {\bibfnamefont {E.~A.}\ \bibnamefont
  {{Lim}}},\ }\bibfield  {title} {\bibinfo {title} {{Stochastic gravitational
  wave production after inflation}},\ }\href
  {https://doi.org/10.1088/1475-7516/2006/04/010} {\bibfield  {journal}
  {\bibinfo  {journal} {\jcap}\ }\textbf {\bibinfo {volume} {4}},\ \bibinfo
  {eid} {010} (\bibinfo {year} {2006})},\ \Eprint
  {https://arxiv.org/abs/astro-ph/0601617} {astro-ph/0601617} \BibitemShut
  {NoStop}%
\bibitem [{\citenamefont {Boyle}\ and\ \citenamefont
  {Buonanno}(2008)}]{boylebuonanno}%
  \BibitemOpen
  \bibfield  {author} {\bibinfo {author} {\bibfnamefont {L.}~\bibnamefont
  {Boyle}}\ and\ \bibinfo {author} {\bibfnamefont {A.}~\bibnamefont
  {Buonanno}},\ }\bibfield  {title} {\bibinfo {title} {Relating gravitational
  wave constraints from primordial nucleosynthesis, pulsar timing, laser
  interferometers, and the {CMB}: implications for the early universe},\
  }\href@noop {} {\bibfield  {journal} {\bibinfo  {journal} {\prd}\ }\textbf
  {\bibinfo {volume} {78}},\ \bibinfo {pages} {043531} (\bibinfo {year}
  {2008})}\BibitemShut {NoStop}%
\bibitem [{\citenamefont {{Witten}}(1984)}]{witten}%
  \BibitemOpen
  \bibfield  {author} {\bibinfo {author} {\bibfnamefont {E.}~\bibnamefont
  {{Witten}}},\ }\bibfield  {title} {\bibinfo {title} {{Cosmic separation of
  phases}},\ }\href {https://doi.org/10.1103/PhysRevD.30.272} {\bibfield
  {journal} {\bibinfo  {journal} {\prd}\ }\textbf {\bibinfo {volume} {30}},\
  \bibinfo {pages} {272} (\bibinfo {year} {1984})}\BibitemShut {NoStop}%
\bibitem [{\citenamefont {Hogan}(1986)}]{hogan}%
  \BibitemOpen
  \bibfield  {author} {\bibinfo {author} {\bibfnamefont {C.~J.}\ \bibnamefont
  {Hogan}},\ }\bibfield  {title} {\bibinfo {title} {{Gravitational radiation
  from cosmological phase transitions}},\ }\href@noop {} {\bibfield  {journal}
  {\bibinfo  {journal} {Mon. Not. Roy. Astron. Soc.}\ }\textbf {\bibinfo
  {volume} {218}},\ \bibinfo {pages} {629} (\bibinfo {year}
  {1986})}\BibitemShut {NoStop}%
\bibitem [{\citenamefont {{Kosowsky}}\ \emph {et~al.}(1992)\citenamefont
  {{Kosowsky}}, \citenamefont {{Turner}},\ and\ \citenamefont
  {{Watkins}}}]{kosowsky}%
  \BibitemOpen
  \bibfield  {author} {\bibinfo {author} {\bibfnamefont {A.}~\bibnamefont
  {{Kosowsky}}}, \bibinfo {author} {\bibfnamefont {M.~S.}\ \bibnamefont
  {{Turner}}},\ and\ \bibinfo {author} {\bibfnamefont {R.}~\bibnamefont
  {{Watkins}}},\ }\bibfield  {title} {\bibinfo {title} {{Gravitational waves
  from first-order cosmological phase transitions}},\ }\href
  {https://doi.org/10.1103/PhysRevLett.69.2026} {\bibfield  {journal} {\bibinfo
   {journal} {Physical Review Letters}\ }\textbf {\bibinfo {volume} {69}},\
  \bibinfo {pages} {2026} (\bibinfo {year} {1992})}\BibitemShut {NoStop}%
\bibitem [{\citenamefont {{Caprini}}\ \emph {et~al.}(2008)\citenamefont
  {{Caprini}}, \citenamefont {{Durrer}},\ and\ \citenamefont
  {{Servant}}}]{caprini1}%
  \BibitemOpen
  \bibfield  {author} {\bibinfo {author} {\bibfnamefont {C.}~\bibnamefont
  {{Caprini}}}, \bibinfo {author} {\bibfnamefont {R.}~\bibnamefont
  {{Durrer}}},\ and\ \bibinfo {author} {\bibfnamefont {G.}~\bibnamefont
  {{Servant}}},\ }\bibfield  {title} {\bibinfo {title} {{Gravitational wave
  generation from bubble collisions in first-order phase transitions: An
  analytic approach}},\ }\href {https://doi.org/10.1103/PhysRevD.77.124015}
  {\bibfield  {journal} {\bibinfo  {journal} {\prd}\ }\textbf {\bibinfo
  {volume} {77}},\ \bibinfo {eid} {124015} (\bibinfo {year} {2008})},\ \Eprint
  {https://arxiv.org/abs/0711.2593} {arXiv:0711.2593 [astro-ph]} \BibitemShut
  {NoStop}%
\bibitem [{\citenamefont {{Bin{\'e}truy}}\ \emph {et~al.}(2012)\citenamefont
  {{Bin{\'e}truy}}, \citenamefont {{Boh{\'e}}}, \citenamefont {{Caprini}},\
  and\ \citenamefont {{Dufaux}}}]{binetruy}%
  \BibitemOpen
  \bibfield  {author} {\bibinfo {author} {\bibfnamefont {P.}~\bibnamefont
  {{Bin{\'e}truy}}}, \bibinfo {author} {\bibfnamefont {A.}~\bibnamefont
  {{Boh{\'e}}}}, \bibinfo {author} {\bibfnamefont {C.}~\bibnamefont
  {{Caprini}}},\ and\ \bibinfo {author} {\bibfnamefont {J.-F.}\ \bibnamefont
  {{Dufaux}}},\ }\bibfield  {title} {\bibinfo {title} {{Cosmological
  backgrounds of gravitational waves and eLISA/NGO: phase transitions, cosmic
  strings and other sources}},\ }\href
  {https://doi.org/10.1088/1475-7516/2012/06/027} {\bibfield  {journal}
  {\bibinfo  {journal} {\jcap}\ }\textbf {\bibinfo {volume} {2012}},\ \bibinfo
  {eid} {027} (\bibinfo {year} {2012})},\ \Eprint
  {https://arxiv.org/abs/1201.0983} {arXiv:1201.0983 [gr-qc]} \BibitemShut
  {NoStop}%
\bibitem [{\citenamefont {{Caprini}}\ \emph {et~al.}(2016)\citenamefont
  {{Caprini}} \emph {et~al.}}]{caprini2}%
  \BibitemOpen
  \bibfield  {author} {\bibinfo {author} {\bibfnamefont {C.}~\bibnamefont
  {{Caprini}}} \emph {et~al.},\ }\bibfield  {title} {\bibinfo {title} {{Science
  with the space-based interferometer eLISA. II: gravitational waves from
  cosmological phase transitions}},\ }\href
  {https://doi.org/10.1088/1475-7516/2016/04/001} {\bibfield  {journal}
  {\bibinfo  {journal} {\jcap}\ }\textbf {\bibinfo {volume} {2016}},\ \bibinfo
  {eid} {001} (\bibinfo {year} {2016})},\ \Eprint
  {https://arxiv.org/abs/1512.06239} {arXiv:1512.06239 [astro-ph.CO]}
  \BibitemShut {NoStop}%
\bibitem [{\citenamefont {{Fitz Axen}}\ \emph {et~al.}(2018)\citenamefont
  {{Fitz Axen}}, \citenamefont {{Banagiri}}, \citenamefont {{Matas}},
  \citenamefont {{Caprini}},\ and\ \citenamefont {{Mandic}}}]{margotpaper}%
  \BibitemOpen
  \bibfield  {author} {\bibinfo {author} {\bibfnamefont {M.}~\bibnamefont
  {{Fitz Axen}}}, \bibinfo {author} {\bibfnamefont {S.}~\bibnamefont
  {{Banagiri}}}, \bibinfo {author} {\bibfnamefont {A.}~\bibnamefont {{Matas}}},
  \bibinfo {author} {\bibfnamefont {C.}~\bibnamefont {{Caprini}}},\ and\
  \bibinfo {author} {\bibfnamefont {V.}~\bibnamefont {{Mandic}}},\ }\bibfield
  {title} {\bibinfo {title} {{Multiwavelength observations of cosmological
  phase transitions using LISA and Cosmic Explorer}},\ }\href
  {https://doi.org/10.1103/PhysRevD.98.103508} {\bibfield  {journal} {\bibinfo
  {journal} {\prd}\ }\textbf {\bibinfo {volume} {98}},\ \bibinfo {eid} {103508}
  (\bibinfo {year} {2018})},\ \Eprint {https://arxiv.org/abs/1806.02500}
  {arXiv:1806.02500 [astro-ph.IM]} \BibitemShut {NoStop}%
\bibitem [{\citenamefont {Caldwell}\ and\ \citenamefont
  {Allen}(1992)}]{caldwellallen}%
  \BibitemOpen
  \bibfield  {author} {\bibinfo {author} {\bibfnamefont {R.}~\bibnamefont
  {Caldwell}}\ and\ \bibinfo {author} {\bibfnamefont {B.}~\bibnamefont
  {Allen}},\ }\bibfield  {title} {\bibinfo {title} {Cosmological constraints on
  cosmic-string gravitational radiation},\ }\href@noop {} {\bibfield  {journal}
  {\bibinfo  {journal} {\prd}\ }\textbf {\bibinfo {volume} {45}},\ \bibinfo
  {pages} {3447} (\bibinfo {year} {1992})}\BibitemShut {NoStop}%
\bibitem [{\citenamefont {{Damour}}\ and\ \citenamefont
  {{Vilenkin}}(2000)}]{DV1}%
  \BibitemOpen
  \bibfield  {author} {\bibinfo {author} {\bibfnamefont {T.}~\bibnamefont
  {{Damour}}}\ and\ \bibinfo {author} {\bibfnamefont {A.}~\bibnamefont
  {{Vilenkin}}},\ }\bibfield  {title} {\bibinfo {title} {{Gravitational Wave
  Bursts from Cosmic Strings}},\ }\href
  {https://doi.org/10.1103/PhysRevLett.85.3761} {\bibfield  {journal} {\bibinfo
   {journal} {Physical Review Letters}\ }\textbf {\bibinfo {volume} {85}},\
  \bibinfo {pages} {3761} (\bibinfo {year} {2000})},\ \Eprint
  {https://arxiv.org/abs/gr-qc/0004075} {gr-qc/0004075} \BibitemShut {NoStop}%
\bibitem [{\citenamefont {{Damour}}\ and\ \citenamefont
  {{Vilenkin}}(2005)}]{DV2}%
  \BibitemOpen
  \bibfield  {author} {\bibinfo {author} {\bibfnamefont {T.}~\bibnamefont
  {{Damour}}}\ and\ \bibinfo {author} {\bibfnamefont {A.}~\bibnamefont
  {{Vilenkin}}},\ }\bibfield  {title} {\bibinfo {title} {{Gravitational
  radiation from cosmic (super)strings: Bursts, stochastic background, and
  observational windows}},\ }\href {https://doi.org/10.1103/PhysRevD.71.063510}
  {\bibfield  {journal} {\bibinfo  {journal} {\prd}\ }\textbf {\bibinfo
  {volume} {71}},\ \bibinfo {eid} {063510} (\bibinfo {year} {2005})},\ \Eprint
  {https://arxiv.org/abs/hep-th/0410222} {hep-th/0410222} \BibitemShut
  {NoStop}%
\bibitem [{\citenamefont {{Siemens}}\ \emph {et~al.}(2007)\citenamefont
  {{Siemens}}, \citenamefont {{Mandic}},\ and\ \citenamefont
  {{Creighton}}}]{cosmstrpaper}%
  \BibitemOpen
  \bibfield  {author} {\bibinfo {author} {\bibfnamefont {X.}~\bibnamefont
  {{Siemens}}}, \bibinfo {author} {\bibfnamefont {V.}~\bibnamefont
  {{Mandic}}},\ and\ \bibinfo {author} {\bibfnamefont {J.}~\bibnamefont
  {{Creighton}}},\ }\bibfield  {title} {\bibinfo {title} {{Gravitational-Wave
  Stochastic Background from Cosmic Strings}},\ }\href
  {https://doi.org/10.1103/PhysRevLett.98.111101} {\bibfield  {journal}
  {\bibinfo  {journal} {Physical Review Letters}\ }\textbf {\bibinfo {volume}
  {98}},\ \bibinfo {eid} {111101} (\bibinfo {year} {2007})},\ \Eprint
  {https://arxiv.org/abs/astro-ph/0610920} {astro-ph/0610920} \BibitemShut
  {NoStop}%
\bibitem [{\citenamefont {{{\"O}lmez}}\ \emph {et~al.}(2010)\citenamefont
  {{{\"O}lmez}}, \citenamefont {{Mandic}},\ and\ \citenamefont
  {{Siemens}}}]{olmez1}%
  \BibitemOpen
  \bibfield  {author} {\bibinfo {author} {\bibfnamefont {S.}~\bibnamefont
  {{{\"O}lmez}}}, \bibinfo {author} {\bibfnamefont {V.}~\bibnamefont
  {{Mandic}}},\ and\ \bibinfo {author} {\bibfnamefont {X.}~\bibnamefont
  {{Siemens}}},\ }\bibfield  {title} {\bibinfo {title} {{Gravitational-wave
  stochastic background from kinks and cusps on cosmic strings}},\ }\href
  {https://doi.org/10.1103/PhysRevD.81.104028} {\bibfield  {journal} {\bibinfo
  {journal} {\prd}\ }\textbf {\bibinfo {volume} {81}},\ \bibinfo {eid} {104028}
  (\bibinfo {year} {2010})},\ \Eprint {https://arxiv.org/abs/1004.0890}
  {arXiv:1004.0890 [astro-ph.CO]} \BibitemShut {NoStop}%
\bibitem [{\citenamefont {Copeland}\ \emph {et~al.}(2004)\citenamefont
  {Copeland}, \citenamefont {Myers},\ and\ \citenamefont
  {Polchinski}}]{polchinski}%
  \BibitemOpen
  \bibfield  {author} {\bibinfo {author} {\bibfnamefont {E.~J.}\ \bibnamefont
  {Copeland}}, \bibinfo {author} {\bibfnamefont {R.~C.}\ \bibnamefont
  {Myers}},\ and\ \bibinfo {author} {\bibfnamefont {J.}~\bibnamefont
  {Polchinski}},\ }\bibfield  {title} {\bibinfo {title} {Cosmic f- and
  d-strings},\ }\href {https://doi.org/10.1088/1126-6708/2004/06/013}
  {\bibfield  {journal} {\bibinfo  {journal} {Journal of High Energy Physics}\
  }\textbf {\bibinfo {volume} {2004}},\ \bibinfo {pages} {013} (\bibinfo {year}
  {2004})}\BibitemShut {NoStop}%
\bibitem [{\citenamefont {{Siemens}}\ \emph {et~al.}(2006)\citenamefont
  {{Siemens}} \emph {et~al.}}]{siemens}%
  \BibitemOpen
  \bibfield  {author} {\bibinfo {author} {\bibfnamefont {X.}~\bibnamefont
  {{Siemens}}} \emph {et~al.},\ }\bibfield  {title} {\bibinfo {title}
  {{Gravitational wave bursts from cosmic (super)strings: Quantitative analysis
  and constraints}},\ }\href {https://doi.org/10.1103/PhysRevD.73.105001}
  {\bibfield  {journal} {\bibinfo  {journal} {\prd}\ }\textbf {\bibinfo
  {volume} {73}},\ \bibinfo {eid} {105001} (\bibinfo {year} {2006})},\ \Eprint
  {https://arxiv.org/abs/gr-qc/0603115} {arXiv:gr-qc/0603115 [gr-qc]}
  \BibitemShut {NoStop}%
\bibitem [{\citenamefont {{Lorenz}}\ \emph {et~al.}(2010)\citenamefont
  {{Lorenz}}, \citenamefont {{Ringeval}},\ and\ \citenamefont
  {{Sakellariadou}}}]{ringeval}%
  \BibitemOpen
  \bibfield  {author} {\bibinfo {author} {\bibfnamefont {L.}~\bibnamefont
  {{Lorenz}}}, \bibinfo {author} {\bibfnamefont {C.}~\bibnamefont
  {{Ringeval}}},\ and\ \bibinfo {author} {\bibfnamefont {M.}~\bibnamefont
  {{Sakellariadou}}},\ }\bibfield  {title} {\bibinfo {title} {{Cosmic string
  loop distribution on all length scales and at any redshift}},\ }\href
  {https://doi.org/10.1088/1475-7516/2010/10/003} {\bibfield  {journal}
  {\bibinfo  {journal} {\jcap}\ }\textbf {\bibinfo {volume} {2010}},\ \bibinfo
  {eid} {003} (\bibinfo {year} {2010})},\ \Eprint
  {https://arxiv.org/abs/1006.0931} {arXiv:1006.0931 [astro-ph.CO]}
  \BibitemShut {NoStop}%
\bibitem [{\citenamefont {{Blanco-Pillado}}\ \emph {et~al.}(2014)\citenamefont
  {{Blanco-Pillado}}, \citenamefont {{Olum}},\ and\ \citenamefont
  {{Shlaer}}}]{olum}%
  \BibitemOpen
  \bibfield  {author} {\bibinfo {author} {\bibfnamefont {J.~J.}\ \bibnamefont
  {{Blanco-Pillado}}}, \bibinfo {author} {\bibfnamefont {K.~D.}\ \bibnamefont
  {{Olum}}},\ and\ \bibinfo {author} {\bibfnamefont {B.}~\bibnamefont
  {{Shlaer}}},\ }\bibfield  {title} {\bibinfo {title} {{Number of cosmic string
  loops}},\ }\href {https://doi.org/10.1103/PhysRevD.89.023512} {\bibfield
  {journal} {\bibinfo  {journal} {\prd}\ }\textbf {\bibinfo {volume} {89}},\
  \bibinfo {eid} {023512} (\bibinfo {year} {2014})},\ \Eprint
  {https://arxiv.org/abs/1309.6637} {arXiv:1309.6637 [astro-ph.CO]}
  \BibitemShut {NoStop}%
\bibitem [{\citenamefont {Abbott}\ \emph
  {et~al.}(2021{\natexlab{a}})\citenamefont {Abbott} \emph
  {et~al.}}]{O3cosmstr}%
  \BibitemOpen
  \bibfield  {author} {\bibinfo {author} {\bibfnamefont {R.}~\bibnamefont
  {Abbott}} \emph {et~al.} (\bibinfo {collaboration} {LIGO Scientific
  Collaboration, Virgo Collaboration, and KAGRA Collaboration}),\ }\bibfield
  {title} {\bibinfo {title} {Constraints on cosmic strings using data from the
  third {A}dvanced {LIGO}--{Virgo} observing run},\ }\href
  {https://doi.org/10.1103/PhysRevLett.126.241102} {\bibfield  {journal}
  {\bibinfo  {journal} {Phys. Rev. Lett.}\ }\textbf {\bibinfo {volume} {126}},\
  \bibinfo {pages} {241102} (\bibinfo {year} {2021}{\natexlab{a}})}\BibitemShut
  {NoStop}%
\bibitem [{\citenamefont {{Jenkins}}\ and\ \citenamefont
  {{Sakellariadou}}(2018)}]{jenkins_cosmstr}%
  \BibitemOpen
  \bibfield  {author} {\bibinfo {author} {\bibfnamefont {A.~C.}\ \bibnamefont
  {{Jenkins}}}\ and\ \bibinfo {author} {\bibfnamefont {M.}~\bibnamefont
  {{Sakellariadou}}},\ }\bibfield  {title} {\bibinfo {title} {{Anisotropies in
  the stochastic gravitational-wave background: Formalism and the cosmic string
  case}},\ }\href {https://doi.org/10.1103/PhysRevD.98.063509} {\bibfield
  {journal} {\bibinfo  {journal} {\prd}\ }\textbf {\bibinfo {volume} {98}},\
  \bibinfo {eid} {063509} (\bibinfo {year} {2018})},\ \Eprint
  {https://arxiv.org/abs/1802.06046} {arXiv:1802.06046} \BibitemShut {NoStop}%
\bibitem [{\citenamefont {Regimbau}\ and\ \citenamefont
  {de~Freitas~Pacheco}(2006)}]{regfrei}%
  \BibitemOpen
  \bibfield  {author} {\bibinfo {author} {\bibfnamefont {T.}~\bibnamefont
  {Regimbau}}\ and\ \bibinfo {author} {\bibfnamefont {J.~A.}\ \bibnamefont
  {de~Freitas~Pacheco}},\ }\bibfield  {title} {\bibinfo {title} {Stochastic
  background from coalescences of neutron star{\textendash}neutron star
  binaries},\ }\href {https://doi.org/10.1086/500190} {\bibfield  {journal}
  {\bibinfo  {journal} {The Astrophysical Journal}\ }\textbf {\bibinfo {volume}
  {642}},\ \bibinfo {pages} {455} (\bibinfo {year} {2006})}\BibitemShut
  {NoStop}%
\bibitem [{\citenamefont {Zhu}\ \emph {et~al.}(2011)\citenamefont {Zhu},
  \citenamefont {Howell}, \citenamefont {Regimbau}, \citenamefont {Blair},\
  and\ \citenamefont {Zhu}}]{zhu_cbc}%
  \BibitemOpen
  \bibfield  {author} {\bibinfo {author} {\bibfnamefont {X.-J.}\ \bibnamefont
  {Zhu}}, \bibinfo {author} {\bibfnamefont {E.}~\bibnamefont {Howell}},
  \bibinfo {author} {\bibfnamefont {T.}~\bibnamefont {Regimbau}}, \bibinfo
  {author} {\bibfnamefont {D.}~\bibnamefont {Blair}},\ and\ \bibinfo {author}
  {\bibfnamefont {Z.-H.}\ \bibnamefont {Zhu}},\ }\bibfield  {title} {\bibinfo
  {title} {Stochastic gravitational wave background from coalescing binary
  black holes},\ }\href {https://doi.org/10.1088/0004-637x/739/2/86} {\bibfield
   {journal} {\bibinfo  {journal} {The Astrophysical Journal}\ }\textbf
  {\bibinfo {volume} {739}},\ \bibinfo {pages} {86} (\bibinfo {year}
  {2011})}\BibitemShut {NoStop}%
\bibitem [{\citenamefont {Marassi}\ \emph
  {et~al.}(2011{\natexlab{a}})\citenamefont {Marassi}, \citenamefont
  {Schneider}, \citenamefont {Corvino}, \citenamefont {Ferrari},\ and\
  \citenamefont {Zwart}}]{marassi_cbc}%
  \BibitemOpen
  \bibfield  {author} {\bibinfo {author} {\bibfnamefont {S.}~\bibnamefont
  {Marassi}}, \bibinfo {author} {\bibfnamefont {R.}~\bibnamefont {Schneider}},
  \bibinfo {author} {\bibfnamefont {G.}~\bibnamefont {Corvino}}, \bibinfo
  {author} {\bibfnamefont {V.}~\bibnamefont {Ferrari}},\ and\ \bibinfo {author}
  {\bibfnamefont {S.~P.}\ \bibnamefont {Zwart}},\ }\bibfield  {title} {\bibinfo
  {title} {Imprint of the merger and ring-down on the gravitational wave
  background from black hole binaries coalescence},\ }\href
  {https://doi.org/10.1103/PhysRevD.84.124037} {\bibfield  {journal} {\bibinfo
  {journal} {Phys. Rev. D}\ }\textbf {\bibinfo {volume} {84}},\ \bibinfo
  {pages} {124037} (\bibinfo {year} {2011}{\natexlab{a}})}\BibitemShut
  {NoStop}%
\bibitem [{\citenamefont {Rosado}(2011)}]{rosado}%
  \BibitemOpen
  \bibfield  {author} {\bibinfo {author} {\bibfnamefont {P.~A.}\ \bibnamefont
  {Rosado}},\ }\bibfield  {title} {\bibinfo {title} {Gravitational wave
  background from binary systems},\ }\href
  {https://doi.org/10.1103/PhysRevD.84.084004} {\bibfield  {journal} {\bibinfo
  {journal} {Phys. Rev. D}\ }\textbf {\bibinfo {volume} {84}},\ \bibinfo
  {pages} {084004} (\bibinfo {year} {2011})}\BibitemShut {NoStop}%
\bibitem [{\citenamefont {Regimbau}\ and\ \citenamefont
  {Mandic}(2008)}]{regman}%
  \BibitemOpen
  \bibfield  {author} {\bibinfo {author} {\bibfnamefont {T.}~\bibnamefont
  {Regimbau}}\ and\ \bibinfo {author} {\bibfnamefont {V.}~\bibnamefont
  {Mandic}},\ }\bibfield  {title} {\bibinfo {title} {Astrophysical sources of a
  stochastic gravitational-wave background},\ }\href
  {https://doi.org/10.1088/0264-9381/25/18/184018} {\bibfield  {journal}
  {\bibinfo  {journal} {Classical and Quantum Gravity}\ }\textbf {\bibinfo
  {volume} {25}},\ \bibinfo {pages} {184018} (\bibinfo {year}
  {2008})}\BibitemShut {NoStop}%
\bibitem [{\citenamefont {Wu}\ \emph {et~al.}(2012)\citenamefont {Wu},
  \citenamefont {Mandic},\ and\ \citenamefont {Regimbau}}]{wu_cbc}%
  \BibitemOpen
  \bibfield  {author} {\bibinfo {author} {\bibfnamefont {C.}~\bibnamefont
  {Wu}}, \bibinfo {author} {\bibfnamefont {V.}~\bibnamefont {Mandic}},\ and\
  \bibinfo {author} {\bibfnamefont {T.}~\bibnamefont {Regimbau}},\ }\bibfield
  {title} {\bibinfo {title} {Accessibility of the gravitational-wave background
  due to binary coalescences to second and third generation gravitational-wave
  detectors},\ }\href {https://doi.org/10.1103/PhysRevD.85.104024} {\bibfield
  {journal} {\bibinfo  {journal} {Phys. Rev. D}\ }\textbf {\bibinfo {volume}
  {85}},\ \bibinfo {pages} {104024} (\bibinfo {year} {2012})}\BibitemShut
  {NoStop}%
\bibitem [{\citenamefont {{Abbott}}\ \emph {et~al.}(2016)\citenamefont
  {{Abbott}}, \citenamefont {others}, \citenamefont {{LIGO Scientific
  Collaboration}},\ and\ \citenamefont {{Virgo
  Collaboration}}}]{GW150914stoch}%
  \BibitemOpen
  \bibfield  {author} {\bibinfo {author} {\bibfnamefont {B.~P.}\ \bibnamefont
  {{Abbott}}}, \bibinfo {author} {\bibnamefont {others}}, \bibinfo {author}
  {\bibnamefont {{LIGO Scientific Collaboration}}},\ and\ \bibinfo {author}
  {\bibnamefont {{Virgo Collaboration}}},\ }\bibfield  {title} {\bibinfo
  {title} {{GW150914: Implications for the Stochastic Gravitational-Wave
  Background from Binary Black Holes}},\ }\href
  {https://doi.org/10.1103/PhysRevLett.116.131102} {\bibfield  {journal}
  {\bibinfo  {journal} {\prl}\ }\textbf {\bibinfo {volume} {116}},\ \bibinfo
  {eid} {131102} (\bibinfo {year} {2016})},\ \Eprint
  {https://arxiv.org/abs/1602.03847} {arXiv:1602.03847 [gr-qc]} \BibitemShut
  {NoStop}%
\bibitem [{\citenamefont {Abbott}\ \emph
  {et~al.}(2018{\natexlab{b}})\citenamefont {Abbott} \emph
  {et~al.}}]{GW170817stoch}%
  \BibitemOpen
  \bibfield  {author} {\bibinfo {author} {\bibfnamefont {B.~P.}\ \bibnamefont
  {Abbott}} \emph {et~al.},\ }\bibfield  {title} {\bibinfo {title} {{GW170817:
  Implications for the Stochastic Gravitational-Wave Background from Compact
  Binary Coalescences}},\ }\href
  {https://doi.org/10.1103/PhysRevLett.120.091101} {\bibfield  {journal}
  {\bibinfo  {journal} {\prl}\ }\textbf {\bibinfo {volume} {120}},\ \bibinfo
  {eid} {091101} (\bibinfo {year} {2018}{\natexlab{b}})},\ \Eprint
  {https://arxiv.org/abs/1710.05837} {arXiv:1710.05837 [gr-qc]} \BibitemShut
  {NoStop}%
\bibitem [{\citenamefont {Cutler}(2002)}]{cutler}%
  \BibitemOpen
  \bibfield  {author} {\bibinfo {author} {\bibfnamefont {C.}~\bibnamefont
  {Cutler}},\ }\bibfield  {title} {\bibinfo {title} {Gravitational waves from
  neutron stars with large toroidal $b$ fields},\ }\href
  {https://doi.org/10.1103/PhysRevD.66.084025} {\bibfield  {journal} {\bibinfo
  {journal} {Phys. Rev. D}\ }\textbf {\bibinfo {volume} {66}},\ \bibinfo
  {pages} {084025} (\bibinfo {year} {2002})}\BibitemShut {NoStop}%
\bibitem [{\citenamefont {Bonazzola}\ and\ \citenamefont
  {Gourgoulhon}(1996)}]{bonazzola}%
  \BibitemOpen
  \bibfield  {author} {\bibinfo {author} {\bibfnamefont {S.}~\bibnamefont
  {Bonazzola}}\ and\ \bibinfo {author} {\bibfnamefont {E.}~\bibnamefont
  {Gourgoulhon}},\ }\bibfield  {title} {\bibinfo {title} {Gravitational waves
  from pulsars: emission by the magnetic field induced distortion},\
  }\href@noop {} {\bibfield  {journal} {\bibinfo  {journal} {Astron. Astrop.}\
  }\textbf {\bibinfo {volume} {312}},\ \bibinfo {pages} {675} (\bibinfo {year}
  {1996})}\BibitemShut {NoStop}%
\bibitem [{\citenamefont {Marassi}\ \emph
  {et~al.}(2011{\natexlab{b}})\citenamefont {Marassi} \emph
  {et~al.}}]{marassi_magnetar}%
  \BibitemOpen
  \bibfield  {author} {\bibinfo {author} {\bibfnamefont {S.}~\bibnamefont
  {Marassi}} \emph {et~al.},\ }\bibfield  {title} {\bibinfo {title}
  {{Stochastic background of gravitational waves emitted by magnetars}},\
  }\href {https://doi.org/10.1111/j.1365-2966.2010.17861.x} {\bibfield
  {journal} {\bibinfo  {journal} {Mon. Not. Roy. Astron. Soc.}\ }\textbf
  {\bibinfo {volume} {411}},\ \bibinfo {pages} {2549} (\bibinfo {year}
  {2011}{\natexlab{b}})},\ \Eprint {https://arxiv.org/abs/1009.1240}
  {arXiv:1009.1240 [astro-ph.CO]} \BibitemShut {NoStop}%
\bibitem [{\citenamefont {Owen}\ \emph {et~al.}(1998)\citenamefont {Owen} \emph
  {et~al.}}]{owen}%
  \BibitemOpen
  \bibfield  {author} {\bibinfo {author} {\bibfnamefont {B.~J.}\ \bibnamefont
  {Owen}} \emph {et~al.},\ }\bibfield  {title} {\bibinfo {title} {Gravitational
  waves from hot young rapidly rotating neutron stars},\ }\href
  {https://doi.org/10.1103/PhysRevD.58.084020} {\bibfield  {journal} {\bibinfo
  {journal} {Phys. Rev. D}\ }\textbf {\bibinfo {volume} {58}},\ \bibinfo
  {pages} {084020} (\bibinfo {year} {1998})}\BibitemShut {NoStop}%
\bibitem [{\citenamefont {Wu}\ \emph {et~al.}(2013)\citenamefont {Wu},
  \citenamefont {Mandic},\ and\ \citenamefont {Regimbau}}]{wu_mag}%
  \BibitemOpen
  \bibfield  {author} {\bibinfo {author} {\bibfnamefont {C.-J.}\ \bibnamefont
  {Wu}}, \bibinfo {author} {\bibfnamefont {V.}~\bibnamefont {Mandic}},\ and\
  \bibinfo {author} {\bibfnamefont {T.}~\bibnamefont {Regimbau}},\ }\bibfield
  {title} {\bibinfo {title} {Accessibility of the stochastic gravitational wave
  background from magnetars to the interferometric gravitational wave
  detectors},\ }\href {https://doi.org/10.1103/PhysRevD.87.042002} {\bibfield
  {journal} {\bibinfo  {journal} {Phys. Rev. D}\ }\textbf {\bibinfo {volume}
  {87}},\ \bibinfo {pages} {042002} (\bibinfo {year} {2013})}\BibitemShut
  {NoStop}%
\bibitem [{\citenamefont {Coward}\ \emph {et~al.}(2002)\citenamefont {Coward},
  \citenamefont {Burman},\ and\ \citenamefont {Blair}}]{SNe}%
  \BibitemOpen
  \bibfield  {author} {\bibinfo {author} {\bibfnamefont {D.~M.}\ \bibnamefont
  {Coward}}, \bibinfo {author} {\bibfnamefont {R.~R.}\ \bibnamefont {Burman}},\
  and\ \bibinfo {author} {\bibfnamefont {D.~G.}\ \bibnamefont {Blair}},\
  }\bibfield  {title} {\bibinfo {title} {{Simulating a stochastic background of
  gravitational waves from neutron star formation at cosmological distances}},\
  }\href {https://doi.org/10.1046/j.1365-8711.2002.04981.x} {\bibfield
  {journal} {\bibinfo  {journal} {Mon. Not. Roy. Astron. Soc.}\ }\textbf
  {\bibinfo {volume} {329}},\ \bibinfo {pages} {411} (\bibinfo {year}
  {2002})}\BibitemShut {NoStop}%
\bibitem [{\citenamefont {Marassi}\ \emph {et~al.}(2009)\citenamefont
  {Marassi}, \citenamefont {Schneider},\ and\ \citenamefont
  {Ferrari}}]{marassi_cc}%
  \BibitemOpen
  \bibfield  {author} {\bibinfo {author} {\bibfnamefont {S.}~\bibnamefont
  {Marassi}}, \bibinfo {author} {\bibfnamefont {R.}~\bibnamefont {Schneider}},\
  and\ \bibinfo {author} {\bibfnamefont {V.}~\bibnamefont {Ferrari}},\
  }\bibfield  {title} {\bibinfo {title} {{Gravitational wave backgrounds and
  the cosmic transition from Population III to Population II stars}},\ }\href
  {https://doi.org/10.1111/j.1365-2966.2009.15120.x} {\bibfield  {journal}
  {\bibinfo  {journal} {Mon. Not. Roy. Astron. Soc.}\ }\textbf {\bibinfo
  {volume} {398}},\ \bibinfo {pages} {293} (\bibinfo {year} {2009})},\ \Eprint
  {https://arxiv.org/abs/0906.0461} {arXiv:0906.0461 [astro-ph.CO]}
  \BibitemShut {NoStop}%
\bibitem [{\citenamefont {Sandick}\ \emph {et~al.}(2006)\citenamefont {Sandick}
  \emph {et~al.}}]{firststars}%
  \BibitemOpen
  \bibfield  {author} {\bibinfo {author} {\bibfnamefont {P.}~\bibnamefont
  {Sandick}} \emph {et~al.},\ }\bibfield  {title} {\bibinfo {title}
  {Gravitational waves from the first stars},\ }\href
  {https://doi.org/10.1103/PhysRevD.73.104024} {\bibfield  {journal} {\bibinfo
  {journal} {Phys. Rev. D}\ }\textbf {\bibinfo {volume} {73}},\ \bibinfo
  {pages} {104024} (\bibinfo {year} {2006})}\BibitemShut {NoStop}%
\bibitem [{\citenamefont {Buonanno}\ \emph {et~al.}(2005)\citenamefont
  {Buonanno} \emph {et~al.}}]{buonanno_cc}%
  \BibitemOpen
  \bibfield  {author} {\bibinfo {author} {\bibfnamefont {A.}~\bibnamefont
  {Buonanno}} \emph {et~al.},\ }\bibfield  {title} {\bibinfo {title}
  {Stochastic gravitational-wave background from cosmological supernovae},\
  }\href {https://doi.org/10.1103/PhysRevD.72.084001} {\bibfield  {journal}
  {\bibinfo  {journal} {Phys. Rev. D}\ }\textbf {\bibinfo {volume} {72}},\
  \bibinfo {pages} {084001} (\bibinfo {year} {2005})}\BibitemShut {NoStop}%
\bibitem [{\citenamefont {Crocker}\ \emph {et~al.}(2015)\citenamefont {Crocker}
  \emph {et~al.}}]{crocker1}%
  \BibitemOpen
  \bibfield  {author} {\bibinfo {author} {\bibfnamefont {K.}~\bibnamefont
  {Crocker}} \emph {et~al.},\ }\bibfield  {title} {\bibinfo {title} {Model of
  the stochastic gravitational-wave background due to core collapse to black
  holes},\ }\href {https://doi.org/10.1103/PhysRevD.92.063005} {\bibfield
  {journal} {\bibinfo  {journal} {Phys. Rev. D}\ }\textbf {\bibinfo {volume}
  {92}},\ \bibinfo {pages} {063005} (\bibinfo {year} {2015})}\BibitemShut
  {NoStop}%
\bibitem [{\citenamefont {Crocker}\ \emph {et~al.}(2017)\citenamefont {Crocker}
  \emph {et~al.}}]{crocker2}%
  \BibitemOpen
  \bibfield  {author} {\bibinfo {author} {\bibfnamefont {K.}~\bibnamefont
  {Crocker}} \emph {et~al.},\ }\bibfield  {title} {\bibinfo {title} {Systematic
  study of the stochastic gravitational-wave background due to stellar core
  collapse},\ }\href {https://doi.org/10.1103/PhysRevD.95.063015} {\bibfield
  {journal} {\bibinfo  {journal} {Phys. Rev. D}\ }\textbf {\bibinfo {volume}
  {95}},\ \bibinfo {pages} {063015} (\bibinfo {year} {2017})}\BibitemShut
  {NoStop}%
\bibitem [{\citenamefont {Finkel}\ \emph {et~al.}(2022)\citenamefont {Finkel}
  \emph {et~al.}}]{finkel}%
  \BibitemOpen
  \bibfield  {author} {\bibinfo {author} {\bibfnamefont {B.}~\bibnamefont
  {Finkel}} \emph {et~al.},\ }\bibfield  {title} {\bibinfo {title} {Stochastic
  gravitational-wave background from stellar core-collapse events},\ }\href
  {https://doi.org/10.1103/PhysRevD.105.063022} {\bibfield  {journal} {\bibinfo
   {journal} {Phys. Rev. D}\ }\textbf {\bibinfo {volume} {105}},\ \bibinfo
  {pages} {063022} (\bibinfo {year} {2022})}\BibitemShut {NoStop}%
\bibitem [{\citenamefont {Contaldi}(2017)}]{Contaldi:2016koz}%
  \BibitemOpen
  \bibfield  {author} {\bibinfo {author} {\bibfnamefont {C.~R.}\ \bibnamefont
  {Contaldi}},\ }\bibfield  {title} {\bibinfo {title} {{Anisotropies of
  Gravitational Wave Backgrounds: A Line Of Sight Approach}},\ }\href
  {https://doi.org/10.1016/j.physletb.2017.05.020} {\bibfield  {journal}
  {\bibinfo  {journal} {Phys. Lett.}\ }\textbf {\bibinfo {volume} {B771}},\
  \bibinfo {pages} {9} (\bibinfo {year} {2017})},\ \Eprint
  {https://arxiv.org/abs/1609.08168} {arXiv:1609.08168 [astro-ph.CO]}
  \BibitemShut {NoStop}%
\bibitem [{\citenamefont {Cusin}\ \emph {et~al.}(2017)\citenamefont {Cusin},
  \citenamefont {Pitrou},\ and\ \citenamefont {Uzan}}]{Cusin:2017fwz}%
  \BibitemOpen
  \bibfield  {author} {\bibinfo {author} {\bibfnamefont {G.}~\bibnamefont
  {Cusin}}, \bibinfo {author} {\bibfnamefont {C.}~\bibnamefont {Pitrou}},\ and\
  \bibinfo {author} {\bibfnamefont {J.-P.}\ \bibnamefont {Uzan}},\ }\bibfield
  {title} {\bibinfo {title} {{Anisotropy of the astrophysical gravitational
  wave background: Analytic expression of the angular power spectrum and
  correlation with cosmological observations}},\ }\href
  {https://doi.org/10.1103/PhysRevD.96.103019} {\bibfield  {journal} {\bibinfo
  {journal} {Phys. Rev. D}\ }\textbf {\bibinfo {volume} {96}},\ \bibinfo
  {pages} {103019} (\bibinfo {year} {2017})},\ \Eprint
  {https://arxiv.org/abs/1704.06184} {arXiv:1704.06184 [astro-ph.CO]}
  \BibitemShut {NoStop}%
\bibitem [{\citenamefont {Cusin}\ \emph
  {et~al.}(2018{\natexlab{a}})\citenamefont {Cusin}, \citenamefont {Pitrou},\
  and\ \citenamefont {Uzan}}]{Cusin:2017mjm}%
  \BibitemOpen
  \bibfield  {author} {\bibinfo {author} {\bibfnamefont {G.}~\bibnamefont
  {Cusin}}, \bibinfo {author} {\bibfnamefont {C.}~\bibnamefont {Pitrou}},\ and\
  \bibinfo {author} {\bibfnamefont {J.-P.}\ \bibnamefont {Uzan}},\ }\bibfield
  {title} {\bibinfo {title} {{The signal of the gravitational wave background
  and the angular correlation of its energy density}},\ }\href
  {https://doi.org/10.1103/PhysRevD.97.123527} {\bibfield  {journal} {\bibinfo
  {journal} {Phys. Rev. D}\ }\textbf {\bibinfo {volume} {97}},\ \bibinfo
  {pages} {123527} (\bibinfo {year} {2018}{\natexlab{a}})},\ \Eprint
  {https://arxiv.org/abs/1711.11345} {arXiv:1711.11345 [astro-ph.CO]}
  \BibitemShut {NoStop}%
\bibitem [{\citenamefont {Cusin}\ and\ \citenamefont
  {Tasinato}(2022)}]{Cusin:2022cbb}%
  \BibitemOpen
  \bibfield  {author} {\bibinfo {author} {\bibfnamefont {G.}~\bibnamefont
  {Cusin}}\ and\ \bibinfo {author} {\bibfnamefont {G.}~\bibnamefont
  {Tasinato}},\ }\bibfield  {title} {\bibinfo {title} {{Doppler boosting the
  stochastic gravitational wave background}},\ }\href
  {https://doi.org/10.1088/1475-7516/2022/08/036} {\bibfield  {journal}
  {\bibinfo  {journal} {JCAP}\ }\textbf {\bibinfo {volume} {08}}\bibfield
  {number} {\bibinfo  {number} { (08)},\ \bibinfo {pages} {036}},\ }\Eprint
  {https://arxiv.org/abs/2201.10464} {arXiv:2201.10464 [astro-ph.CO]}
  \BibitemShut {NoStop}%
\bibitem [{\citenamefont {Chung}\ \emph {et~al.}(2022)\citenamefont {Chung},
  \citenamefont {Jenkins}, \citenamefont {Romano},\ and\ \citenamefont
  {Sakellariadou}}]{Chung:2022xhv}%
  \BibitemOpen
  \bibfield  {author} {\bibinfo {author} {\bibfnamefont {A.~K.-W.}\
  \bibnamefont {Chung}}, \bibinfo {author} {\bibfnamefont {A.~C.}\ \bibnamefont
  {Jenkins}}, \bibinfo {author} {\bibfnamefont {J.~D.}\ \bibnamefont
  {Romano}},\ and\ \bibinfo {author} {\bibfnamefont {M.}~\bibnamefont
  {Sakellariadou}},\ }\bibfield  {title} {\bibinfo {title} {{Targeted search
  for the kinematic dipole of the gravitational-wave background}},\ }\href
  {https://doi.org/10.1103/PhysRevD.106.082005} {\bibfield  {journal} {\bibinfo
   {journal} {Phys. Rev. D}\ }\textbf {\bibinfo {volume} {106}},\ \bibinfo
  {pages} {082005} (\bibinfo {year} {2022})},\ \Eprint
  {https://arxiv.org/abs/2208.01330} {arXiv:2208.01330 [gr-qc]} \BibitemShut
  {NoStop}%
\bibitem [{\citenamefont {Geller}\ \emph {et~al.}(2018)\citenamefont {Geller},
  \citenamefont {Hook}, \citenamefont {Sundrum},\ and\ \citenamefont
  {Tsai}}]{Geller:2018mwu}%
  \BibitemOpen
  \bibfield  {author} {\bibinfo {author} {\bibfnamefont {M.}~\bibnamefont
  {Geller}}, \bibinfo {author} {\bibfnamefont {A.}~\bibnamefont {Hook}},
  \bibinfo {author} {\bibfnamefont {R.}~\bibnamefont {Sundrum}},\ and\ \bibinfo
  {author} {\bibfnamefont {Y.}~\bibnamefont {Tsai}},\ }\bibfield  {title}
  {\bibinfo {title} {{Primordial Anisotropies in the Gravitational Wave
  Background from Cosmological Phase Transitions}},\ }\href
  {https://doi.org/10.1103/PhysRevLett.121.201303} {\bibfield  {journal}
  {\bibinfo  {journal} {Phys. Rev. Lett.}\ }\textbf {\bibinfo {volume} {121}},\
  \bibinfo {pages} {201303} (\bibinfo {year} {2018})},\ \Eprint
  {https://arxiv.org/abs/1803.10780} {arXiv:1803.10780 [hep-ph]} \BibitemShut
  {NoStop}%
\bibitem [{\citenamefont {Cusin}\ \emph
  {et~al.}(2018{\natexlab{b}})\citenamefont {Cusin}, \citenamefont {Dvorkin},
  \citenamefont {Pitrou},\ and\ \citenamefont {Uzan}}]{Cusin:2018rsq}%
  \BibitemOpen
  \bibfield  {author} {\bibinfo {author} {\bibfnamefont {G.}~\bibnamefont
  {Cusin}}, \bibinfo {author} {\bibfnamefont {I.}~\bibnamefont {Dvorkin}},
  \bibinfo {author} {\bibfnamefont {C.}~\bibnamefont {Pitrou}},\ and\ \bibinfo
  {author} {\bibfnamefont {J.-P.}\ \bibnamefont {Uzan}},\ }\bibfield  {title}
  {\bibinfo {title} {{First predictions of the angular power spectrum of the
  astrophysical gravitational wave background}},\ }\href
  {https://doi.org/10.1103/PhysRevLett.120.231101} {\bibfield  {journal}
  {\bibinfo  {journal} {Phys. Rev. Lett.}\ }\textbf {\bibinfo {volume} {120}},\
  \bibinfo {pages} {231101} (\bibinfo {year} {2018}{\natexlab{b}})},\ \Eprint
  {https://arxiv.org/abs/1803.03236} {arXiv:1803.03236 [astro-ph.CO]}
  \BibitemShut {NoStop}%
\bibitem [{\citenamefont {Cusin}\ \emph
  {et~al.}(2019{\natexlab{a}})\citenamefont {Cusin}, \citenamefont {Dvorkin},
  \citenamefont {Pitrou},\ and\ \citenamefont {Uzan}}]{Cusin:2019jpv}%
  \BibitemOpen
  \bibfield  {author} {\bibinfo {author} {\bibfnamefont {G.}~\bibnamefont
  {Cusin}}, \bibinfo {author} {\bibfnamefont {I.}~\bibnamefont {Dvorkin}},
  \bibinfo {author} {\bibfnamefont {C.}~\bibnamefont {Pitrou}},\ and\ \bibinfo
  {author} {\bibfnamefont {J.-P.}\ \bibnamefont {Uzan}},\ }\bibfield  {title}
  {\bibinfo {title} {{Properties of the stochastic astrophysical gravitational
  wave background: astrophysical sources dependencies}},\ }\href
  {https://doi.org/10.1103/PhysRevD.100.063004} {\bibfield  {journal} {\bibinfo
   {journal} {Phys. Rev. D}\ }\textbf {\bibinfo {volume} {100}},\ \bibinfo
  {pages} {063004} (\bibinfo {year} {2019}{\natexlab{a}})},\ \Eprint
  {https://arxiv.org/abs/1904.07797} {arXiv:1904.07797 [astro-ph.CO]}
  \BibitemShut {NoStop}%
\bibitem [{\citenamefont {Jenkins}\ \emph {et~al.}(2018)\citenamefont
  {Jenkins}, \citenamefont {Sakellariadou}, \citenamefont {Regimbau},\ and\
  \citenamefont {Slezak}}]{Jenkins:2018uac}%
  \BibitemOpen
  \bibfield  {author} {\bibinfo {author} {\bibfnamefont {A.~C.}\ \bibnamefont
  {Jenkins}}, \bibinfo {author} {\bibfnamefont {M.}~\bibnamefont
  {Sakellariadou}}, \bibinfo {author} {\bibfnamefont {T.}~\bibnamefont
  {Regimbau}},\ and\ \bibinfo {author} {\bibfnamefont {E.}~\bibnamefont
  {Slezak}},\ }\bibfield  {title} {\bibinfo {title} {{Anisotropies in the
  astrophysical gravitational-wave background: Predictions for the detection of
  compact binaries by {LIGO} and {Virgo}}},\ }\href
  {https://doi.org/10.1103/PhysRevD.98.063501} {\bibfield  {journal} {\bibinfo
  {journal} {Phys. Rev.}\ }\textbf {\bibinfo {volume} {D98}},\ \bibinfo {pages}
  {063501} (\bibinfo {year} {2018})},\ \Eprint
  {https://arxiv.org/abs/1806.01718} {arXiv:1806.01718 [astro-ph.CO]}
  \BibitemShut {NoStop}%
\bibitem [{\citenamefont {Jenkins}\ \emph
  {et~al.}(2019{\natexlab{a}})\citenamefont {Jenkins}, \citenamefont
  {O'Shaughnessy}, \citenamefont {Sakellariadou},\ and\ \citenamefont
  {Wysocki}}]{Jenkins:2018kxc}%
  \BibitemOpen
  \bibfield  {author} {\bibinfo {author} {\bibfnamefont {A.~C.}\ \bibnamefont
  {Jenkins}}, \bibinfo {author} {\bibfnamefont {R.}~\bibnamefont
  {O'Shaughnessy}}, \bibinfo {author} {\bibfnamefont {M.}~\bibnamefont
  {Sakellariadou}},\ and\ \bibinfo {author} {\bibfnamefont {D.}~\bibnamefont
  {Wysocki}},\ }\bibfield  {title} {\bibinfo {title} {{Anisotropies in the
  astrophysical gravitational-wave background: The impact of black hole
  distributions}},\ }\href {https://doi.org/10.1103/PhysRevLett.122.111101}
  {\bibfield  {journal} {\bibinfo  {journal} {Phys. Rev. Lett.}\ }\textbf
  {\bibinfo {volume} {122}},\ \bibinfo {pages} {111101} (\bibinfo {year}
  {2019}{\natexlab{a}})},\ \Eprint {https://arxiv.org/abs/1810.13435}
  {arXiv:1810.13435 [astro-ph.CO]} \BibitemShut {NoStop}%
\bibitem [{\citenamefont {Jenkins}\ and\ \citenamefont
  {Sakellariadou}(2018)}]{Jenkins:2018lvb}%
  \BibitemOpen
  \bibfield  {author} {\bibinfo {author} {\bibfnamefont {A.~C.}\ \bibnamefont
  {Jenkins}}\ and\ \bibinfo {author} {\bibfnamefont {M.}~\bibnamefont
  {Sakellariadou}},\ }\bibfield  {title} {\bibinfo {title} {{Anisotropies in
  the stochastic gravitational-wave background: Formalism and the cosmic string
  case}},\ }\href {https://doi.org/10.1103/PhysRevD.98.063509} {\bibfield
  {journal} {\bibinfo  {journal} {Phys. Rev.}\ }\textbf {\bibinfo {volume}
  {D98}},\ \bibinfo {pages} {063509} (\bibinfo {year} {2018})},\ \Eprint
  {https://arxiv.org/abs/1802.06046} {arXiv:1802.06046 [astro-ph.CO]}
  \BibitemShut {NoStop}%
\bibitem [{\citenamefont {Cusin}\ \emph {et~al.}(2020)\citenamefont {Cusin},
  \citenamefont {Dvorkin}, \citenamefont {Pitrou},\ and\ \citenamefont
  {Uzan}}]{Cusin:2019jhg}%
  \BibitemOpen
  \bibfield  {author} {\bibinfo {author} {\bibfnamefont {G.}~\bibnamefont
  {Cusin}}, \bibinfo {author} {\bibfnamefont {I.}~\bibnamefont {Dvorkin}},
  \bibinfo {author} {\bibfnamefont {C.}~\bibnamefont {Pitrou}},\ and\ \bibinfo
  {author} {\bibfnamefont {J.-P.}\ \bibnamefont {Uzan}},\ }\bibfield  {title}
  {\bibinfo {title} {{Stochastic gravitational wave background anisotropies in
  the mHz band: astrophysical dependencies}},\ }\href
  {https://doi.org/10.1093/mnrasl/slz182} {\bibfield  {journal} {\bibinfo
  {journal} {Mon. Not. Roy. Astron. Soc.}\ }\textbf {\bibinfo {volume} {493}},\
  \bibinfo {pages} {L1} (\bibinfo {year} {2020})},\ \Eprint
  {https://arxiv.org/abs/1904.07757} {arXiv:1904.07757 [astro-ph.CO]}
  \BibitemShut {NoStop}%
\bibitem [{\citenamefont {Jenkins}\ and\ \citenamefont
  {Sakellariadou}(2019)}]{Jenkins:2019uzp}%
  \BibitemOpen
  \bibfield  {author} {\bibinfo {author} {\bibfnamefont {A.~C.}\ \bibnamefont
  {Jenkins}}\ and\ \bibinfo {author} {\bibfnamefont {M.}~\bibnamefont
  {Sakellariadou}},\ }\bibfield  {title} {\bibinfo {title} {{Shot noise in the
  astrophysical gravitational-wave background}},\ }\href
  {https://doi.org/10.1103/PhysRevD.100.063508} {\bibfield  {journal} {\bibinfo
   {journal} {Phys. Rev.}\ }\textbf {\bibinfo {volume} {D100}},\ \bibinfo
  {pages} {063508} (\bibinfo {year} {2019})},\ \Eprint
  {https://arxiv.org/abs/1902.07719} {arXiv:1902.07719 [astro-ph.CO]}
  \BibitemShut {NoStop}%
\bibitem [{\citenamefont {Jenkins}\ \emph
  {et~al.}(2019{\natexlab{b}})\citenamefont {Jenkins}, \citenamefont {Romano},\
  and\ \citenamefont {Sakellariadou}}]{Jenkins:2019nks}%
  \BibitemOpen
  \bibfield  {author} {\bibinfo {author} {\bibfnamefont {A.~C.}\ \bibnamefont
  {Jenkins}}, \bibinfo {author} {\bibfnamefont {J.~D.}\ \bibnamefont
  {Romano}},\ and\ \bibinfo {author} {\bibfnamefont {M.}~\bibnamefont
  {Sakellariadou}},\ }\bibfield  {title} {\bibinfo {title} {{Estimating the
  angular power spectrum of the gravitational-wave background in the presence
  of shot noise}},\ }\href {https://doi.org/10.1103/PhysRevD.100.083501}
  {\bibfield  {journal} {\bibinfo  {journal} {Phys. Rev.}\ }\textbf {\bibinfo
  {volume} {D100}},\ \bibinfo {pages} {083501} (\bibinfo {year}
  {2019}{\natexlab{b}})},\ \Eprint {https://arxiv.org/abs/1907.06642}
  {arXiv:1907.06642 [astro-ph.CO]} \BibitemShut {NoStop}%
\bibitem [{\citenamefont {Cusin}\ \emph
  {et~al.}(2019{\natexlab{b}})\citenamefont {Cusin}, \citenamefont {Durrer},\
  and\ \citenamefont {Ferreira}}]{Cusin:2018avf}%
  \BibitemOpen
  \bibfield  {author} {\bibinfo {author} {\bibfnamefont {G.}~\bibnamefont
  {Cusin}}, \bibinfo {author} {\bibfnamefont {R.}~\bibnamefont {Durrer}},\ and\
  \bibinfo {author} {\bibfnamefont {P.~G.}\ \bibnamefont {Ferreira}},\
  }\bibfield  {title} {\bibinfo {title} {{Polarization of a stochastic
  gravitational wave background through diffusion by massive structures}},\
  }\href {https://doi.org/10.1103/PhysRevD.99.023534} {\bibfield  {journal}
  {\bibinfo  {journal} {Phys. Rev. D}\ }\textbf {\bibinfo {volume} {99}},\
  \bibinfo {pages} {023534} (\bibinfo {year} {2019}{\natexlab{b}})},\ \Eprint
  {https://arxiv.org/abs/1807.10620} {arXiv:1807.10620 [astro-ph.CO]}
  \BibitemShut {NoStop}%
\bibitem [{\citenamefont {Pitrou}\ \emph {et~al.}(2020)\citenamefont {Pitrou},
  \citenamefont {Cusin},\ and\ \citenamefont {Uzan}}]{Pitrou:2019rjz}%
  \BibitemOpen
  \bibfield  {author} {\bibinfo {author} {\bibfnamefont {C.}~\bibnamefont
  {Pitrou}}, \bibinfo {author} {\bibfnamefont {G.}~\bibnamefont {Cusin}},\ and\
  \bibinfo {author} {\bibfnamefont {J.-P.}\ \bibnamefont {Uzan}},\ }\bibfield
  {title} {\bibinfo {title} {{Unified view of anisotropies in the astrophysical
  gravitational-wave background}},\ }\href
  {https://doi.org/10.1103/PhysRevD.101.081301} {\bibfield  {journal} {\bibinfo
   {journal} {Phys. Rev. D}\ }\textbf {\bibinfo {volume} {101}},\ \bibinfo
  {pages} {081301} (\bibinfo {year} {2020})},\ \Eprint
  {https://arxiv.org/abs/1910.04645} {arXiv:1910.04645 [astro-ph.CO]}
  \BibitemShut {NoStop}%
\bibitem [{\citenamefont {Alonso}\ \emph
  {et~al.}(2020{\natexlab{a}})\citenamefont {Alonso}, \citenamefont {Cusin},
  \citenamefont {Ferreira},\ and\ \citenamefont {Pitrou}}]{Alonso:2020mva}%
  \BibitemOpen
  \bibfield  {author} {\bibinfo {author} {\bibfnamefont {D.}~\bibnamefont
  {Alonso}}, \bibinfo {author} {\bibfnamefont {G.}~\bibnamefont {Cusin}},
  \bibinfo {author} {\bibfnamefont {P.~G.}\ \bibnamefont {Ferreira}},\ and\
  \bibinfo {author} {\bibfnamefont {C.}~\bibnamefont {Pitrou}},\ }\bibfield
  {title} {\bibinfo {title} {{Detecting the anisotropic astrophysical
  gravitational wave background in the presence of shot noise through
  cross-correlations}},\ }\href {https://doi.org/10.1103/PhysRevD.102.023002}
  {\bibfield  {journal} {\bibinfo  {journal} {Phys. Rev. D}\ }\textbf {\bibinfo
  {volume} {102}},\ \bibinfo {pages} {023002} (\bibinfo {year}
  {2020}{\natexlab{a}})},\ \Eprint {https://arxiv.org/abs/2002.02888}
  {arXiv:2002.02888 [astro-ph.CO]} \BibitemShut {NoStop}%
\bibitem [{\citenamefont {Yang}\ \emph {et~al.}(2020)\citenamefont {Yang},
  \citenamefont {Mandic}, \citenamefont {Scarlata},\ and\ \citenamefont
  {Banagiri}}]{Yang:2020usq}%
  \BibitemOpen
  \bibfield  {author} {\bibinfo {author} {\bibfnamefont {K.~Z.}\ \bibnamefont
  {Yang}}, \bibinfo {author} {\bibfnamefont {V.}~\bibnamefont {Mandic}},
  \bibinfo {author} {\bibfnamefont {C.}~\bibnamefont {Scarlata}},\ and\
  \bibinfo {author} {\bibfnamefont {S.}~\bibnamefont {Banagiri}},\ }\bibfield
  {title} {\bibinfo {title} {{Searching for Cross-Correlation Between
  Stochastic Gravitational Wave Background and Galaxy Number Counts}},\ }\href
  {https://doi.org/10.1093/mnras/staa3159} {\bibfield  {journal} {\bibinfo
  {journal} {Mon. Not. Roy. Astron. Soc.}\ }\textbf {\bibinfo {volume} {500}},\
  \bibinfo {pages} {1666} (\bibinfo {year} {2020})},\ \Eprint
  {https://arxiv.org/abs/2007.10456} {arXiv:2007.10456 [astro-ph.CO]}
  \BibitemShut {NoStop}%
\bibitem [{\citenamefont {Capurri}\ \emph {et~al.}(2021)\citenamefont
  {Capurri}, \citenamefont {Lapi}, \citenamefont {Baccigalupi}, \citenamefont
  {Boco}, \citenamefont {Scelfo},\ and\ \citenamefont
  {Ronconi}}]{Capurri:2021zli}%
  \BibitemOpen
  \bibfield  {author} {\bibinfo {author} {\bibfnamefont {G.}~\bibnamefont
  {Capurri}}, \bibinfo {author} {\bibfnamefont {A.}~\bibnamefont {Lapi}},
  \bibinfo {author} {\bibfnamefont {C.}~\bibnamefont {Baccigalupi}}, \bibinfo
  {author} {\bibfnamefont {L.}~\bibnamefont {Boco}}, \bibinfo {author}
  {\bibfnamefont {G.}~\bibnamefont {Scelfo}},\ and\ \bibinfo {author}
  {\bibfnamefont {T.}~\bibnamefont {Ronconi}},\ }\bibfield  {title} {\bibinfo
  {title} {{Intensity and anisotropies of the stochastic gravitational wave
  background from merging compact binaries in galaxies}},\ }\href
  {https://doi.org/10.1088/1475-7516/2021/11/032} {\bibfield  {journal}
  {\bibinfo  {journal} {JCAP}\ }\textbf {\bibinfo {volume} {11}},\ \bibinfo
  {pages} {032}},\ \Eprint {https://arxiv.org/abs/2103.12037} {arXiv:2103.12037
  [gr-qc]} \BibitemShut {NoStop}%
\bibitem [{\citenamefont {Mukherjee}\ \emph {et~al.}(2022)\citenamefont
  {Mukherjee}, \citenamefont {Krolewski}, \citenamefont {Wandelt},\ and\
  \citenamefont {Silk}}]{Mukherjee:2022afz}%
  \BibitemOpen
  \bibfield  {author} {\bibinfo {author} {\bibfnamefont {S.}~\bibnamefont
  {Mukherjee}}, \bibinfo {author} {\bibfnamefont {A.}~\bibnamefont
  {Krolewski}}, \bibinfo {author} {\bibfnamefont {B.~D.}\ \bibnamefont
  {Wandelt}},\ and\ \bibinfo {author} {\bibfnamefont {J.}~\bibnamefont
  {Silk}},\ }\bibfield  {title} {\bibinfo {title} {{Cross-correlating dark
  sirens and galaxies: measurement of $H_0$ from {GWTC}-3 of
  {LIGO}-{V}irgo-{KAGRA}}},\ }\href@noop {} {\  (\bibinfo {year} {2022})},\
  \Eprint {https://arxiv.org/abs/2203.03643} {arXiv:2203.03643 [astro-ph.CO]}
  \BibitemShut {NoStop}%
\bibitem [{\citenamefont {Marin}\ \emph {et~al.}(2013)\citenamefont {Marin}
  \emph {et~al.}}]{WiggleZ:2013kor}%
  \BibitemOpen
  \bibfield  {author} {\bibinfo {author} {\bibfnamefont {F.~A.}\ \bibnamefont
  {Marin}} \emph {et~al.} (\bibinfo {collaboration} {WiggleZ}),\ }\bibfield
  {title} {\bibinfo {title} {{The WiggleZ Dark Energy Survey: constraining
  galaxy bias and cosmic growth with 3-point correlation functions}},\ }\href
  {https://doi.org/10.1093/mnras/stt520} {\bibfield  {journal} {\bibinfo
  {journal} {Mon. Not. Roy. Astron. Soc.}\ }\textbf {\bibinfo {volume} {432}},\
  \bibinfo {pages} {2654} (\bibinfo {year} {2013})},\ \Eprint
  {https://arxiv.org/abs/1303.6644} {arXiv:1303.6644 [astro-ph.CO]}
  \BibitemShut {NoStop}%
\bibitem [{\citenamefont {Rassat}\ \emph {et~al.}(2008)\citenamefont {Rassat},
  \citenamefont {Amara}, \citenamefont {Amendola}, \citenamefont {Castander},
  \citenamefont {Kitching}, \citenamefont {Kunz}, \citenamefont {Refregier},
  \citenamefont {Wang},\ and\ \citenamefont {Weller}}]{Rassat:2008ja}%
  \BibitemOpen
  \bibfield  {author} {\bibinfo {author} {\bibfnamefont {A.}~\bibnamefont
  {Rassat}}, \bibinfo {author} {\bibfnamefont {A.}~\bibnamefont {Amara}},
  \bibinfo {author} {\bibfnamefont {L.}~\bibnamefont {Amendola}}, \bibinfo
  {author} {\bibfnamefont {F.~J.}\ \bibnamefont {Castander}}, \bibinfo {author}
  {\bibfnamefont {T.}~\bibnamefont {Kitching}}, \bibinfo {author}
  {\bibfnamefont {M.}~\bibnamefont {Kunz}}, \bibinfo {author} {\bibfnamefont
  {A.}~\bibnamefont {Refregier}}, \bibinfo {author} {\bibfnamefont
  {Y.}~\bibnamefont {Wang}},\ and\ \bibinfo {author} {\bibfnamefont
  {J.}~\bibnamefont {Weller}},\ }\bibfield  {title} {\bibinfo {title}
  {{Deconstructing Baryon Acoustic Oscillations: A Comparison of Methods}},\
  }\href@noop {} {\  (\bibinfo {year} {2008})},\ \Eprint
  {https://arxiv.org/abs/0810.0003} {arXiv:0810.0003 [astro-ph]} \BibitemShut
  {NoStop}%
\bibitem [{\citenamefont {Ain}\ \emph {et~al.}(2015)\citenamefont {Ain},
  \citenamefont {Dalvi},\ and\ \citenamefont {Mitra}}]{folding}%
  \BibitemOpen
  \bibfield  {author} {\bibinfo {author} {\bibfnamefont {A.}~\bibnamefont
  {Ain}}, \bibinfo {author} {\bibfnamefont {P.}~\bibnamefont {Dalvi}},\ and\
  \bibinfo {author} {\bibfnamefont {S.}~\bibnamefont {Mitra}},\ }\bibfield
  {title} {\bibinfo {title} {{Fast Gravitational Wave Radiometry using Data
  Folding}},\ }\href {https://doi.org/10.1103/PhysRevD.92.022003} {\bibfield
  {journal} {\bibinfo  {journal} {Phys. Rev.}\ }\textbf {\bibinfo {volume}
  {D92}},\ \bibinfo {pages} {022003} (\bibinfo {year} {2015})},\ \Eprint
  {https://arxiv.org/abs/1504.01714} {arXiv:1504.01714 [gr-qc]} \BibitemShut
  {NoStop}%
\bibitem [{\citenamefont {Collaboration}\ \emph {et~al.}(2022)\citenamefont
  {Collaboration}, \citenamefont {Collaboration},\ and\ \citenamefont
  {Collaboration}}]{folded_data}%
  \BibitemOpen
  \bibfield  {author} {\bibinfo {author} {\bibfnamefont {L.~S.}\ \bibnamefont
  {Collaboration}}, \bibinfo {author} {\bibfnamefont {V.}~\bibnamefont
  {Collaboration}},\ and\ \bibinfo {author} {\bibfnamefont {K.}~\bibnamefont
  {Collaboration}},\ }\bibfield  {title} {\bibinfo {title} {{Folded data for
  first three observing runs of {A}dvanced {LIGO} and {A}dvanced {Virgo}}},\
  }\href {https://doi.org/10.5281/zenodo.6326656} {10.5281/zenodo.6326656}
  (\bibinfo {year} {2022})\BibitemShut {NoStop}%
\bibitem [{\citenamefont {Thrane}\ \emph {et~al.}(2009)\citenamefont {Thrane},
  \citenamefont {Ballmer}, \citenamefont {Romano}, \citenamefont {Mitra},
  \citenamefont {Talukder}, \citenamefont {Bose},\ and\ \citenamefont
  {Mandic}}]{Thrane:2009fp}%
  \BibitemOpen
  \bibfield  {author} {\bibinfo {author} {\bibfnamefont {E.}~\bibnamefont
  {Thrane}}, \bibinfo {author} {\bibfnamefont {S.}~\bibnamefont {Ballmer}},
  \bibinfo {author} {\bibfnamefont {J.~D.}\ \bibnamefont {Romano}}, \bibinfo
  {author} {\bibfnamefont {S.}~\bibnamefont {Mitra}}, \bibinfo {author}
  {\bibfnamefont {D.}~\bibnamefont {Talukder}}, \bibinfo {author}
  {\bibfnamefont {S.}~\bibnamefont {Bose}},\ and\ \bibinfo {author}
  {\bibfnamefont {V.}~\bibnamefont {Mandic}},\ }\bibfield  {title} {\bibinfo
  {title} {{Probing the anisotropies of a stochastic gravitational-wave
  background using a network of ground-based laser interferometers}},\ }\href
  {https://doi.org/10.1103/PhysRevD.80.122002} {\bibfield  {journal} {\bibinfo
  {journal} {Phys. Rev.}\ }\textbf {\bibinfo {volume} {D80}},\ \bibinfo {pages}
  {122002} (\bibinfo {year} {2009})},\ \Eprint
  {https://arxiv.org/abs/0910.0858} {arXiv:0910.0858 [astro-ph.IM]}
  \BibitemShut {NoStop}%
\bibitem [{\citenamefont {{Romano}}\ and\ \citenamefont
  {{Cornish}}(2017)}]{romanocornish}%
  \BibitemOpen
  \bibfield  {author} {\bibinfo {author} {\bibfnamefont {J.~D.}\ \bibnamefont
  {{Romano}}}\ and\ \bibinfo {author} {\bibfnamefont {N.~J.}\ \bibnamefont
  {{Cornish}}},\ }\bibfield  {title} {\bibinfo {title} {{Detection methods for
  stochastic gravitational-wave backgrounds: a unified treatment}},\ }\href
  {https://doi.org/10.1007/s41114-017-0004-1} {\bibfield  {journal} {\bibinfo
  {journal} {Living Reviews in Relativity}\ }\textbf {\bibinfo {volume} {20}},\
  \bibinfo {eid} {2} (\bibinfo {year} {2017})},\ \Eprint
  {https://arxiv.org/abs/1608.06889} {arXiv:1608.06889 [gr-qc]} \BibitemShut
  {NoStop}%
\bibitem [{\citenamefont {Mitra}\ \emph {et~al.}(2008)\citenamefont {Mitra}
  \emph {et~al.}}]{Mitra:2007mc}%
  \BibitemOpen
  \bibfield  {author} {\bibinfo {author} {\bibfnamefont {S.}~\bibnamefont
  {Mitra}} \emph {et~al.},\ }\bibfield  {title} {\bibinfo {title}
  {{Gravitational wave radiometry: Mapping a stochastic gravitational wave
  background}},\ }\href {https://doi.org/10.1103/PhysRevD.77.042002} {\bibfield
   {journal} {\bibinfo  {journal} {Phys. Rev. D}\ }\textbf {\bibinfo {volume}
  {77}},\ \bibinfo {pages} {042002} (\bibinfo {year} {2008})},\ \Eprint
  {https://arxiv.org/abs/0708.2728} {arXiv:0708.2728 [gr-qc]} \BibitemShut
  {NoStop}%
\bibitem [{\citenamefont {{Allen}}\ and\ \citenamefont
  {{Romano}}(1999)}]{allenromano}%
  \BibitemOpen
  \bibfield  {author} {\bibinfo {author} {\bibfnamefont {B.}~\bibnamefont
  {{Allen}}}\ and\ \bibinfo {author} {\bibfnamefont {J.~D.}\ \bibnamefont
  {{Romano}}},\ }\bibfield  {title} {\bibinfo {title} {{Detecting a stochastic
  background of gravitational radiation: Signal processing strategies and
  sensitivities}},\ }\href {https://doi.org/10.1103/PhysRevD.59.102001}
  {\bibfield  {journal} {\bibinfo  {journal} {\prd}\ }\textbf {\bibinfo
  {volume} {59}},\ \bibinfo {eid} {102001} (\bibinfo {year}
  {1999})}\BibitemShut {NoStop}%
\bibitem [{\citenamefont {Panda}\ \emph {et~al.}(2019)\citenamefont {Panda}
  \emph {et~al.}}]{Panda:2019hyg}%
  \BibitemOpen
  \bibfield  {author} {\bibinfo {author} {\bibfnamefont {S.}~\bibnamefont
  {Panda}} \emph {et~al.},\ }\bibfield  {title} {\bibinfo {title} {{Stochastic
  gravitational wave background mapmaking using regularized deconvolution}},\
  }\href {https://doi.org/10.1103/PhysRevD.100.043541} {\bibfield  {journal}
  {\bibinfo  {journal} {Phys. Rev. D}\ }\textbf {\bibinfo {volume} {100}},\
  \bibinfo {pages} {043541} (\bibinfo {year} {2019})},\ \Eprint
  {https://arxiv.org/abs/1905.08276} {arXiv:1905.08276 [gr-qc]} \BibitemShut
  {NoStop}%
\bibitem [{\citenamefont {Agarwal}\ \emph {et~al.}(2021)\citenamefont {Agarwal}
  \emph {et~al.}}]{Agarwal:2021gvz}%
  \BibitemOpen
  \bibfield  {author} {\bibinfo {author} {\bibfnamefont {D.}~\bibnamefont
  {Agarwal}} \emph {et~al.},\ }\bibfield  {title} {\bibinfo {title} {{Upper
  limits on persistent gravitational waves using folded data and the full
  covariance matrix from {A}dvanced {LIGO}\textquoteright{}s first two
  observing runs}},\ }\href {https://doi.org/10.1103/PhysRevD.104.123018}
  {\bibfield  {journal} {\bibinfo  {journal} {Phys. Rev. D}\ }\textbf {\bibinfo
  {volume} {104}},\ \bibinfo {pages} {123018} (\bibinfo {year} {2021})},\
  \Eprint {https://arxiv.org/abs/2105.08930} {arXiv:2105.08930 [gr-qc]}
  \BibitemShut {NoStop}%
\bibitem [{\citenamefont {Floden}\ \emph {et~al.}(2022)\citenamefont {Floden}
  \emph {et~al.}}]{Floden:2022scq}%
  \BibitemOpen
  \bibfield  {author} {\bibinfo {author} {\bibfnamefont {E.}~\bibnamefont
  {Floden}} \emph {et~al.},\ }\bibfield  {title} {\bibinfo {title} {{Angular
  resolution of the search for anisotropic stochastic gravitational-wave
  background with terrestrial gravitational-wave detectors}},\ }\href
  {https://doi.org/10.1103/PhysRevD.106.023010} {\bibfield  {journal} {\bibinfo
   {journal} {Phys. Rev. D}\ }\textbf {\bibinfo {volume} {106}},\ \bibinfo
  {pages} {023010} (\bibinfo {year} {2022})},\ \Eprint
  {https://arxiv.org/abs/2203.17141} {arXiv:2203.17141 [astro-ph.IM]}
  \BibitemShut {NoStop}%
\bibitem [{\citenamefont {Ain}\ \emph {et~al.}(2018)\citenamefont {Ain},
  \citenamefont {Suresh},\ and\ \citenamefont {Mitra}}]{pystoch}%
  \BibitemOpen
  \bibfield  {author} {\bibinfo {author} {\bibfnamefont {A.}~\bibnamefont
  {Ain}}, \bibinfo {author} {\bibfnamefont {J.}~\bibnamefont {Suresh}},\ and\
  \bibinfo {author} {\bibfnamefont {S.}~\bibnamefont {Mitra}},\ }\bibfield
  {title} {\bibinfo {title} {{Very fast stochastic gravitational wave
  background map making using folded data}},\ }\href
  {https://doi.org/10.1103/PhysRevD.98.024001} {\bibfield  {journal} {\bibinfo
  {journal} {Phys. Rev.}\ }\textbf {\bibinfo {volume} {D98}},\ \bibinfo {pages}
  {024001} (\bibinfo {year} {2018})},\ \Eprint
  {https://arxiv.org/abs/1803.08285} {arXiv:1803.08285 [gr-qc]} \BibitemShut
  {NoStop}%
\bibitem [{\citenamefont {Suresh}\ \emph {et~al.}(2021)\citenamefont {Suresh},
  \citenamefont {Ain},\ and\ \citenamefont {Mitra}}]{pystoch_sph}%
  \BibitemOpen
  \bibfield  {author} {\bibinfo {author} {\bibfnamefont {J.}~\bibnamefont
  {Suresh}}, \bibinfo {author} {\bibfnamefont {A.}~\bibnamefont {Ain}},\ and\
  \bibinfo {author} {\bibfnamefont {S.}~\bibnamefont {Mitra}},\ }\bibfield
  {title} {\bibinfo {title} {Unified mapmaking for an anisotropic stochastic
  gravitational wave background},\ }\href
  {https://doi.org/10.1103/PhysRevD.103.083024} {\bibfield  {journal} {\bibinfo
   {journal} {Phys. Rev. D}\ }\textbf {\bibinfo {volume} {103}},\ \bibinfo
  {pages} {083024} (\bibinfo {year} {2021})}\BibitemShut {NoStop}%
\bibitem [{\citenamefont {Abbott}\ \emph
  {et~al.}(2021{\natexlab{b}})\citenamefont {Abbott}, \citenamefont {Abbott}
  \emph {et~al.}}]{O3directional}%
  \BibitemOpen
  \bibfield  {author} {\bibinfo {author} {\bibfnamefont {R.}~\bibnamefont
  {Abbott}}, \bibinfo {author} {\bibfnamefont {T.~D.}\ \bibnamefont {Abbott}},
  \emph {et~al.} (\bibinfo {collaboration} {LIGO Scientific Collaboration,
  Virgo Collaboration, and KAGRA Collaboration}),\ }\bibfield  {title}
  {\bibinfo {title} {Search for anisotropic gravitational-wave backgrounds
  using data from advanced ligo and advanced virgo's first three observing
  runs},\ }\href {https://doi.org/10.1103/PhysRevD.104.022005} {\bibfield
  {journal} {\bibinfo  {journal} {Phys. Rev. D}\ }\textbf {\bibinfo {volume}
  {104}},\ \bibinfo {pages} {022005} (\bibinfo {year}
  {2021}{\natexlab{b}})}\BibitemShut {NoStop}%
\bibitem [{\citenamefont {{Agarwal}}\ \emph {et~al.}(2023)\citenamefont
  {{Agarwal}}, \citenamefont {{Suresh}}, \citenamefont {{Mitra}},\ and\
  \citenamefont {{Ain}}}]{Agarwal:2023lzz}%
  \BibitemOpen
  \bibfield  {author} {\bibinfo {author} {\bibfnamefont {D.}~\bibnamefont
  {{Agarwal}}}, \bibinfo {author} {\bibfnamefont {J.}~\bibnamefont {{Suresh}}},
  \bibinfo {author} {\bibfnamefont {S.}~\bibnamefont {{Mitra}}},\ and\ \bibinfo
  {author} {\bibfnamefont {A.}~\bibnamefont {{Ain}}},\ }\bibfield  {title}
  {\bibinfo {title} {{Angular power spectra of anisotropic stochastic
  gravitational wave background: developing statistical methods and analyzing
  data from ground-based detectors}},\ }\href
  {https://doi.org/10.48550/arXiv.2302.12516} {\bibfield  {journal} {\bibinfo
  {journal} {arXiv e-prints}\ ,\ \bibinfo {eid} {arXiv:2302.12516}} (\bibinfo
  {year} {2023})},\ \Eprint {https://arxiv.org/abs/2302.12516}
  {arXiv:2302.12516 [gr-qc]} \BibitemShut {NoStop}%
\bibitem [{\citenamefont {{Ahumada}}\ \emph {et~al.}(2020)\citenamefont
  {{Ahumada}} \emph {et~al.}}]{SDSS_DR16}%
  \BibitemOpen
  \bibfield  {author} {\bibinfo {author} {\bibfnamefont {R.}~\bibnamefont
  {{Ahumada}}} \emph {et~al.},\ }\bibfield  {title} {\bibinfo {title} {{The
  16th Data Release of the Sloan Digital Sky Surveys: First Release from the
  APOGEE-2 Southern Survey and Full Release of eBOSS Spectra}},\ }\href
  {https://doi.org/10.3847/1538-4365/ab929e} {\bibfield  {journal} {\bibinfo
  {journal} {The Astrophysical Journal Supplement}\ }\textbf {\bibinfo {volume}
  {249}},\ \bibinfo {eid} {3} (\bibinfo {year} {2020})},\ \Eprint
  {https://arxiv.org/abs/1912.02905} {arXiv:1912.02905 [astro-ph.GA]}
  \BibitemShut {NoStop}%
\bibitem [{\citenamefont {{Gorski}}\ \emph {et~al.}(1999)\citenamefont
  {{Gorski}}, \citenamefont {{Wandelt}}, \citenamefont {{Hansen}},
  \citenamefont {{Hivon}},\ and\ \citenamefont {{Banday}}}]{Gorski:1999rt}%
  \BibitemOpen
  \bibfield  {author} {\bibinfo {author} {\bibfnamefont {K.~M.}\ \bibnamefont
  {{Gorski}}}, \bibinfo {author} {\bibfnamefont {B.~D.}\ \bibnamefont
  {{Wandelt}}}, \bibinfo {author} {\bibfnamefont {F.~K.}\ \bibnamefont
  {{Hansen}}}, \bibinfo {author} {\bibfnamefont {E.}~\bibnamefont {{Hivon}}},\
  and\ \bibinfo {author} {\bibfnamefont {A.~J.}\ \bibnamefont {{Banday}}},\
  }\bibfield  {title} {\bibinfo {title} {{The HEALPix Primer}},\ }\href@noop {}
  {\bibfield  {journal} {\bibinfo  {journal} {arXiv e-prints}\ ,\ \bibinfo
  {eid} {astro-ph/9905275}} (\bibinfo {year} {1999})},\ \Eprint
  {https://arxiv.org/abs/astro-ph/9905275} {arXiv:astro-ph/9905275 [astro-ph]}
  \BibitemShut {NoStop}%
\bibitem [{\citenamefont {Alonso}\ \emph
  {et~al.}(2020{\natexlab{b}})\citenamefont {Alonso}, \citenamefont {Contaldi},
  \citenamefont {Cusin}, \citenamefont {Ferreira},\ and\ \citenamefont
  {Renzini}}]{Alonso:2020rar}%
  \BibitemOpen
  \bibfield  {author} {\bibinfo {author} {\bibfnamefont {D.}~\bibnamefont
  {Alonso}}, \bibinfo {author} {\bibfnamefont {C.~R.}\ \bibnamefont
  {Contaldi}}, \bibinfo {author} {\bibfnamefont {G.}~\bibnamefont {Cusin}},
  \bibinfo {author} {\bibfnamefont {P.~G.}\ \bibnamefont {Ferreira}},\ and\
  \bibinfo {author} {\bibfnamefont {A.~I.}\ \bibnamefont {Renzini}},\
  }\bibfield  {title} {\bibinfo {title} {{Noise angular power spectrum of
  gravitational wave background experiments}},\ }\href
  {https://doi.org/10.1103/PhysRevD.101.124048} {\bibfield  {journal} {\bibinfo
   {journal} {Phys. Rev. D}\ }\textbf {\bibinfo {volume} {101}},\ \bibinfo
  {pages} {124048} (\bibinfo {year} {2020}{\natexlab{b}})},\ \Eprint
  {https://arxiv.org/abs/2005.03001} {arXiv:2005.03001 [astro-ph.CO]}
  \BibitemShut {NoStop}%
\bibitem [{\citenamefont {Bruni}\ \emph {et~al.}(2012)\citenamefont {Bruni},
  \citenamefont {Crittenden}, \citenamefont {Koyama}, \citenamefont {Maartens},
  \citenamefont {Pitrou},\ and\ \citenamefont {Wands}}]{Bruni:2011ta}%
  \BibitemOpen
  \bibfield  {author} {\bibinfo {author} {\bibfnamefont {M.}~\bibnamefont
  {Bruni}}, \bibinfo {author} {\bibfnamefont {R.}~\bibnamefont {Crittenden}},
  \bibinfo {author} {\bibfnamefont {K.}~\bibnamefont {Koyama}}, \bibinfo
  {author} {\bibfnamefont {R.}~\bibnamefont {Maartens}}, \bibinfo {author}
  {\bibfnamefont {C.}~\bibnamefont {Pitrou}},\ and\ \bibinfo {author}
  {\bibfnamefont {D.}~\bibnamefont {Wands}},\ }\bibfield  {title} {\bibinfo
  {title} {{Disentangling non-Gaussianity, bias and GR effects in the galaxy
  distribution}},\ }\href {https://doi.org/10.1103/PhysRevD.85.041301}
  {\bibfield  {journal} {\bibinfo  {journal} {Phys. Rev. D}\ }\textbf {\bibinfo
  {volume} {85}},\ \bibinfo {pages} {041301} (\bibinfo {year} {2012})},\
  \Eprint {https://arxiv.org/abs/1106.3999} {arXiv:1106.3999 [astro-ph.CO]}
  \BibitemShut {NoStop}%
\bibitem [{\citenamefont {et~al.}(2010)}]{ET}%
  \BibitemOpen
  \bibfield  {author} {\bibinfo {author} {\bibfnamefont {M.~P.}\ \bibnamefont
  {et~al.}},\ }\bibfield  {title} {\bibinfo {title} {The einstein telescope: a
  third-generation gravitational wave observatory},\ }\href
  {http://iopscience.iop.org/0264-9381/27/19/194002/} {\bibfield  {journal}
  {\bibinfo  {journal} {Classical and Quantum Gravity}\ }\textbf {\bibinfo
  {volume} {27}} (\bibinfo {year} {2010})}\BibitemShut {NoStop}%
\bibitem [{\citenamefont {Abbott}\ \emph {et~al.}(2017)\citenamefont {Abbott}
  \emph {et~al.}}]{CE}%
  \BibitemOpen
  \bibfield  {author} {\bibinfo {author} {\bibfnamefont {B.}~\bibnamefont
  {Abbott}} \emph {et~al.},\ }\bibfield  {title} {\bibinfo {title} {Exploring
  the sensitivity of next generation gravitational wave detectors},\ }\href
  {https://doi.org/10.1088/1361-6382/aa51f4} {\bibfield  {journal} {\bibinfo
  {journal} {Classical and Quantum Gravity}\ }\textbf {\bibinfo {volume}
  {34}},\ \bibinfo {pages} {044001} (\bibinfo {year} {2017})}\BibitemShut
  {NoStop}%
\bibitem [{\citenamefont {{Banagiri}}\ \emph {et~al.}(2020)\citenamefont
  {{Banagiri}}, \citenamefont {{Mandic}}, \citenamefont {{Scarlata}},\ and\
  \citenamefont {{Yang}}}]{Sharan}%
  \BibitemOpen
  \bibfield  {author} {\bibinfo {author} {\bibfnamefont {S.}~\bibnamefont
  {{Banagiri}}}, \bibinfo {author} {\bibfnamefont {V.}~\bibnamefont
  {{Mandic}}}, \bibinfo {author} {\bibfnamefont {C.}~\bibnamefont
  {{Scarlata}}},\ and\ \bibinfo {author} {\bibfnamefont {K.~Z.}\ \bibnamefont
  {{Yang}}},\ }\bibfield  {title} {\bibinfo {title} {{Measuring angular N-point
  correlations of binary black-hole merger gravitational-wave events with
  hierarchical Bayesian inference}},\ }\href@noop {} {\bibfield  {journal}
  {\bibinfo  {journal} {arXiv e-prints}\ ,\ \bibinfo {eid} {arXiv:2006.00633}}
  (\bibinfo {year} {2020})},\ \Eprint {https://arxiv.org/abs/2006.00633}
  {arXiv:2006.00633 [astro-ph.CO]} \BibitemShut {NoStop}%
\end{thebibliography}%
\end{document}